\documentclass[namedreferences]{SolarPhysics}
\usepackage[optionalrh,solaromanenum]{spr-sola-addons} 
\usepackage{graphicx}                    
\usepackage{color}                       
\usepackage{url}                         
\usepackage{mathrsfs}
\bibpunct{(}{)}{;}{}{,}{,}
\usepackage[margin=1.25in]{geometry}
\usepackage[pdfborder={0 0 0 },urlcolor=blue,breaklinks]{hyperref}
\usepackage{listings}

\ifx \urlurl    \undefined \def \urlurl#1{\href{http://#1}{\textsf{#1}}}\fi
\ifx \doiurl    \undefined \def \doiurl#1{\href{http://dx.doi.org/#1}{\textsf{\textsf{DOI}}}}\fi
\ifx \adsurl    \undefined \def \adsurl#1{\href{http://adsabs.harvard.edu/abs/#1}{\textsf{\textsf{ADS}}}}\fi
\ifx \arxivurl  \undefined \def \arxivurl#1{\href{http://arxiv.org/abs/#1}{\textsf{\textsf{arXiv}}}}\fi

\def\revtxtb{}

\def\revtxtr{}

\begin{document}
\begin{article}


\begin{opening}

\title{On-Orbit Performance of the \textit{Helioseismic and Magnetic Imager}
       Instrument onboard the \textit{Solar Dynamics Observatory}}
\author{
J.T.~\surname{Hoeksema}$^{1}$\sep
C.S.~\surname{Baldner}$^{1}$\sep
R.I.~\surname{Bush}$^{1}$\sep
J.~\surname{Schou}$^{2}$\sep
P.H.~\surname{Scherrer}$^{1}$\sep
}
\institute{$^{1}$ W.W. Hansen Experimental Physics Laboratory, Stanford University, Stanford,
CA 94305 USA
        Email: \href{mailto:jthoeksema@sun.stanford.edu}{jthoeksema@sun.stanford.edu} \\
        $^{2}$ Max-Planck-Institut f\"ur Sonnensystemforschung, Justus-von-Liebig-Weg 3,
37077 G\"ottingen, Germany\\ }

\date{Received: 7 December 2017; Revised: 26 January 2018}
\runningauthor{J.T. Hoeksema {\textit {et al.}}}
\runningtitle{HMI On-Orbit Performance}

\begin{abstract}

The \textit{Helioseismic and Magnetic Imager} (HMI) instrument is a major
component of NASA's \textit{Solar Dynamics Observatory} (SDO) spacecraft.
Since commencement of full regular science operations on 1 May 2010, HMI has
operated with remarkable continuity, \textit{e.g.} during the more than five years
of the SDO prime mission that ended 30 September 2015, HMI collected 
98.4\,\% of all possible 45-second velocity maps; minimizing gaps in these 
full-disk Dopplergrams is crucial for helioseismology. HMI velocity, intensity, and
magnetic-field measurements are used in numerous investigations, so understanding
the quality of the data is important. This article describes the calibration
measurements used to track the performance of the HMI instrument, and it details
trends in important instrument parameters during the prime mission. Regular
calibration sequences provide information used to improve and update the
calibration of HMI data. The set-point temperature of the instrument front
window and optical bench is adjusted regularly to maintain instrument focus,
and changes in the temperature-control scheme have been made to improve
stability in the observable quantities. The exposure time has been changed to
compensate for a 15\,\% decrease in instrument throughput.  Measurements of
the performance of the shutter and tuning mechanisms show that they are
aging as expected and continue to perform according to specification.
Parameters of the tunable-optical-filter elements are regularly adjusted
to account for drifts in the central wavelength. Frequent measurements of
changing CCD-camera characteristics, such as gain and flat field, are used to
calibrate the observations. Infrequent expected events, such as eclipses, transits,
and spacecraft off-points interrupt regular instrument operations and provide the
opportunity to perform additional calibration. Onboard instrument anomalies
are rare and seem to occur quite uniformly in time. The instrument
continues to perform very well.

\end{abstract}


\end{opening}

\section{Introduction}
\label{sec:intro}

The \textit{Solar Dynamics Observatory} (SDO) with the \textit{Helioseismic
and Magnetic Imager} (HMI) instrument onboard was launched 11 February 2010
to provide the observations necessary to understand the sources of solar
variability and its impact on the terrestrial environment \citep{Pesnell2012,
Scherrer2012}. Since 1 May 2010, HMI has observed the full disk of the Sun
almost continuously to measure the velocity, intensity, and magnetic field
in the photosphere \citep{Schou2012}. As of October 2016, nearly 1100 refereed 
articles had made use of HMI data. 
This article describes how the instrument has performed.

HMI operates using two $4096 \times 4096$ CCD cameras to take sequences of
polarized filtergrams of the photosphere. The full-disk images, tuned to six
wavelengths across the Fe\,{\sc i} 6173.3433\,\AA\ spectral line in each of
six polarization states, are downlinked and combined to determine the basic
HMI observable quantities: Doppler velocity, line-of-sight magnetic field,
line width, line depth, continuum intensity, and the Stokes polarization
parameters \citep{Couvidat2016}. More advanced products computed from those
observables include vector magnetograms \citep{Hoeksema2014} and 
subsurface-flow maps \citep{JZhao2012}.

\subsection{HMI Filtergram Data Processing and Calibration}

SDO data are collected continuously at a ground station in White Sands,
New Mexico, and the HMI and \textit{Atmospheric Imaging Assembly} (AIA)
housekeeping and science-data telemetry packets are transfered in near real
time to the Joint Science Operations Center (JSOC) Science Data Processing
facility at Stanford University. The HMI processing pipeline produces several
levels of data products from the incoming 55 megabit-per-second data stream.

The raw HMI bit stream is initially converted into Level-0 images (\textsf{Lev0}), 
and all of the relevant metadata are extracted. 

Image-specific calibrations are applied during the creation of the Level-1
filtergrams. One of the main objectives of this article is to describe
those calibrations and the on-orbit measurements made to enable them. 
CCD overscan rows and columns 
{\revtxtr (extra values returned for pixels that are not part of the image)}
are removed from the imagest this stage,
the CCD dark current is subtracted, and a flat field is applied.
A limb-finder
algorithm estimates the Sun-center location and the solar radius of each image. 
Another software module is applied to detect cosmic-ray hits
and identify bad pixels. The resulting polarized filtergram images, with
their lists of bad pixels, are termed Level-1 data (\textsf{Lev1}).

Other corrections (for image distortion, wavelength differences, and
polarization cross talk) are made later, at the point when filtergrams are
combined during the computation of the scientific observables as described
by \citet{Couvidat2016}. However, the calibration observations that enable
those calibrations are described here.

Initial calibrations of HMI were carried out before launch to assess the
performance of the wavelength-filter system \citep{Couvidat2012}, polarization
system \citep{Schou-Polarization2012}, and imaging optics \citep{Wachter2012}.
Here we detail how the instrument has been operated, monitored, 
calibrated, and adjusted since launch.
\citet{Schou2012}, \citet{CRW2012}, and \citet{Couvidat2016} describe the
HMI data processing required to compute the observable quantities from the
filtergrams. Additional systematic calibration issues determined after
launch are addressed by \citet{Liu2012} (line-of-sight magnetic field),
\citet{Hoeksema2014} and \citet{Bobra2014} (vector magnetic field), and
\citet{Kuhn2012} (limb shape).

\subsection{Overall HMI Data Recovery}

An important requirement for HMI is high observing continuity, the strongest 
driver being the need for precise determination of solar-oscillation frequencies 
for helioseismology. 

After two and a half months of commissioning, the HMI instrument 
formally began full science operations on 1 May 2010, although some data products 
are available prior to that date. Since then HMI has operated
almost continuously. Most interruptions are either planned, in order to 
accommodate spacecraft operations and calibrations, or due to unavoidable 
seasonal eclipses that are a consequence of the SDO geosynchronous orbit.

HMI acquired more than 112 million images from 1 May 2010 to 31 December 2016.
Table \ref{table:lev0_statistics} reports the total number of Level-0 images, as well 
as the numbers of 4K\,$\times$\,4K images that are missing or partially recovered. 
Images deliberately not collected during the dark phase of eclipses
are not reported as missing in the table.
About 1.19\,\% of the images were taken with the Image Stabilization System
(ISS) turned off during some spacecraft maneuvers and around the time of
eclipses.

\begin{table}[htb]
\begin{tabular}{rrr}
\hline
Parameter & Number of Images & Percentage\\[4pt]
\hline
Total HMI Exposures   & 112,043,265 &  \\ 
Missing Images &     61,563 & 0.055\,\% \\
Partial Images &     23,698 & 0.021\,\% \\
\hline
\end{tabular}
\caption{HMI Level-0 Image Recovery Completeness; 1 May 2010 -- 31 December 2016}
\label{table:lev0_statistics}
\end{table}

A more relevant statistic may be the number of Dopplergrams
recovered during the mission. Dopplergrams, one of the prime HMI \textit{observables},
are computed every 45 seconds using filtergrams obtained by one of the two HMI cameras.
This camera is variously referred to as the front camera, the Doppler camera, or Camera 2.
The other camera is called the side camera, vector camera, or Camera 1.
As shown in Table \ref{table:dopplergram_recovery}, more than 98\,\% of all possible
Dopplergrams have been recovered during the first five years of the mission.
An overall assessment of the quality of each Dopplergram appears in the 
\textsf{QUALITY} keyword.  A zero value for \textsf{QUALITY} indicates that there are
no known issues with the data; these are reported as {\em good} in Table
\ref{table:dopplergram_recovery}.  In fact all HMI data products at every
processing level include a \textsf{QUALITY} assessment.  Each bit in the 
\textsf{QUALITY} keyword indicates an issue that might affect the data. The top bit
indicates the data are missing and other non-zero bits indicate lesser quality
or explain why data are not present. This is discussed further in Section
\ref{sec:Quality} and detailed {\revtxtr in Appendix
\ref{sec:Lev0Qual}, \ref{sec:Lev1Qual}, and \ref{sec:S_ObsQual}.}
Because sensitivity to various subtle differences in the
data collection and processing vary depending on the analysis, the instrument
conditions, data-processing details, calibration-procedure versions, and a
host of other quantities are all available in keywords.

Table \ref{table:72day_recovery} in Appendix \ref{sec:Appendix1} provides
details of the Dopplergram recovery rate for each of the first 37 72-day
intervals. The lowest percentages ordinarily occur during eclipse seasons
in Spring and Fall. The lowest was 96.45\,\% in June--August 2016. The highest
was 99.87\,\% in November 2016--January 2017.

\begin{table}[htb]
\begin{tabular}{rrr}
\hline
Parameter & Value & Fraction \\[3pt]
\hline
Possible 45s Time Slots & 4,679,040 & 100.0\,\% \\
Good Dopplergrams & 4,505,062 & 96.28\,\% \\
Lower-Quality Dopplergrams & 95,484 & 2.04\,\% \\
Missing Dopplergrams & 78,494 & 1.68\,\% \\
\hline
\end{tabular}
\caption{Recovery of 45-second HMI Dopplergrams; 1 May 2010 to 31 December 2016
}
\label{table:dopplergram_recovery}
\end{table}

\begin{figure}[htb]
\centering
\includegraphics[width=\textwidth]{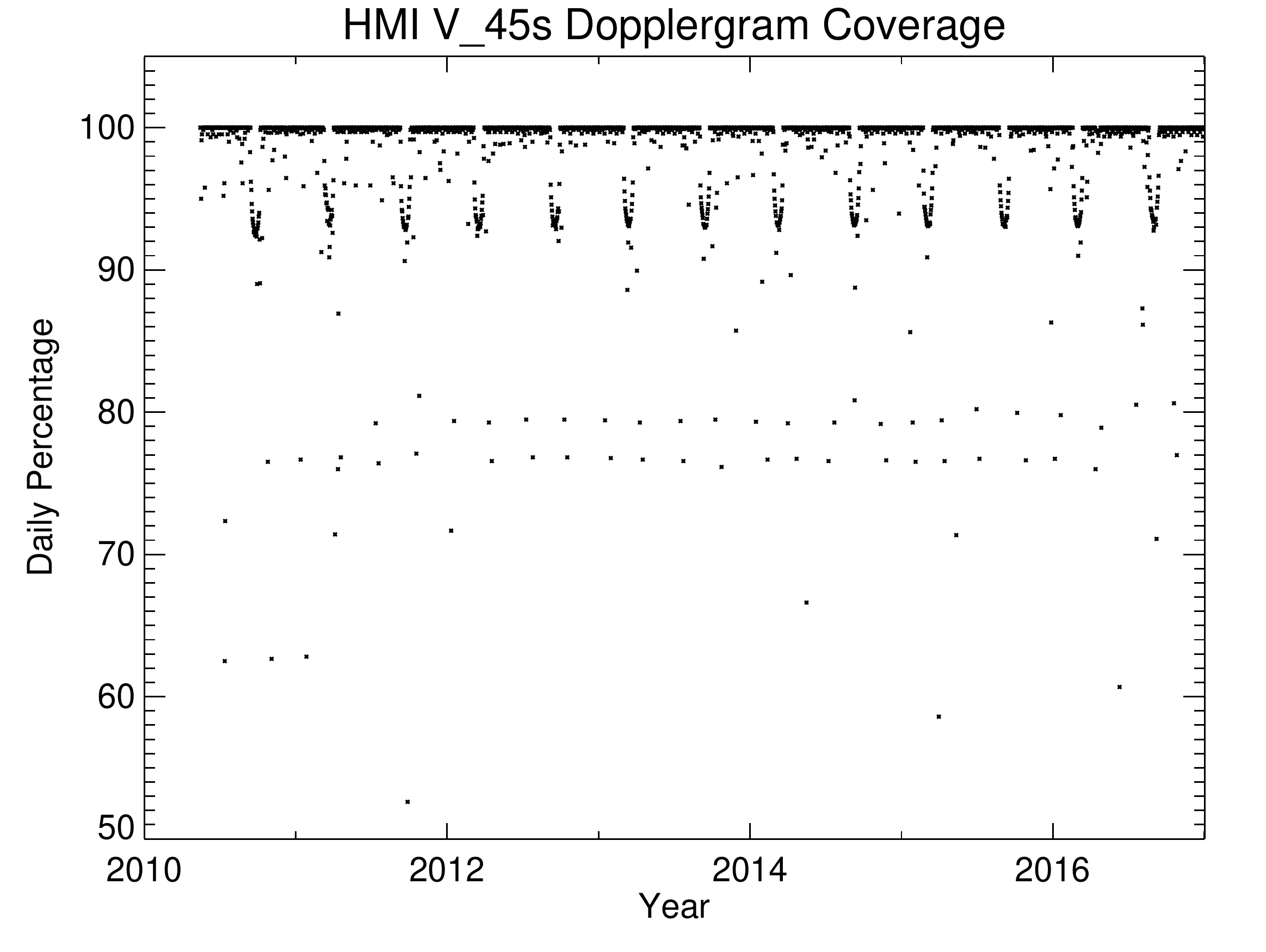}
\caption{HMI Dopplergram recovery during the mission. The daily percentage of all possible 
good-quality 45-second Dopplergrams recovered is plotted as a function of time from 1 May 2010 to 
31 December 2016. On only 79 days were fewer than 90\,\% of all possible Dopplergrams recovered
{\revtxtr and only five days had less than 50\,\% coverage.}}
\label{figure:dopplergram_recovery}
\end{figure}

Figure \ref{figure:dopplergram_recovery} shows the percentage of the 1920
possible 45-second Dopplergrams recovered each day.  Most days are nearly
perfect; only 321 had less than 95\,\% recovery.  The semi-annual eclipse
seasons can be seen as U-shaped dips to below 95\,\% that extend over a couple
weeks each Spring and Fall when the Earth comes between the spacecraft and
the Sun for up to 72 minutes each day. Gaps that can last as long as several
hours occur regularly on a few days each quarter when spacecraft operations are
scheduled. Occasional dips are deeper when there are special calibrations. On
a few occasions there have been instrument or spacecraft anomalies that have
taken longer to recover from. Section \ref{sec:eventsanomalies} provides
more information about such events.

\subsection{Outline}

The purpose of this article is to explain the observations used to calibrate
the HMI filtergrams and to characterize the basic performance of the
HMI instrument after launch and how it changes with time. This includes
consideration of quantities such as throughput, focus, wavelength, and
overall data capture, as well as trends in important instrument parameters,
such as camera operation, shutter and tuning-motor performance, and subsystem
temperatures.

Section \ref{sec:calibobs} describes the routine calibration observations
made in order to monitor and optimize the operation of the
instrument. Section \ref{sec:trending} explains various measurements that
show how the instrument has changed over time or responded to events. Section
\ref{sec:opticsfilters} addresses the calibration of the optics and filter
systems. In Section \ref{sec:level1} we describe the Level-1 processing
that produces calibrated filtergrams from Level-0 images, principally the
calibrations related to the CCD cameras (flat fields and bad pixels) but
also single-pixel corrections for transient problems, such as those caused by
cosmic rays. This section also summarizes how characteristics of the image
and information about the processing are documented in keywords and encoded in the bits of the
\textsf{QUALITY} and \textsf{CALVER} keywords. The implications of events (such
as the semiannual eclipses) and occasonal onboard anomalies are covered
in Section \ref{sec:eventsanomalies}.  Section \ref{sec:conclusion} gives
a summary and discussion of HMI performance. Appendices provide an additional
level of detail about observing sequences used for both primary observing and
for calibrations {\revtxtr as well as } annotated
descriptions of more of the keywords for Level-0 and Level-1 filtergrams.

\section{On-Orbit Calibration Observations}
\label{sec:calibobs}

A variety of calibration observations are taken on a regular basis to monitor the 
evolution of the HMI instrument and maintain optimal performance. This section 
describes the daily, weekly, bi-weekly, and occasional calibration sequences. 

{\revtxtb HMI acquires data using a Framelist Timeline Specification (FTS), 
or framelist. The FTS defines the filter tuning, polarization state, focus, and timing
of each filtergram to be executed in a sequence. 
The FTS ID is stored in the Level-0 and Level-1 keyword \textsf{HFTSACID}.
A roster of the most common frame lists appears in a table 
in Appendix \ref{sec:MainSequences}; more complete listings are provided 
in Appendix \ref{sec:FTSTables}.
The FTS IDs for standard calibration sequences are indicated. 

Standard HMI observations were initially obtained with a framelist called Mod C that 
repeated every 135 seconds. Mod L, a 90-second FTS, replaced Mod C on 13 April 2016.
The two versions of Mod C have FTS ID 1001 or 1021; the Mod L \textsf{HFTSACID} is 1022. 
Some calibration framelists changed when the standard sequences changed.
}

\subsection{Twice-Daily Calibration Sequences}

Twice a day, starting at 06~UT and 18~UT, the regular observing sequence is
interrupted to run a calibration that includes eight non-standard filtergrams.
At these times, near local Noon and Midnight in the orbit, the spacecraft is
close to zero radial velocity with respect to the Sun (the exact time varies
throughout the year). The sequence consists of four (nearly) true continuum
images (tuned such that the filter pass bands are
about 344\,m\AA\ away from the Fe~{\sc i} line center at rest)
taken in two different polarizations, two Calmode images (that is,
images taken with the instrument completely defocused in calibration mode),
and two dark frames. The continuum frames are not used for calibration purposes
but have been used for some scientific investigations. The Calmode images
are used to track the evolution of the throughput of the optical system;
the dark images are used to create mean dark frames four times a year (see
\ref{sec:level1}). 
The normal line-of-sight observing sequence in Camera 2 is minimally disturbed.
During mod-C (135-second cadence) operations, the FTS ID was
2021; under current mod-L operations the FTS ID is 2042. 

\subsection{Weekly Focus Sweeps and PZT Offpoints}

Additional calibration sequences are run every week, typically on Tuesdays
and Wednesdays around 19:00~UT, although they are sometimes rescheduled or
canceled due to conflicts with other events.

\begin{figure}[htb]
\centering
\includegraphics[width=\textwidth]{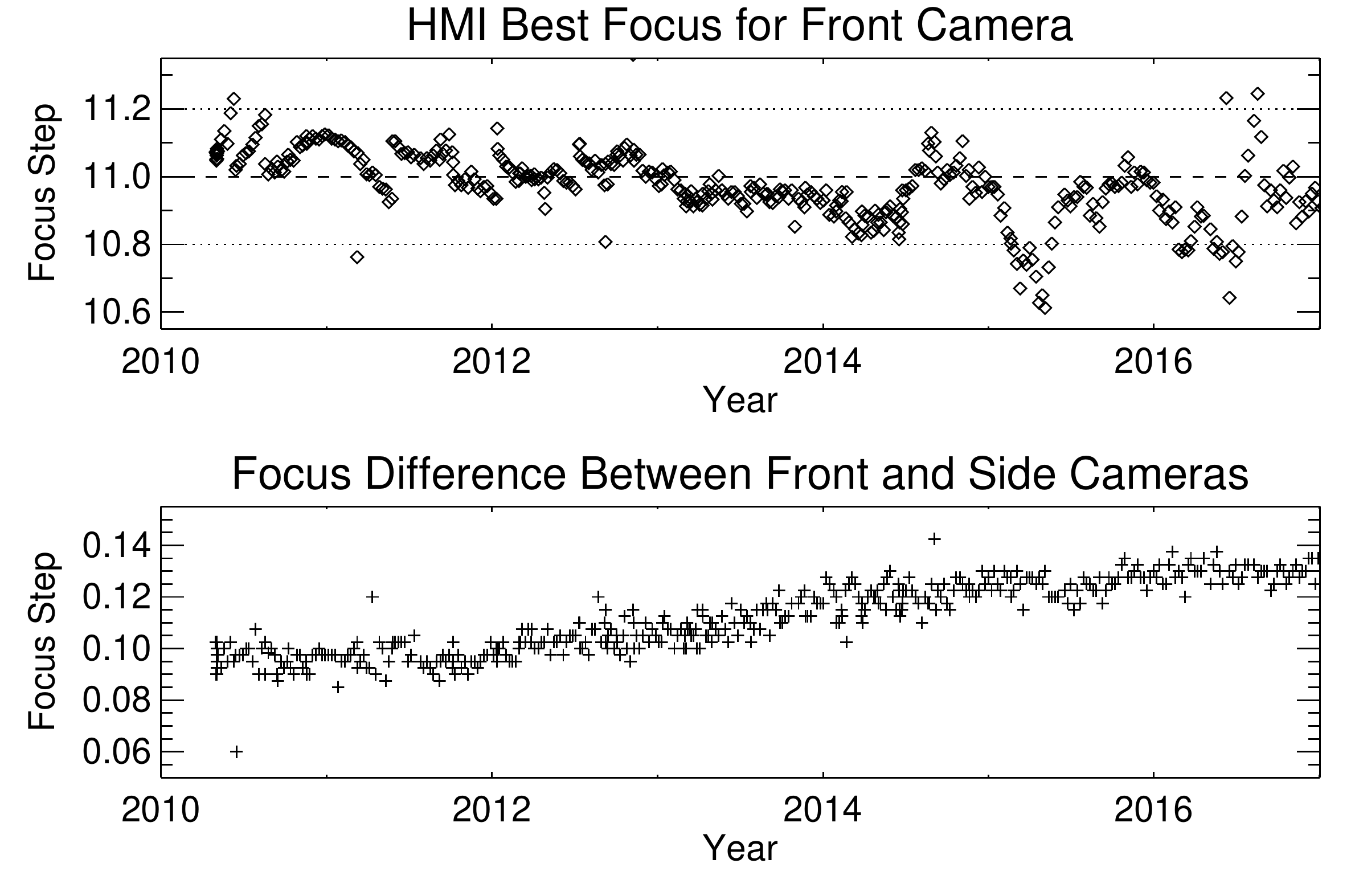}
\caption{Focus trend observed from the start of the prime mission on 1 May 2010 through the 
end of 2016 for the HMI front cameras (top), and the difference in best focus between the 
front and side cameras (bottom). The temperature of the front window is periodically adjusted 
to keep the focus near step 11.}
\label{figure:best_focus}
\end{figure}

Once per week a focus sweep is taken to determine the instrument's best focus.
Two different sequences are used, run on alternate weeks: a full sweep that
takes continuum-tuned images at all HMI focus positions
({\revtxtb FTS ID 3020, 3040)}
and a reduced sweep
that only uses the seven focus positions around the best-focus position
{\revtxtb (FTS ID 3023, 3043)}. 
The calibration images are processed to determine the focus-block setting that
results in the highest image contrast and therefore the optimal focus. 
Results from these weekly measurements are used to adjust
the front-window temperatures to maintain best focus as consistently as
possible. The mission-long HMI focus-trend plot for the front camera is
presented in the upper panel of Figure \ref{figure:best_focus}. The lower
panel shows the difference between best-focus position for the front and side
cameras. The focus is measured in units of focus steps that are equivalent
to 1.04\,mm at the CCD camera, about 2/3 of one depth of field.

The focus of the two cameras is not identical because of differences in
the two light paths. The causes of the relative drift of about 0.03 focus steps over the
course of the mission is not fully understood, but could be due to a small 
(30 micron) change in the relative positions of the CCD detectors due to 
thermal expansion of the optics package.

Another set of calibration images are taken with the Sun deliberately driven
off-center using the Image Stabilization System (ISS). Rather than operating with the
normal closed-loop control, the piezo-electric transducers (PZTs) on the guide
mirror are driven in a pre-set pattern to move the solar image around on
the CCDs. The purpose of these observations is to measure the flat field
of each CCD 
{\revtxtb (FTS ID 3021, 3022, 3041, 3042)}. 
This is described further in Section \ref{sec:Flatfield}.

\subsection{Bi-weekly Detune Sequence}
\label{sec:Detune}

Every other week a 60-frame detune sequence is taken to monitor
changes in the instrument wavelength-tuning positions and to update the
filter-transmission profiles. For the first three months of the regular
mission the detune sequence was run weekly. In this sequence the filter
elements are deliberately not co-tuned, \textit{i.e.} they are tuned to a series of
54 different wavelength combinations. The detunes are used to monitor 
the wavelength drift of the tunable elements. The sequence is taken 
in calibration mode (Calmode). In Calmode the entrance pupil of the telescope
is imaged on the CCDs. The Calmode detunes have been used to determine profiles
for the entire duration of the mission. Six dark frames are also collected.
The results of these detunes and the periodic adjustments to the best tuning
are discussed in Section \ref{sec:Tuning}. The current FTS ID of this sequence
is 3027.

 \subsection{Occasional Calibrations}
\label{sec:Occasional}

Other calibrations are performed on a less regular basis during 
spacecraft maneuvers that interrupt regular science observations, but
provide opportunities to operate the instrument in a unique and useful mode.
These include times when SDO is deliberately pointed away from the Sun (offpoints)
and times when the spacecraft is rolled from its normal orientation 
with respect to the solar rotation axis (rolls).

\subsubsection{Offpoint Flat Fields}
\label{sec:Offpoint}

Spacecraft offpoint maneuvers are used by all three instruments on SDO for
various calibrations. While some procedures are not useful for HMI calibration,
quarterly offpoints are used to generate better flat fields. Twenty-two
pointings are used, and HMI takes a sequence of continuum-tuned images at
a single polarization with a set of varying focus positions. The offpoint
flat fields are discussed in more detail in Section \ref{sec:Flatfield}. The
current FTS ID for offpoint flat fields is 4031.

\subsubsection{Roll Calibrations}
\label{sec:Rolls}

Roll maneuvers are ordinarily performed twice per year, typically after the eclipse
seasons in April and October, when the SDO spacecraft is rotated $360^\circ$
around the Sun--spacecraft line. The spacecraft pauses every
$22.5^\circ$ for approximately twelve minutes. When rolled, the light rays
from parts of the solar disk having different rotational velocities take
different paths through the instrument filters. 
This allows us to calibrate the wavelength dependence of the filters
\citep{Couvidat2016}. 
Data taken during these rolls can be also used
for (among other things) measuring optical distortion and the shape 
of the Sun's limb \citep[\textit{e.g.}][]{Kuhn2012}. 

Additional roll angles were measured during commissioning in April 2010.
A special roll calibration was performed on 23\,--\,24 March 2016 when SDO was
rolled 180$^\circ$ from its normal orientation for twenty-four hours. 
During this interval HMI took detunes every three hours in both normal focus (Obsmode) 
and completely defocused (Calmode). The FTS IDs for these detunes are 3086 and 3087.
The same sets of detunes were taken with the spacecraft in the normal orientation the day before. 
Analysis verified that the Lyot and Michelson filter-element details 
(as well as daily temperature variations of the front window) 
contribute to the 24-hour calibration variations.

\subsubsection{Other Special Calibrations}
\label{sec:OtherCal}

SDO has observed two planetary transits since the beginning of the prime mission: one of 
Venus and one of Mercury. These transits are useful for calibrating the 
instrument roll angle, point spread function, and distortion correction \citep[][]{Couvidat2016}. During each 
transit a non-standard observing sequence was run. The line-of-sight observables, taken 
from the front camera, were produced as normal, but the side camera took continuum-tuned 
filtergrams in four polarization states for the Venus transit and one polarization for the 
Mercury transit. The FTS IDs for Venus and Mercury were 4035 and 4039, respectively.

\section{Trending}
\label{sec:trending}

It is essential to track the evolution of environmental conditions impacting
the HMI observables. This helps with the early detection of problems, characterization 
of instrument changes and degradation, and the
adjustment of the data calibration to maintain the best observables quality
possible. Temperatures and voltages are monitored continuously by an autonomous
system, and SDO staff are alerted if specified limits are reached. In addition,
personnel check the values and trends of various components of the system
several times each day to look for odd behavior or to spot problems before
they reach cautionary limits. The first two subsections focus primarily on
long-term temperatures trends measured in the instrument over the course
of the mission and on typical daily variations observed during the month of July
2015. The final subsection explains how the plate scale varies in response
to temperature changes and how instrument calibration is affected by it.

\subsection{Long-term Instrument Temperature Trends}
\label{sec:Temperature}

\begin{figure}[!ht]
\centering
\includegraphics[height=0.8\textheight]{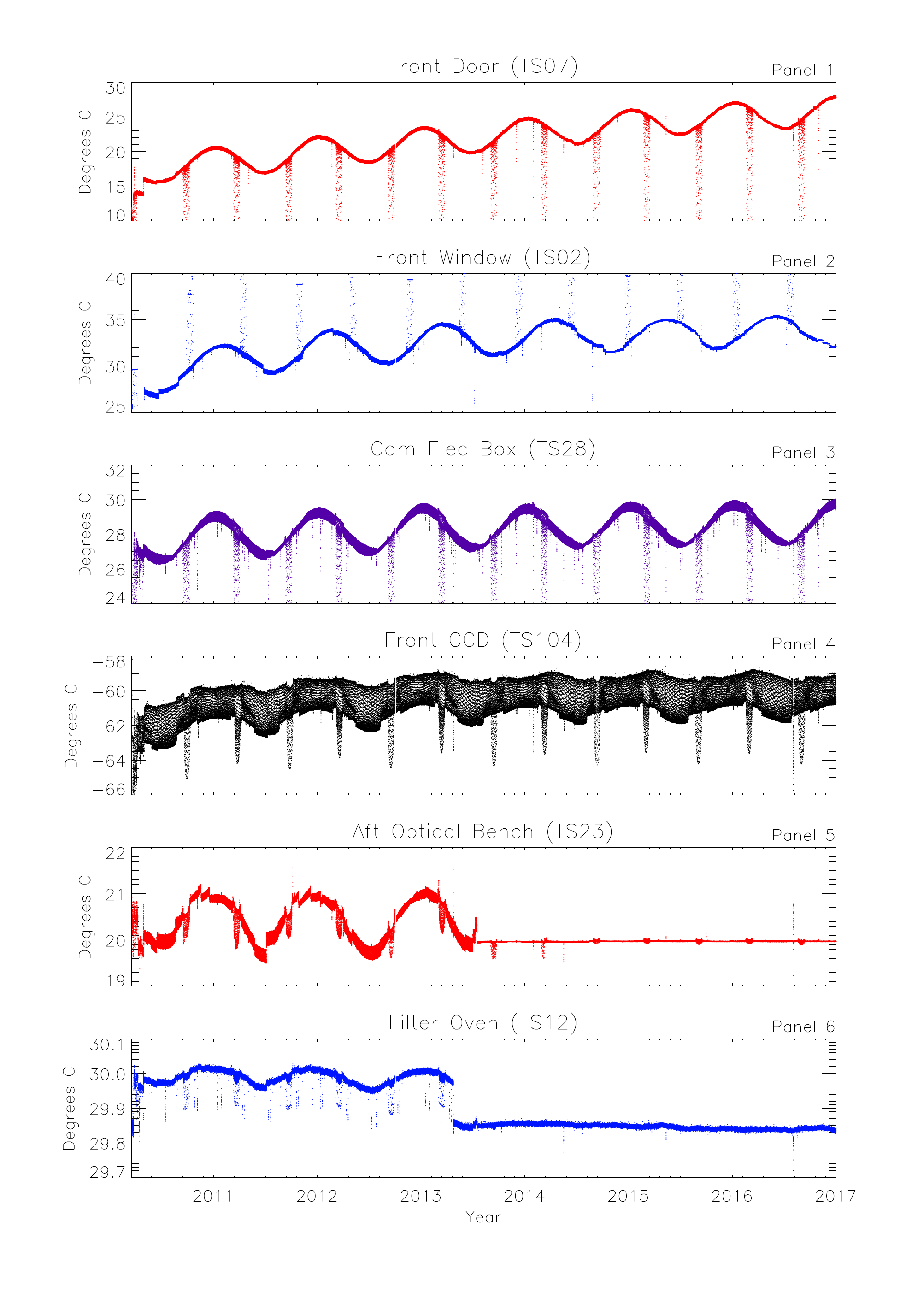}
\caption{HMI instrument subsystem temperatures from 1 March 2010 through
31 December 2016. The points are 30-minute averages of eight-second
telemetry measurements sampled every three hours. The panels show the
temperatures of the Front Door (top panel), Front-Window Mounting Ring
(Panel 2), Front-Camera Electronic Box (CEB, Panel 3), Front CCD (computed
from 16-second telemetry), Aft Optical Bench (Panel 5), and Filter Oven
(bottom panel). Note the different temperature ranges, particularly for the
tightly controlled filter oven and nearby optical bench. Annual variations
and semi-annual eclipse-season perturbations are visible on the longer
term. The first HMI processor reboot occurred on 20 April 2013. The thermal
control scheme for elements of the optics package changed on 16 July 2013
and 25 February 2014. Daily differences between Noon and Midnight dominate
the short-term variations. Systematic daily variations (see next figure)
produce what look like multiple lines in the three-hour samples shown here. }
\label{figure:hmi_temps_mission}
\end{figure}

Numerous temperature sensors placed throughout the instrument monitor HMI's
response to every aspect of its thermal environment 
\citep[see Appendix B and supplementary material in][for thermistor locations]{Schou2012}. 
Figure \ref{figure:hmi_temps_mission} shows temperatures at six representative
locations in the instrument. Three-hour samples of thirty-minute averages of quantities 
measured every eight seconds highlight long-term variations. The six locations illustrate 
the variations of different subsystems with varying levels of themal control: the front door, 
the mounting ring of the front window, the front-camera electronics box (CEB), the front CCD, 
the optical bench, and the filter oven. The front window and last three have the greatest 
measurable impact on the observables.

The top panel shows the temperature of the front door from 1 March 2010
through the end of 2016. The front door is outside the optics package and its
temperature is essentially uncontrolled, except that it is in thermal contact with other
controlled parts of the instrument. There is a jump just before the start of the prime
mission in early 2010 when the initial operating temperature was set. The
most obvious features are the regular annual variation of about 4\,K
due to the change in Sun--SDO distance and transient decreases during the
twice-annual SDO eclipse season. The instrument was designed to operate
near room temperature. The equilibrium temperature has increased by about
9\,K since the start of the mission. This is due to changes in
reflectance\,/\,absorbtion of the front-door surface and to deliberate temperature changes 
in the nearby front window (see discussion in Section \ref{sec:Platescale}.

The temperature at the bottom of the front-window mounting ring 
(Temperature Sensor 02, TS02), shown 
in Panel 2, is not directly controlled; instead, the thermistor is
attached to the edge of the front window opposite the sensor used
to control the temperature. The front-window temperature has been allowed
to increase by about 5\,K since 2010 in order to keep the focus of the
instrument constant. Unlike most other locations, the front-window temperature
increases during eclipses because the heaters are turned up to keep
thermal gradients in the front window small so that the post-eclipse recovery
is shortened (Section \ref{sec:eclipses}).

The front-camera electonics box (CEB, TS28 in Panel 3) is mounted on the front of the
HMI optics package. It too shows variations with annual periodicity (about
3\,K) and exhibits short strong dips during eclipses (third panel).
The temperature runs a little hotter than most of the optics package
because the camera electronics generate heat that is not fully dissapated by
its own dedicated radiator. The average CEB temperature increases a couple degrees in
the first two years, but has been relatively stable thereafter. Shorter-term
24-hour variability is discussed in the next section.

Each CCD detector has its own large radiator on the outboard surface of the
instrument that is sheltered from direct solar radiation; it faces solar South
(perpendicular to the Sun--spacecraft line) and has a nearly unobstructed
view of cold space, except for the Earth. The CCD temperature is kept very
low to minimize dark current. The fourth panel shows the temperature at the
front CCD detector (TS104, determined from averages of temperature readings
made every 16 seconds). The annual variation in temperature is smaller;
shorter-term variations dominate. \citet{Couvidat2016} determined an intensity
sensitivity of 0.25\,\% per degree.

Panel 5 shows a temperature measured on the optical bench inside
the optics package (TS23). During the first three years of operation,
the temperature was controlled by specifying a specific power input from
the internal heaters. The constant overall duty cycle of the heaters was
occasionally adjusted, but there was no active on-board control. Consequently
the temperature varied with the overall equilibrium temperature of the
instrument and an annual variation of about 1\,K was apparent. On 16 July 2013
the scheme was changed to turn the heaters on at a specified duty cycle only
when the temperature goes below a set minimum. Subsequently the temperature
variation has been greatly reduced,
and even the response to eclipses has significantly diminished. A consequence 
of this is discussed in Section \ref{sec:Platescale}.

The bottom panel of Figure \ref{figure:hmi_temps_mission} shows the temperature
measured on the outside of the tightly controlled filter oven. The oven is
kept warmer than the rest of the optics package so that its temperature can
be more precisely controlled. The specification for thermal control of the
filters is 0.01\,K per hour. While the specification is more than met within
the oven, an annual peak-to-peak variation of about 0.05\,K remained at the
externally mounted sensor. On 20 April 2013 the HMI processor was rebooted
for the first time, and that eliminated a small amount of current that had
been flowing in the redundant oven-thermal-control system. The internal oven
temperature did {\em not} change, but the temperature measured at the external
sensor did because the gradient between the oven and the rest of the instrument
was altered. The 16 July 2013 change to the optical-bench thermal-control
scheme nearly eliminated the annual variation.  Inside the oven the annual
variation was attenuated by a factor of two to three (not shown).

\subsection{Short-Term Instrument Temperature Trends}
\label{sec:shortT}

\begin{figure}[!ht]
\centering
\includegraphics[height=0.8\textheight]{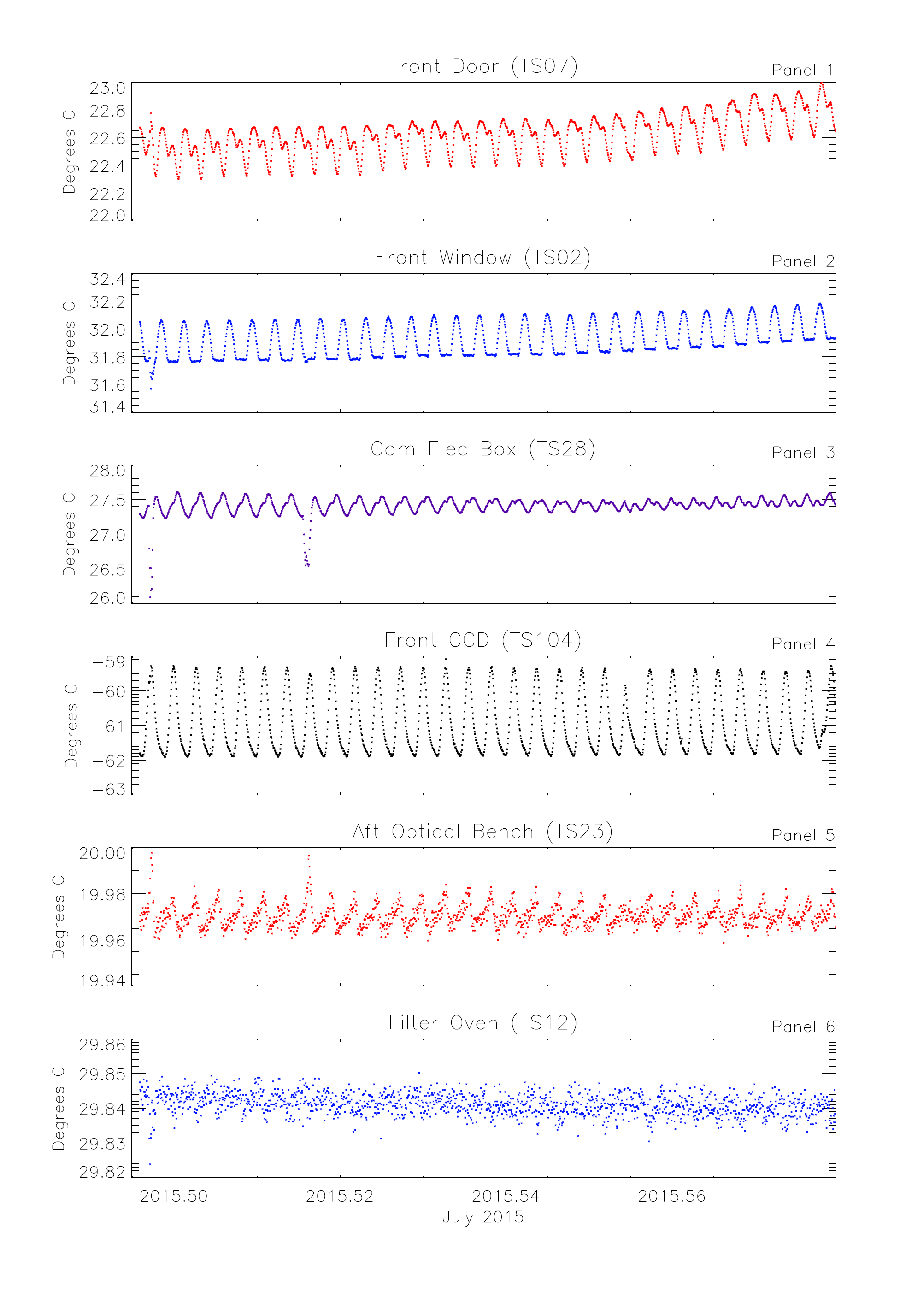}
\caption{HMI instrument subsystem temperatures for the month
of July 2015. Data are 30-minute averages and highlight the daily variations.
Panels show temperatures for the Front Door (top), Front-window Mounting Ring (Panel 2), 
CEB (3), Front CCD (4), Optical Bench (5), and Filter Oven (bottom). The temperatures of the 
front door, CCD, and CEB are not actively controlled. The CCD radiators are oriented 
to see (mostly) dark, cold space. {\revtxtr The temperature of the front-window mounting
ring at the sensor (TS02) shown in Panel 2 remains constant during only part of the day.}
The door and electronics box show more complex daily patterns due to varying exposure 
to the Earth and other environmental factors. }
\label{figure:hmi_temps_month}
\end{figure}

Figure \ref{figure:hmi_temps_month} shows temperatures measured at the same 
locations in the instrument for the month of July 2015 -- after the changes 
in the temperature control scheme. 
{\revtxtr This month is fairly typical and was selected because it has
a few interesting features that can be examined in a little more depth.} 
Averages have been made for 30 minutes (225
eight-second measurements) to highlight shorter-term variations and reduce
noise. Unless there is some anomalous event, measurements of variations
on time scales less than 30 minutes may not be meaningful because the
digitization interval (about 0.05\,K, depending on gain) and read
noise (the standard deviation of five-minute averages is about 0.03\,K) are
larger than the actual short-term variability in most instrument temperatures.

Because of its $28^\circ$ inclined geosynchronous orbit (up to about $52^\circ$
to the ecliptic), the environment of the spacecraft changes with a 24-hour
period and the relative viewing angles of the Earth and Moon at a
particular time of day change during the month and year. 
{\revtxtr The orbit was chosen so that the spacecraft remains}
near $100^\circ$\,W
longitude, in constant view of the ground station in White Sands, NM. Eclipses
occur only during the Spring and Fall when the spacecraft passes near the 
Equator at local Midnight. {\revtxtr The eclipse dates change as the orbit slowly precesses.}

The top panel of Figure \ref{figure:hmi_temps_month} shows daily variations of
the uncontrolled front-door temperature (TS07). Short-term temperature variations are
dominated by changes in the spacecraft environment, primarily the view of the
Earth, and by thermal changes elsewhere in the instrument. The maximum
daily temperature occurs shortly after 06 UT, local Midnight at the ground station, 
when Earth is closest to the Sun--SDO line. A second smaller maximum appears a
little less than 12 hours later in phase with the temperature maximum of the
CCD camera (discussed below). The temperature minimum is fairly sharp and occurs
near 0 UT, which is dusk at the spacecraft. The daily temperature range is
about 0.3\,K.

The temperature of the front window is controlled {\revtxtr using measurements from}  
a sensor (TS01) located
on the mounting ring opposite the one shown in the second panel (TS02). There
is a temperature gradient across the front window. During the first half
of the day (0--12 UT) the Earth is in view of the front window, so it radiates
less energy. As a result the temperature at TS02 rises due to the change in
gradient across the window. During the other half of the day the window cools
more efficiently, the gradient changes, and the temperature at TS02 is
better regulated. The front door (shown in the top panel) is close to the
front window, so it is affected by the thermal control of the front window.

The front-camera electronics box (Panel 3) is mounted on the front of the
instrument. It is insulated from direct Sun and has a shield / radiator
mounted perpendicular to the Sun--SDO line. Changing views of the Earth
affect the amount of heat that is absorbed and also affect the temperatures
of other parts of the spacecraft in its field of view. The daily thermal
variation of the CEB is more complicated; it shows profile features of both
the front window and the CCD (Panel 4).

The CCD temperatures are not actively controlled, but they are kept very
cold using independent large radiators mounted on the outboard side of
the instrument, ordinarily facing solar South (TS04, shown in Panel 4 of 
Figure \ref{figure:hmi_temps_month}). The visibility of the Earth from the
radiators changes significantly during the 24-hour orbit, and the daily
CCD temperature variation is fairly large: nearly 3\,K. The phase of
the environmental variation shifts through the year. SDO is located below
Earth's Equator at local Noon during half of the year and above it during
the other half; eclipses occur during the transition. In July the fairly
sharp daily temperature profile of the CCD peaks at local Noon (about 20
UT) when the Earth is near the anti-sunward direction and most visible to
the radiators. Whatever causes the variation in the CCD temperature also
affects other external, uncontrolled parts of the instrument, as seen in
Panels 1 and 3. Multiple lines appear in the corresponding panels of Figure
\ref{figure:hmi_temps_mission} because of the three-hour sampling of the
systematic daily temperature profile.

The optical-bench temperature is controlled using measurements made at
a particular location; Panel 5 shows that the temperature measured at a
nearby location on the optical bench varies within a range of 0.02\,K. The
temperature has a sawtooth daily profile and peaks each day at the same time
as the CCD detector.

The filter oven is thermally isolated from the rest of the instrument, has a
long thermal time constant, and varies in temperature by less than 0.01\,K
with only a very weak daily pattern (TS12, in the bottom panel). Remaining
variations at the surface of the oven shown here are consistent with read
noise of the sensors.

There are a couple of interesting features of note during the month. On 1
July and 8 July there are clear offsets in the front-camera electronics-box
temperature (Panel 3) that can also be seen to varying degrees in the optical bench,
front window, and front door (Panels 5, 2, and 1 respectively). 
On 1 July SDO performed a ``cruciform maneuver"
for the purpose of calibrating the EVE instrument. Over the course of
about 4.5 hours the spacecraft was pointed to 112 different locations up
to $3.05^\circ$ away from the Sun along two orthogonal directions and that
caused small changes in the temperatures. On 8 July small offpoints of
the spacecraft were made to determine AIA and HMI offset flat fields. The
corresponding temperature perturbations were smaller.

Careful inspection shows that on 22 July the front-CCD temperature profile was
unusual (Panel 4). Small perturbations in the optical-bench and camera-electronics-box
temperatures (Panels 5 and 3) can also be perceived. These occurred during a spacecraft-roll 
maneuver performed for HMI calibration (see Section \ref{sec:Rolls}).
During the roll, the Sun--Earth pointing is maintained, but the spacecraft
is oriented with solar North at sixteen different roll angles. The change
in roll changes the viewing angle of the Earth from the HMI radiators.

\subsection{Plate Scale}
\label{sec:Platescale}

\begin{figure}[htb]
\centering
\includegraphics[origin=lb,width=\textwidth]{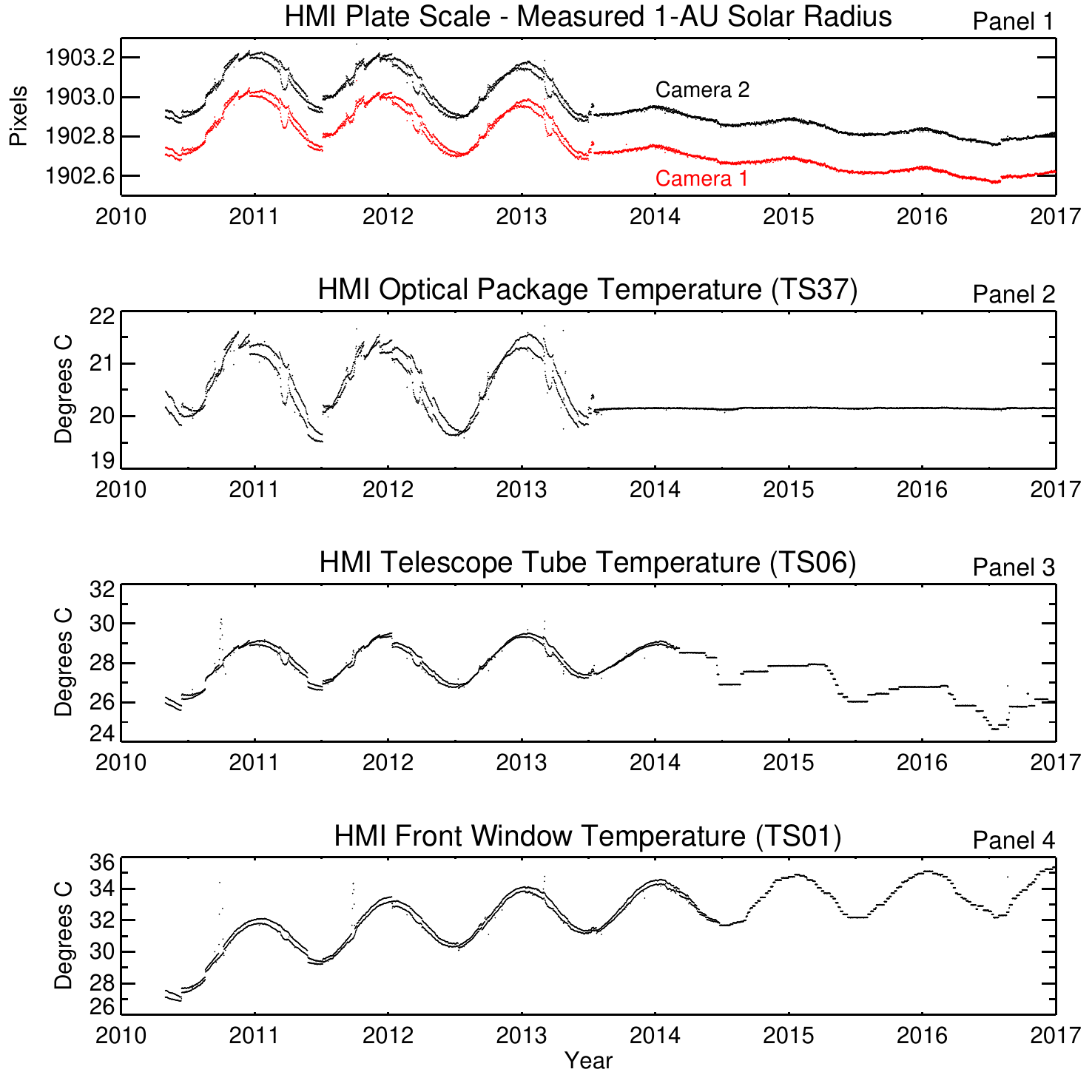}
\caption{Variation of the HMI plate scale (\textsf{cdelt1}) with time (top
panel) compared to three different instrument temperatures. The solar radius
has already been normalized to 1 AU using known geometric parameters. 
Camera~2 is shown in black; {\revtxtr slightly cooler} Camera~1 in red. The second
panel shows the temperature measured by a representative temperature sensor (TS37) in the
HMI optics package. Panel 3 shows the temperature of the telescope
tube. The bottom panel shows the front-window temperature. In each panel two
values are shown for each day, one measured near the orbital perihelion and 
the other near aphelion, which correspond roughly to daily extremes in the
instrument temperature. 
}
\label{figure:plate_scale}
\end{figure}

The plate scale is set by the mechanical and optical properties of the
telescope and is measured by determining the observed radius of the solar
image in CCD pixels and applying a geometric correction to normalize the value
to 1 AU. The HMI plate scale correlates strongly with the temperature
of the HMI optics package and to a lesser degree with the telescope-tube
temperature, as shown in Figure \ref{figure:plate_scale}. 

The pronounced annual periodicity present during the first three years is
due to temperature drift of the HMI instrument caused by the change
in irradiance due to variation in the Sun--spacecraft distance. 
{\revtxtr Daily variations are driven primarily by changes in the 
spacecraft environment related to the SDO orbit.}

In the early years, when the instrument temperature varied by a little more
than a degree during the course of a year, the measured radius varied by about
0.3 pixels (0.15 arc seconds). As described in Section \ref{sec:Temperature},
the temperature-control scheme for the optics package was changed on
16 July 2013 to reduce variations in the temperature. The
variations in plate scale were greatly reduced. Similar changes were made to the
temperature-control scheme for the telescope tube and front window on 25
February 2014. Since then more frequent temperature adjustments have been
made to keep the focus of the instrument in the proper range. The gradual
long-term decrease in the measured solar radius may be related to changes in
the front-window temperature (which affects magnification), tube temperature
(which affects the distance between lens and image), or other factors.

Using HMI data collected during the 2012 Venus transit, \citet{Emilio2015} 
derived a 1-AU solar radius in the continuum wing of the line of 
$959.57 \pm 0.02$ arc seconds, equivalent to $695,946 \pm 15$\,km.
Similarly, \citet{Couvidat2016} found that the image of the Sun
is a little larger than expected. For the image scale the ratio of their best
estimate to that in the headers is $0.99992053$. Consequently, we conclude
that for the HMI spectral line the reference radius of the Sun (keyword 
\textsf{RSUN\_REF}) should be decreased by about 55\,km to 695,944,685\,m.
\section{Optics and Filter Issues}
\label{sec:opticsfilters}

This section describes calibrations and observations made to assess 
optical performance of the HMI instrument and elements of the 
filter system. A more complete discussion of the filter calibration is 
found in \citet{Couvidat2016}.

\subsection{Instrument Throughput Changes}
\label{sec:throughput}

\begin{figure}[htb]
\centering
\includegraphics[width=0.98\textwidth]{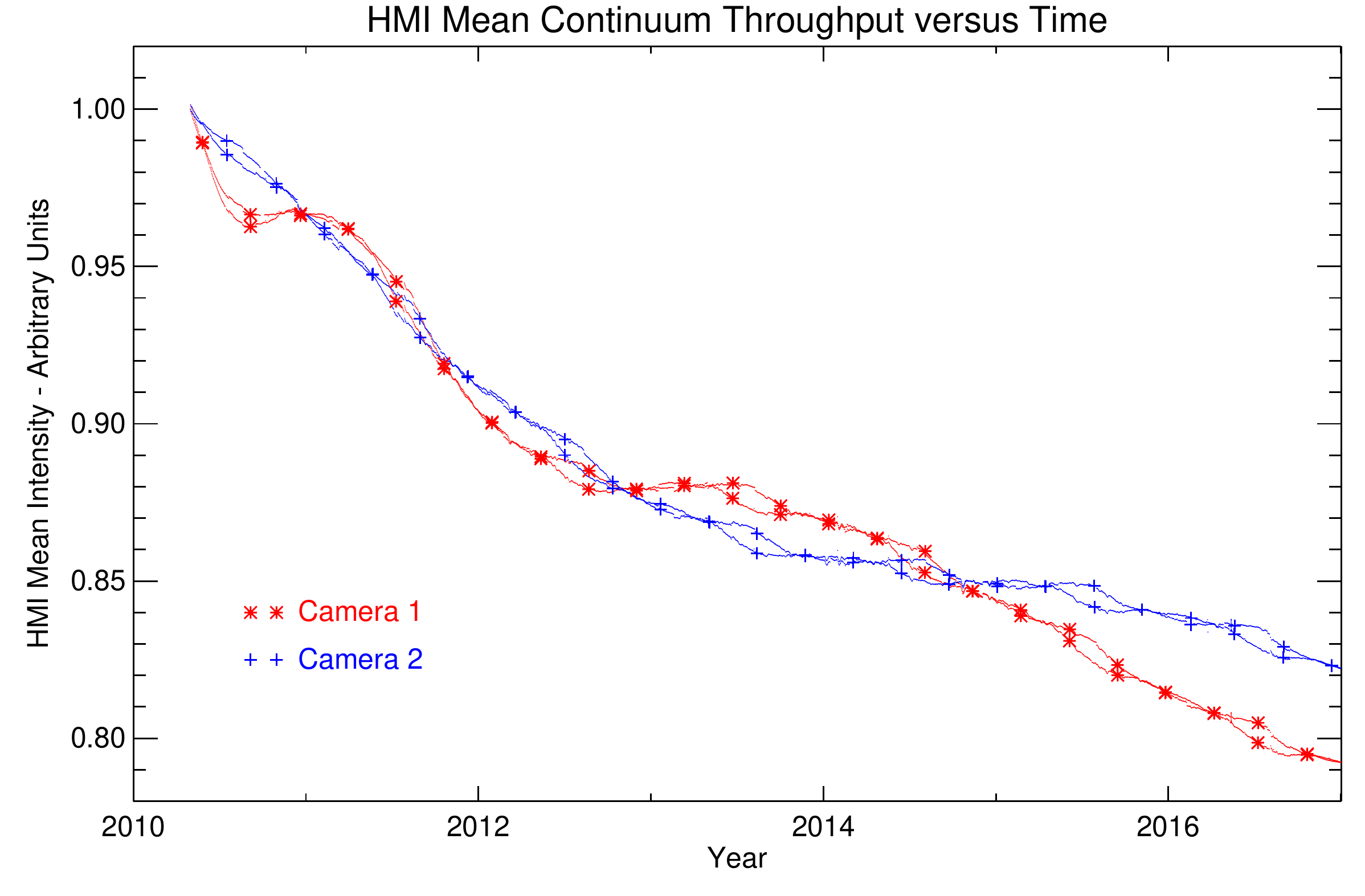}
\caption{Evolution of the end-to-end instrument throughput during the
SDO mission. The average on-disk solar continuum intensity measured with
Camera~1 (Camera~2) is plotted as a function of time in red (blue). {\revtxtr
The throughput of Camera~1 had decreased a little more than 20\,\% by the
end of 2016.} The continuum intensity is measured during the twice-daily
calibration sequences at about 06~UT and 18~UT. {\revtxtr Symbols highlight
06~UT and 18~UT measurements approximately every 200 days for each camera.}
Short-term differences in a single camera primarily reflect temperature
changes due to solar-irradiance and thermal-environment variations.  Values,
normalized to the intensity of the first image, have been corrected for the
Sun--SDO distance and exposure time. Values have also been empirically adjusted
to compensate for a permanent change in image crop radius on 28 January 2015.
}
\label{figure:intensity_decrease}
\end{figure}

The instrument throughput has been slowly decreasing since launch. 
Figure \ref{figure:intensity_decrease} shows the average solar intensity 
measured in twice-daily full-disk continuum exposures (Frame ID = 10000) for each camera.
The \textsf{DATAMEAN} values have been corrected for exposure time, the Sun--SDO distance, 
and for a one-time change in the image crop radius at 19:51 UT on 28 January 2015.
The exponentially decreasing decay rate observed in both cameras is generally 
consistent with expected effects of radiation damage darkening the front window. 
Short-term variations of a single camera or between the cameras is likely due to the 
changing thermal environment. \citet{Couvidat2016} measured a temperature sensitivity \
of -0.25\,\% per degree in Camera~2, but as shown in Panel 4 of 
Figure \ref{figure:hmi_temps_mission}, except for regular daily and annual changes,
the nominal temperature measured near the CCD has not changed much over the course of the mission.
The origins of the long-term differences between the two cameras are not understood. 
The local-Noon--Midnight asymmetry (6\,--\,18~UT) is greatest in the middle of the year
when Earth is South of SDO and thus most visible to the radiators at local Midnight.

The gradual decrease in instrument throughput requires occasional exposure-time
increases to maintain a roughly uniform signal intensity. Since launch
the exposure duration has been increased three times, in each instance by
five milliseconds, as shown in Table \ref{table:exposure_adjustments}. There is
still sufficient margin in the timing of the camera image taking to compensate
for further throughput decreases; the current mode of operation allows for
exposures of up to 430\,ms without compromising the basic 45-second cadence.

\begin{table}[htb]
\begin{tabular}{ccc}
\hline
Date & Front Camera (1) & Side Camera (2) \\
\hline
01 May 2010 & 125 ms & 115 ms \\
13 Jul 2011 & 130 ms & 120 ms \\
16 Jan 2013 & 135 ms & 125 ms \\
15 Jan 2015 & 140 ms & 130 ms \\
\hline
\end{tabular}
\caption{ HMI Camera Exposure-Time Adjustments}
\label{table:exposure_adjustments}
\end{table}

HMI observables are computed from sums and differences of filtergrams, so
exposure-time uncertainty contributes directly to errors in the measured
quantities. A mechanical shutter motor controls the exposure time by
rotating the cut-out sector of an opaque disk into place, with a pause in
the {\it open} position for a specified time. 
{\revtxtb The shutter is located in the observing beam near an image of the
pupil when in Obsmode.  The mechanical exposure time can be specified with
precision of about 120 microseconds and has an observed standard deviation
of 13.2 microseconds, about a part in 10,000 of the nominal exposure. The
difference between the commanded and actual exposure time is determined
with precision of one microsecond and accuracy better than 4 microseconds
using integral detectors to determine the precise times that the leading and
trailing edges of the open sector rotate past each of three characteristic
locations in the beam.}
The actual exposure time is used in the analysis.  Typical exposures are 115 --
140 milliseconds.  The four-microsecond exposure-time knowledge is a part in
30,000 of the nominal exposure time. This is a factor of three or more better
than what is required to beat the photon noise level for global averages of
the mean magnetic field and the large-scale velocity for low-spatial-degree
helioseismology.  The SDO/HMI exposure time is monitored far more closely
than it was for the Solar and Heliospheric Observatory / Michelson Doppler Imager
\citep[SOHO/MDI,][]{Scherrer1995} and has much less variability. See 
Appendix \ref{sec:FigureSupplement} for a plot of the mechanical exposure quality.

\subsection{Distortion}
\label{sec:Distortion}

Image distortion arises because of small imperfections in the optics,
including the optics that move to tune the instrument. 
{\revtxtb The distortion map determined prior to launch for each camera
\citep[See Figures 7 and 8 of][]{Wachter2012} has been characterized using
Zernike polynomials.  The fitted instrumental-distortion correction is applied
to each Level-1 filtergram.  The maximum displacement before correction is
less than 2 pixels and occurs near the top and bottom of the CCD camera;
the mean residual distortion after correction is $0.043 \pm 0.005$ pixels.
Differences between the front and side cameras are of order 0.2 pixels.}
\citet{Couvidat2016} analyzed HMI images taken during the Venus transit
of 6 June 2012 and found that all along the path of the planet, the
distortion-corrected observed position agreed with the ephemeris coordinates
to better than 0.1 pixels (0.05 arcseconds).

\subsection{P-angle}
\label{sec:pangle}

The roll angle of the solar image relative to the instrument is commonly called the p-angle
(not to be confused with the position angle determined for Earth-based observations).
In the case of HMI, the {\it top} of the CCD is nominally near the solar South Pole, so the
WCS standard \textsf{CROTA2} keyword that gives the angle between heliographic north and CCD 
coordinates typically has a value very close to  $180^\circ$. For HMI the p-angle = $180 - $\textsf{CROTA2}.

\citet{Couvidat2016} reported on a careful analysis of both the absolute p-angle based on 
observations of the 6 June 2012 Venus transit and the relative p-angle of the two cameras based on
comparison of near-simultaneous images obtained by the two cameras in July 2012. They find that the 
p-angle for the front-camera is $-0.0135 ^\circ$ and for the side camera $+0.0702 ^\circ$.
The difference in p-angle between the two cameras is $0.0837 ^\circ$ 
with a constant drift rate of $-0.00020 ^\circ$~year$^{-1}$ during the SDO prime mission.
The drift is probably due to curing of materials used to mount the CCDs or to thermal
changes.

The absolute p-angle was also determined by \citet{Liang2017} for the Mercury transit
using the same methods used by \citet{Couvidat2016}. However, the much smaller size of Mercury 
meant that no annulus extraction was done. They found that the values for Camera~1 changed from
-0.0140 to -0.0114 (+0.0026) and those from Camera~2 from +0.0712 to +0.0735 (+0.0023). Given
the size of the residuals seen by \citet{Couvidat2016}, the difference does not appear to be
significant.

\subsection{Camera Differences}
\label{sec:cameradifferences}

The front and side cameras of HMI are not identical and their images exhibit
slightly different properties, for example in their focus, alignment,
and the occurence of bad pixels. Of course to some degree the temperature
and radiation environments of the two cameras also differ.  Although the
CCD radiators are adjacent and on the same solar-south-facing side of the
instrument, the radiators for the camera-electronics packages have different
geometries. The only significant differences in the optical paths are due
to a beam splitter, fold mirrors, and shutters that direct the light to the
two cameras after all of the other optics. Since 13 April 2016, filtergrams
from the two cameras have been combined to compute the vector magnetic
field \citep{Couvidat2016,Hoeksema2014}.  Figure \ref{figure:best_focus}
shows that there is only a small drift in focus difference between the two cameras
during the lifetime of the mission, probably due to aging of materials that
affect the CCD mounting position or to thermal drifts.

\subsection{ISS performance}
\label{sec:ISSPerf}

Basic spacecraft pointing information is provided by three Inertial Reference Units (IRUs).
The spacecraft relies on signals from AIA for more fine-guiding information.
Small, rapid pointing variations are driven by movements of mechanisms throughout
the spacecraft. The HMI image stabilization system (ISS) uses a tip--tilt
mirror to remove fine-scale jitter measured at a primary image plane in the
instrument. The ISS measures the solar-limb position using four orthogonal
detectors to sense image motion on the limb. The HMI guiding mirror has
a three-point piezo-electric transducer (PZT) actuator to compensate for
position errors in the observed limb position. The ISS holds the image 
location constant to about 0.025 arcseconds (a twentieth of a pixel) 
with a frequency roll-off of a factor of two at about 50 Hz \citep{Schou2012}. 
The PZTs nominally operate at about 35\,V, and there is a superposed
annual period of amplitude about 5\,--\,10 V associated with variations in the
spacecraft thermal environment and size of the solar image. The nominal set point
can also change when the instrument legs are moved to recenter the image 
(approximately monthly).

\begin{figure}
\label{fig:pztb}
\centering
\includegraphics[width=0.95\textwidth]{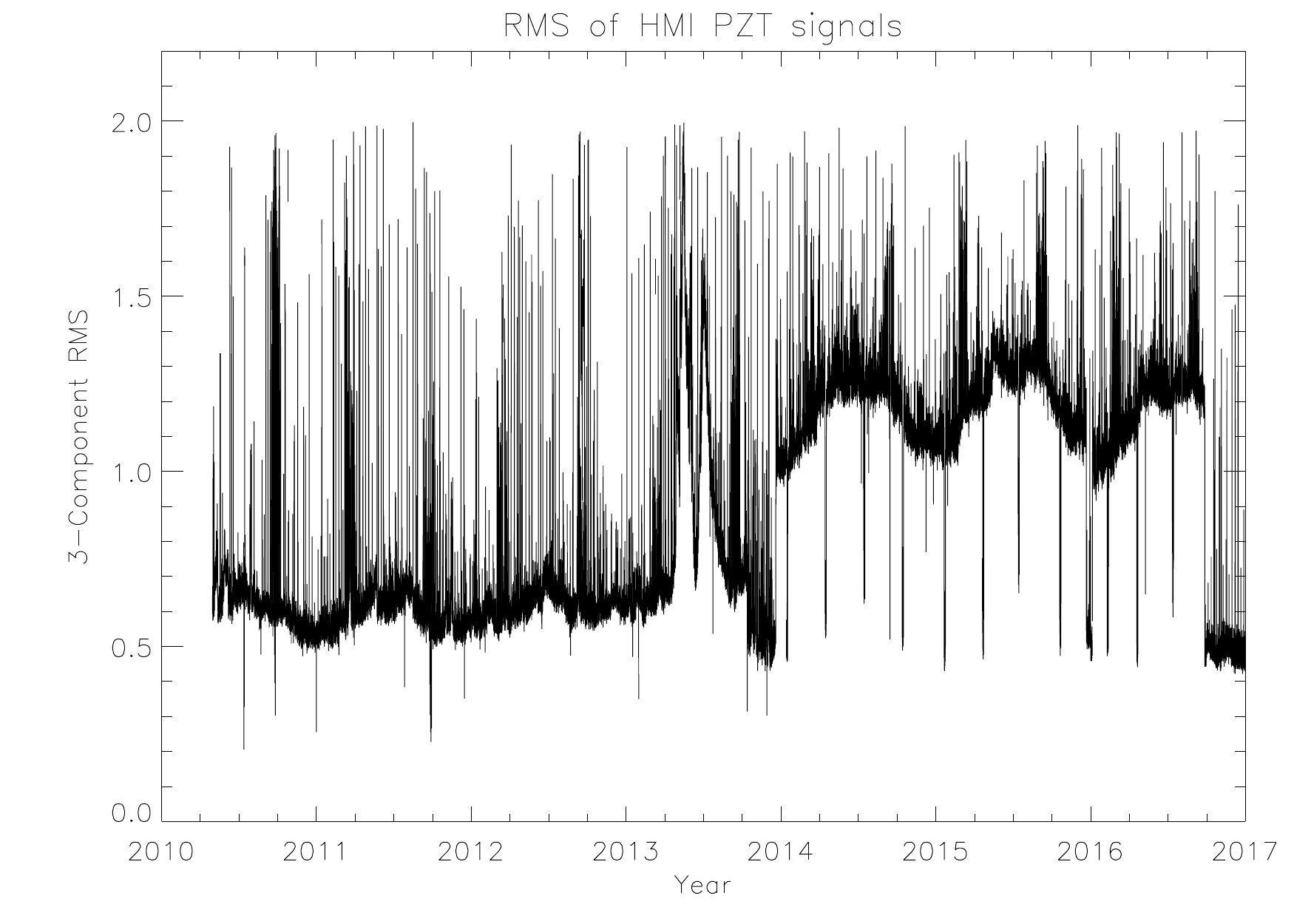}
\caption{Voltage variations of the Image Stabilization System (ISS)
\textit{versus} time. HMI uses three PZTs to control the guiding mirror based
on an error signal determined by limb sensors. The RMS of the voltage over
an hour is an indication of the pointing jitter for which the system must
compensate. The plot shows the RMS of the {\revtxtr three one-hour-RMS} values
\textit{versus} time. The SDO pointing was fairly stable until mid-2013,
when the performance of one of three Inertial Reference Units started to
deteriorate. A new mode using just two IRUs commenced in October 2013. The
operating temperature of the IRU wheels was changed in September 2016 and
the spacecraft pointing stability improved noticeably. For clarity, values
outside the range 0.2\,--\,2.0 are omitted.}
\label{figure:PZT_RMS}
\end{figure}

The RMS voltage variation for each PZT computed over an hour is of order half
a volt, with occasional spikes when spacecraft mechanisms are active. The RMS
value of the three computed PZT-RMS values is an indicator of the magnitude
of the jitter signal. Figure \ref{figure:PZT_RMS} shows the hour-averaged 
three-PZT RMS value of the ISS voltages from 1 May 2010 to the end of 2016.

Regular large-amplitude spikes are due to brief weekly and biweekly
excursions when the instrument is intentionally pointed away from Sun center
for calibrations. Also visible are regular intervals of increased RMS each
Spring and Fall during eclipse season. The ISS control loop is ordinarily 
turned off around eclipse times and during spacecraft off points.

SDO is equipped with three Inertial Reference Units (IRUs) to provide
information to help keep the solar pointing stable; however, the operation
of the IRUs has changed during the mission. The IRUs were operated at a
temperature that was colder than optimal during most of the mission due
to concerns about potentially deleterious effects of their heaters on the
spacecraft battery. As a result there was some jitter introduced by the
wheels. In 2013, the performance of IRU-1 began to deteriorate more rapidly,
and on 12 October 2013 the current draw increased sharply. The next day IRU-1
was removed from the control loop and it was powered down in December 2013. Since
that time SDO has operated with only two IRUs. In early 2015 IRU-2 exhibited
early signs of similar behavior. A test in late 2015 showed that increasing
the IRU temperature eliminated the worrying symptoms of IRU-2 and improved
overall jitter levels. After careful analysis of the effects on the battery,
the IRU temperatures were raised on 16 September 2016. The decrease in the
jitter signal is apparent in Figure \ref{figure:PZT_RMS}. These changes in
operation of the spacecraft IRU units have had no apparent effect on the
final performance of the ISS system, nor have they been detected in the HMI
science products, except for an increase in five-minute power in the 
full-disk intensity means {\revtxtr between October 2013 and September 2016}
(R. Howe, private communication 2016) and in 
local-correlation-tracking results (B. L\"{o}ptien, private communication, 2015)
that may be due to jitter in the spacecraft roll angle.

\subsection{HMI Filter Element Wavelength Drift and Tuning Changes}
\label{sec:Tuning}

\begin{figure}[htb]
\centering
\includegraphics[width=0.9\textwidth]{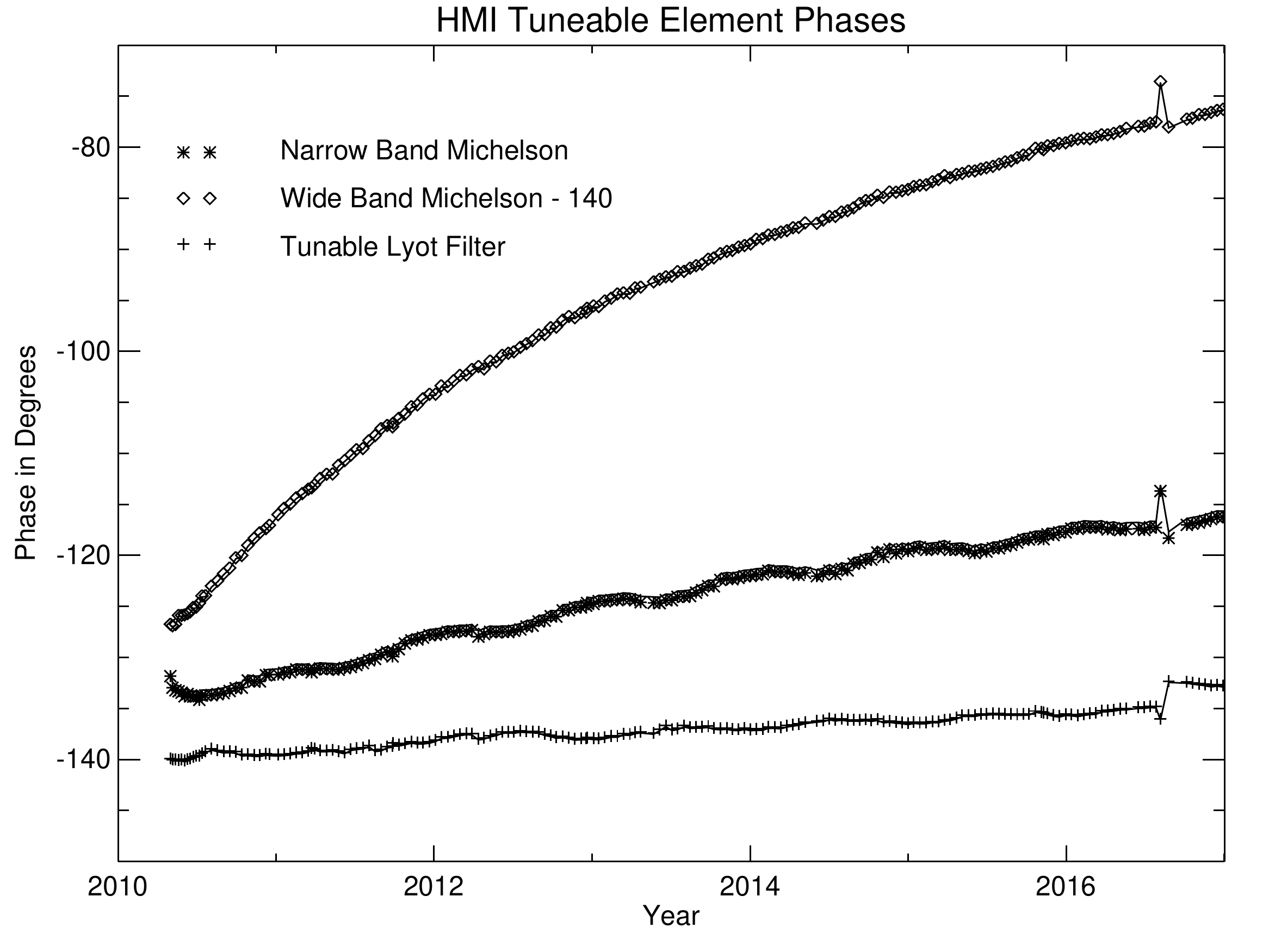}
\caption{Wavelength drift of the HMI tunable elements determined during
regularly scheduled detunes. The phase for each element has an arbitrary zero
and $360^\circ$ corresponds to the full FSR of the element. The tuneable Lyot
element (``+"-symbols) drifts slowly with time. The narrow-band (NB) Michelson
(``*"-symbols) drifts just a little more rapidly. The wide-band Michelson
(diamonds, offset in the plot by $-140^\circ$) has the largest drift, about
an eighth of a FSR during the mission. A spacecraft anomaly on 2 August 2016
resulted in an extended loss of thermal control that had lasting effects,
particularly on the Lyot filter phase. Symbols show the fit determined with
images from Camera~2 and the connected solid lines show Camera~1; the difference 
is very small. A handful of anomalous fits are not shown.}
\label{figure:phase_drift} 
\end{figure}

HMI uses a series of filters to select the wavelength of each filtergram.
The entrance window and broad-band blocking filter are followed by a five-stage
Lyot filter and two Michelson interferometers. The final stage of the Lyot
(E1) and the Michelsons are tuneable. The nominal wavelength of each tuneable
element is set by rotating a half-wave plate. Rotation of the wave plate by $90^\circ$ scans the
element through its free spectral range (FSR). For convenience, the wavelength
tuning is characterized in terms of the phase within the FSR. \textit{I.e.} scanning
$360^\circ$ in phase tunes through the entire spectral range of the element,
so each $1.5^\circ$ step of the hollow-core motor that holds the wave plate
changes the phase by six degrees.

The central wavelengths of the filter elements drift with time. The wavelength
of each of the three tuneable elements can be determined from the biweekly
detune calibration sequences described in Section \ref{sec:Detune}. A relative
minimum in intensity occurs when an element is tuned to the spectral-line
center. The average phases of the HMI tunable elements change slowly with
time, as can be seen in Figure \ref{figure:phase_drift}. No correction has
been made for the motion of the spacecraft since the detunes are ordinarily
taken when the Sun--SDO velocity is small.

It is important to cotune the filter elements to the same wavelength and to
keep the wavelength range over which the filtergrams are taken centered
on the Fe\,{\sc i} spectral line. The observed drifts warrant regular re-tuning of
the instrument. The wide-band (WB) Michelson exhibits a stronger
time-dependence, whose origin is thought to be the glue holding the mirrors
in the two legs; it is believed that the glue in the vacuum leg has expanded
or contracted with time. A similar issue was encountered by SOHO/MDI. 
The rate of change in the WB Michelson phase is slowing down. 
The instrument tuning has been adjusted about once per year, 
as indicated in Table \ref{table:Retunes}.\footnote{See 
\urlurl{jsoc.stanford.edu/doc/data/hmi/hmi\_retuning.txt}}
The table also indicates the Wavelength Tuning ID number (WTID)
and the specific index positions of the three tuning motors.

\begin{table}[htb]
\caption{Dates of HMI Retunings}
\begin{tabular}{c c c c c}
	       & Wavelength & \multicolumn{3}{c}{Reference Tuning Position} \\
Retuning Date  & Tuning ID  & & & \\
and TAI Time   & (WTID)     & Lyot/E1 & Wideband & Narrowband \\
\hline
 30 Apr 2010 22:24 & 10 & 36 & 58 & 82 \\
 13 Dec 2010 19:45 & 11 & 37 & 56 & 82 \\
 13 Jul 2011 18:35 & 14 & 37 & 54 & 82 \\
 18 Jan 2012 18:15 & 17 & 37 & 53 & 81 \\
 14 Mar 2013 06:42 & 20 & 37 & 52 & 81 \\
 15 Jan 2014 19:13 & 23 & 37 & 51 & 80 \\
 08 Apr 2015 18:51 & 26 & 37 & 50 & 80 \\
 27 Apr 2016 18:56 & 29 & 37 & 50 & 79 \\
 19 Apr 2017 19:58 & 31 & 38 & 49 & 79 \\
\end{tabular}
\label{table:Retunes}
\end{table}

\begin{figure}[htb]
\centering
\includegraphics[width=\textwidth]{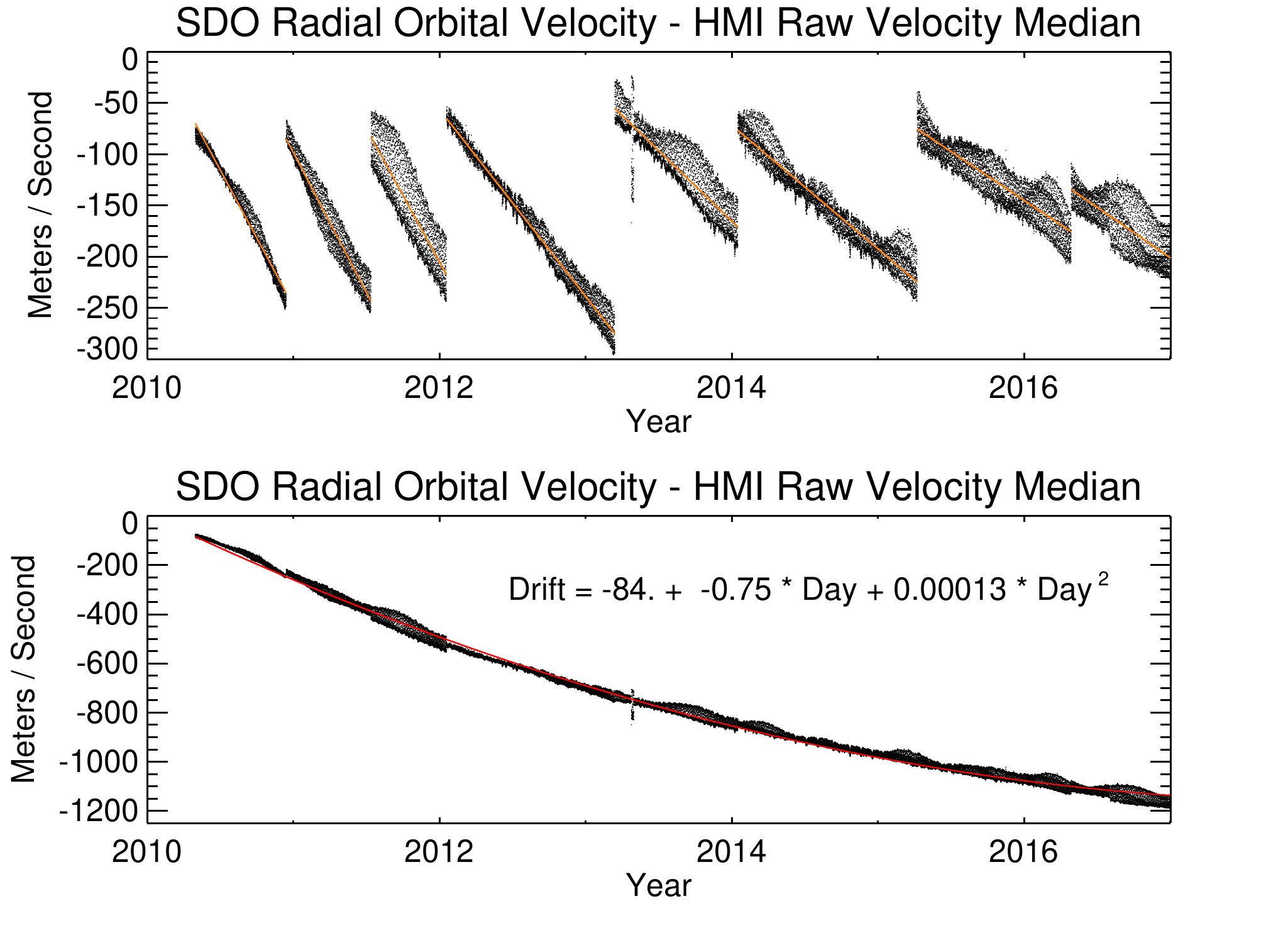}
\caption{Velocity drift of the HMI observable. The top panel shows the difference 
between the known Sun--SDO velocity and the median uncorrected velocity determined from 
an HMI Dopplergram. The drift in the measured velocity is due to the drift of the
HMI filter elements. Breaks in the curve occur when the filter tuning is changed.
The bottom panel shows the same thing, but with the velocity offset due to the retuning
removed. A polynomial fit to the velocity drift is given, which indicates the 
drift was initially slowing by -0.75 m\,s$^{-1}$ per day.}
\label{fig:velocitydrift}
\end{figure}

If the instrument were tuned and calibrated perfectly, the measured median
velocity of the Sun would be nearly the same as the Sun--SDO velocity. Figure
\ref{fig:velocitydrift} plots the difference between those two quantities,
demonstrating the effect of the slowly changing wavelength and the effects
of compensating changes in the HMI filter tuning. The Sun--SDO velocity is
known to a few mm\,s$^{-1}$ and the baseline zero offset is due to the nominal
tuning of the instrument. The daily scatter is due to the effects of changes
in the instrument environment and to actual solar signals that appear in
the median-velocity signal. Changes in the short-term noise level arise
from changes in sensitivity and imprefections in calibration discussed
elsewhere. The upper panel shows that the residual velocity decreases with
time at a significant rate and that the rate seems to slow with time. The
tuning has been adjusted regularly to keep the offset from zero less than
about 300 m\,s$^{-1}$. The bottom panel adds back in the velocity offset 
due to the changes in the tuning, as determined by matching the endpoints 
of the linear fit for each subset. A quadratic fit matches the curve very 
well and shows that the overall drift in meters per second is 
$-84 - 0.75 D + 0.00013 D^2$ for $D$ measured in days from the start of 
the prime mission.

The constant and evolving spatial characteristics of the HMI filter elements are 
described in considerable detail by \citet{Couvidat2016} and \citet{Couvidat2012}.
\section{Level-1 Corrections: Camera and Detector}
\label{sec:level1}

The Data Capture System (DCS) at Stanford's Joint Science Operations Center
(JSOC) receives raw science data directly from the SDO Ground Station;
housekeeping and other spacecraft data come via the mission operation center
at NASA/Goddard. The image data are extracted, combined with the appropriate
metadata, and packaged as image files. These raw, uncorrected filtergrams
are referred to as Level-0 data, and they are typically available within three
minutes of the image acquisition onboard the spacecraft. The first stage of
data processing applied to these images at the JSOC, which includes overscan
row removal, dark-current and flat-field correction, and cosmic-ray detection,
as well as added metadata, generates Level-1 data. This processing is done
twice: once as quickly as possible to generate the near-real-time (NRT)
data for use in space-weather applications, and then a second time, 
typically four days later, with occasional ground-based transmission
gaps filled and with better calibrations to generate the definitive Level-1
data. The Level-1 processing is described in this section.

\subsection{Dark-Current Correction}

Dark frames are taken with each camera twice a day as part of the 
calibration sequences started at 06:00~UT and 18:00~UT. Zero-length 
pedestal-current {\revtxtr (bias)} measurements are not taken; the CCD bias and dark 
current are measured together, and we do not distinguish between them.
The measured dark current in both cameras has been extremely
stable over the course of the mission, with average dark values of 122
counts and 131 counts for Cameras~1 and 2, respectively. To minimize the
impact of photon noise on the dark correction, average dark frames
are generated from the individual darks every three months, and it is these
averages that are used in the Level-1 processing. There is a diurnal variation
in the temperatures of the CCDs which likely gives rise to a small variation
in CCD dark signal, but this is not currently measured or corrected for.
In principle, data from the overscan area could provide additional information about 
dark current and other parameters for each image.

\subsection{Flat Field Correction}
\label{sec:Flatfield}

\begin{figure}[htb]
\centering
\includegraphics[width=\textwidth]{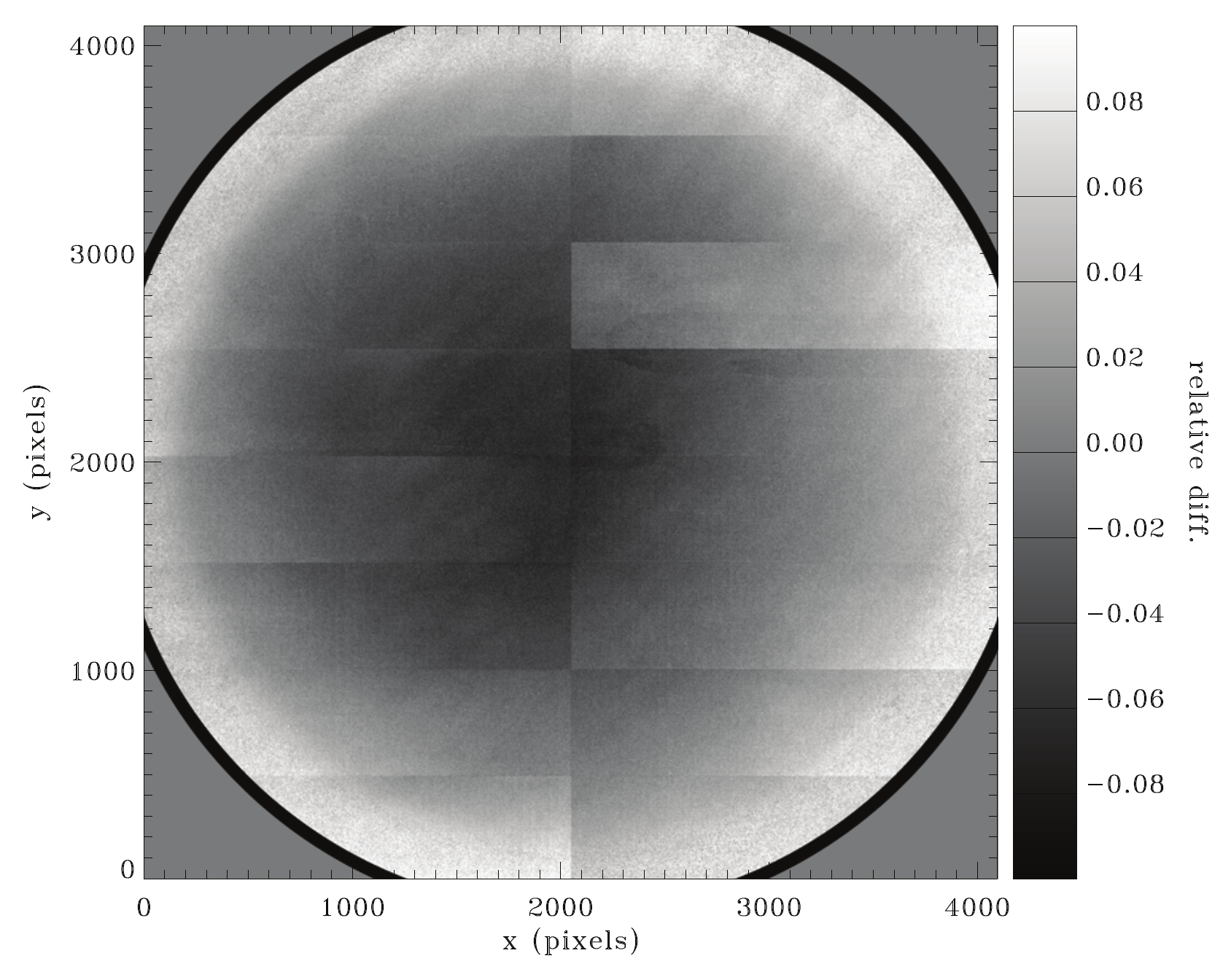}
\caption{Relative differences between a flat field from 23 January 2015 and
one from 1 March 2010. Both flat fields are for Camera 2.}
\label{figure:flat_diff}
\end{figure}

Pixel-to-pixel gain variations in the CCD detectors are corrected using
flat fields measured for each camera. Because there is no way to illuminate the
CCDs on orbit with a sufficiently uniform light source, the pixel gains are
determined by shifting the solar image to various locations on the CCDs. The procedure for using
these images to determine the flat field is described by \citet{Kuhn1991},
\citet{Toussaint2003}, and \citet{Wachter2012}. The solar image can be
shifted in two ways, and both are used in determining HMI flat fields. First,
the entire spacecraft can be slewed to a set of off-points. This is done
quarterly, and it involves nine off-point positions in a cruciform pattern.
The entire maneuver takes approximately two hours and forty minutes. The
second method uses the instrument's image stabilization system (ISS)
to shift the image. The piezo-electric transducers (PZTs) in the ISS are
activated to tilt the ISS mirror to a predetermined set of offsets.  
PZT flat fields are performed weekly to provide a good measure of 
small-spatial-scale sensitivity, whereas the quarterly offpoints provide a better 
large-scale flat field. The flat fields of both cameras have evolved
slowly over the course of the mission. The difference between the front
camera flat field at the begining and end of the prime mission is shown in
Figure \ref{figure:flat_diff}.

A different method of generating flat fields, using the rotation of the Sun to
smooth out inhomogeneities in the solar image, has also been implemented. The
algorithm used to calculate rotational flat fields is described by \citet{Wachter2009}.
Rotational flat fields are expensive to compute and are not used in the current 
Level-1 HMI data, since they provide only a small improvement over the PZT method.

\subsection{Bad Pixels and Cosmic Rays}

Each filtergram taken by HMI has a number of bad pixels that must be
identified and properly treated. There are a very small number
of totally bad pixels -- none in Camera~1 and just three in Camera~2. 
In addition, pixels from the quarterly off-point flat fields with
gains less than 50\,\% of the average gain are considered to be permanently
bad and are identified as such in each filtergram. The list of such pixels
is propagated into each Level-1 filtergram record. Camera~1 has 45 pixels
flagged as permanently bad, and this has been consistent since the beginning
of science operations. The number of bad pixels in Camera~2 increased
from 31 to 34 over the course of the prime mission. As with Camera~1, 
pixels flagged as bad are consistent from off-point to off-point.

Transient events (cosmic rays) account for the remainder of the
bad pixels in each filtergram. Cosmic ray hits are first detected by applying a
high-pass filter to each filtergram and flagging pixels that exceed a
certain threshold. In the production code, this threshold is 10.5 times the
variance in the center of the image. These pixels are included in the
Level-1 bad-pixel list. Cosmic rays are detected out to 0.98 of the solar radius, 
even though image statistics are computed to 0.99. This may be adjusted in
the near future.

A second cosmic-ray-detection algorithm is employed {\em after} individual
Level-1 filtergrams are generated. Run daily as part of the rotational
flat-field module, the algorithm identifes bad pixels in tracked locations
based on intensity variance over about 20 minutes. False identifications
in the initial per-filtergram detection algorithm are sometimes found. The
results for each image are logged, but they are not easy to recover. The
higher-level processing modules that combine multiple filtergrams to calculate
the observables \citep{Couvidat2016} exclude the bad pixels from the temporal
and spatial interpolation. This second cosmic-ray detection is not run for
HMI-NRT observables.

\begin{figure}[htb]
\centering
\includegraphics[width=\textwidth]{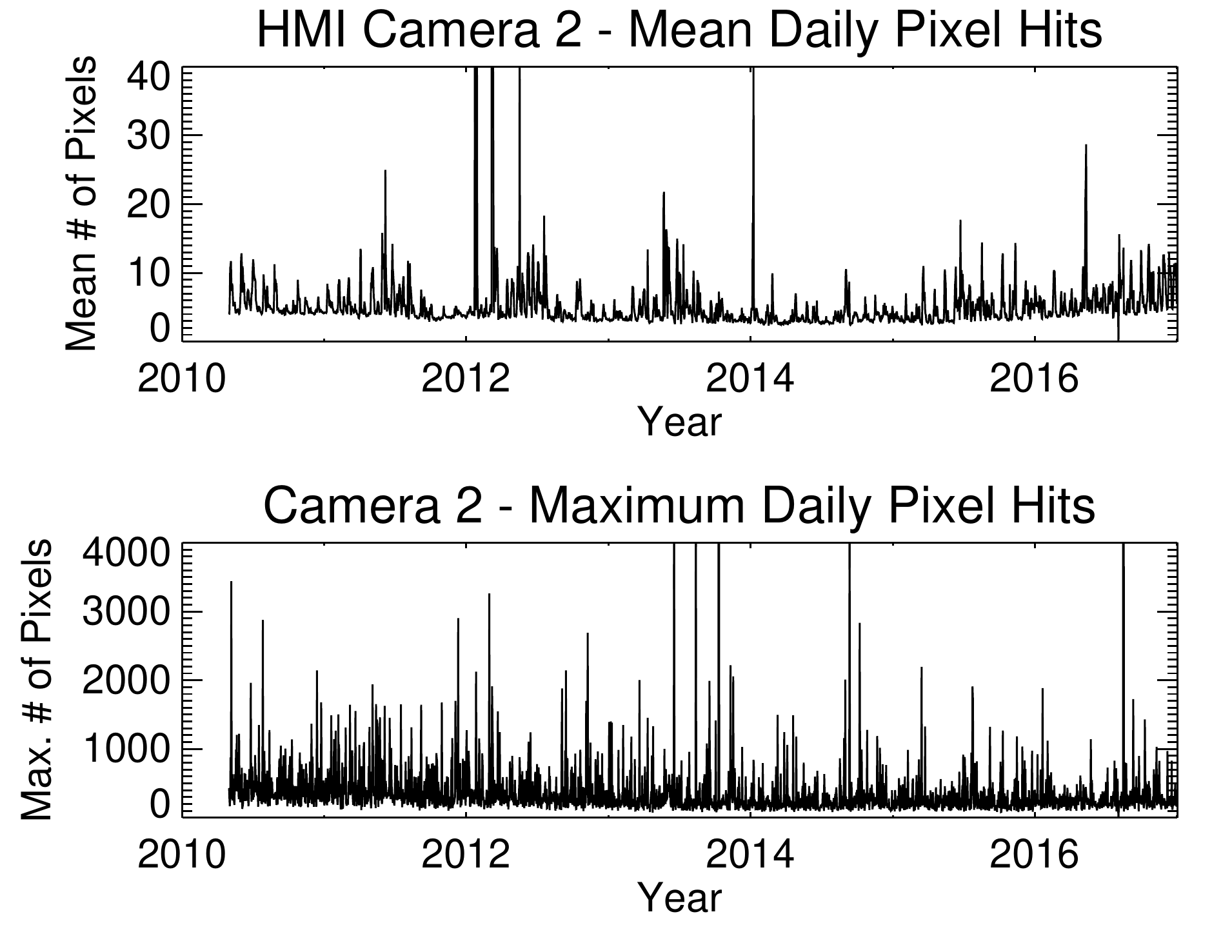}
\caption{Daily mean and maximum number of bad pixels per image as a function of time for Camera 2.}
\label{fig:Badpixels}
\end{figure}

The number of pixels removed due to cosmic rays varies throughout the year
and with solar activity. Figure \ref{fig:Badpixels} shows the daily mean and
maximum number of pixel hits in Camera 2. Camera 2, mounted on the Sun-facing
side of the instrument, generally takes roughly twice as many hits as the
other camera.

\subsection{Solar-Radius Correction for Height of Formation}

\begin{figure}
\centering
\includegraphics[width=0.9\textwidth]{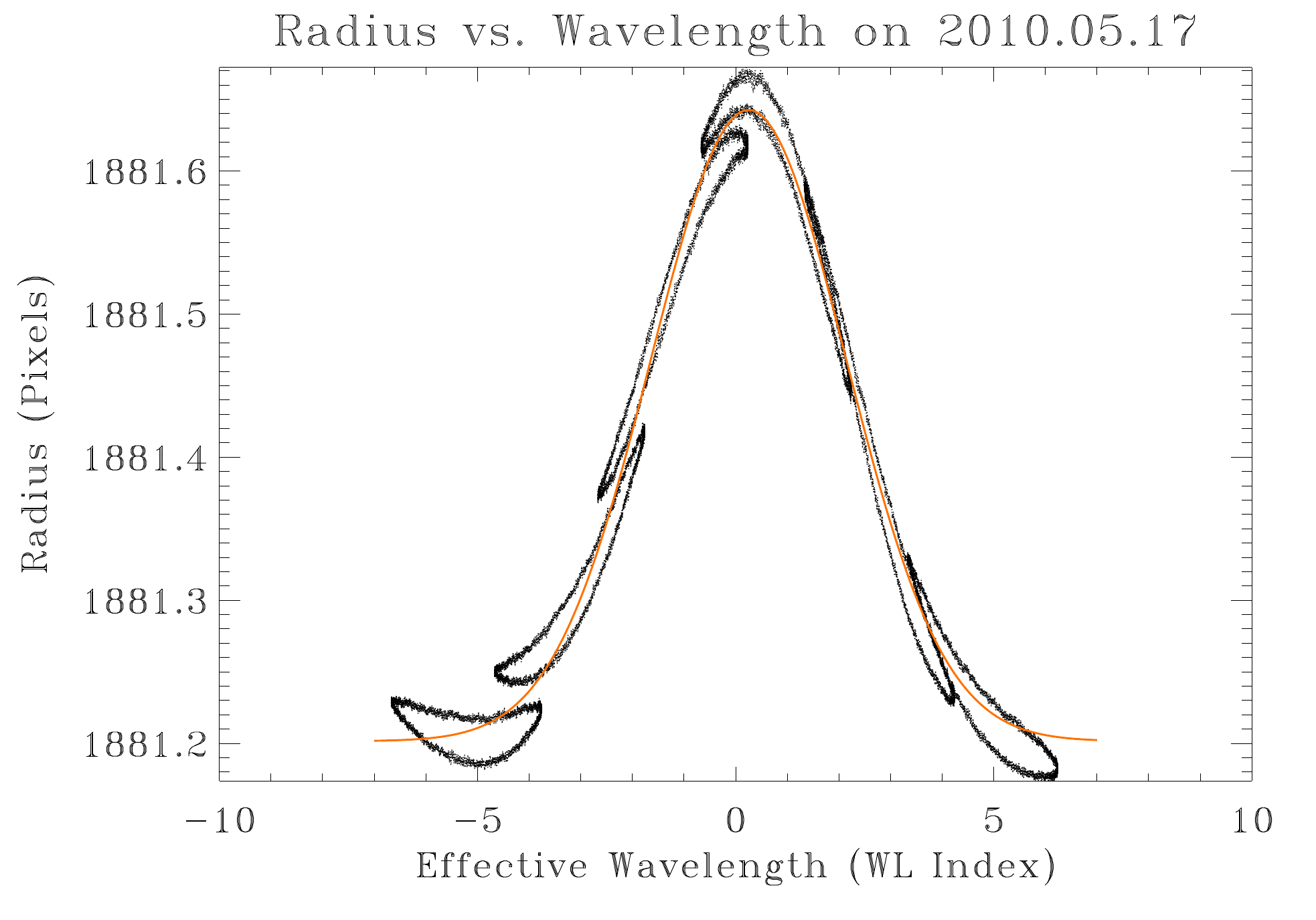}
\caption{Solar radius returned by the limb finder as a function of the
effective wavelength at which the image is taken. Each of the six closed 
loops shows the radius determined for a particular tuning of the HMI wavelength 
filter system over the course of 17 May 2010, as the solar line shifts relative to 
HMI during the orbit. The hysteresis arises because of temperature changes in 
the instrument correlated with orbital position. The solid line is the Gaussian 
fit described in the text computed for this particular day.}
\label{fig:RadiusWavelength}
\end{figure}

The height of formation near the 6173 \AA~Fe\,{\sc i} spectral line
changes with wavelength by a few hundred kilometers \citep{Fleck2011,
Emilio2015}. Because the standard HMI observing sequence samples the solar 
Fe\,{\sc i} line at six wavelengths separated by about $68.8$ m\AA, the apparent
size of the Sun varies with wavelength by as much as half a pixel. Figure
\ref{fig:RadiusWavelength} shows the measured solar radius as a function of
the wavelength index, where each index step corresponds to a nominal $34.4$
m\AA~HMI tuning-motor increment relative to line center.

Even though the location of the solar limb depends on wavelength, the 
physical scale of the image does not change. To account for this properly,
the radius returned by the limb finder is adjusted for use later in the
processing pipeline when filtergrams are resized. Specifically, the values
returned by the limb finder (\textsf{X0\_LF, Y0\_LF}, and \textsf{RSUN\_LF})
are corrected for the wavelength dependence in the keywords \textsf{CRPIX1,
CRPIX2}, and \textsf{R\_SUN}.

The limb-finder radius is reduced by a wavelength-dependent quantity 
\begin{equation}
\Delta R = A \, \mathrm{exp}(-(wl_x - wl_0)^2/wl_w), 
\end{equation}
where $wl_x=wl-$\textsf{OBS\_VR}/$dvdw$, $wl$ is the integer wavelength index
of the image relative to the index of the center wavelength, \textsf{OBS\_VR}
is the known Sun--SDO radial velocity, and 
dvdw = $\delta\lambda / \lambda \times c = 0.0344 / 6173.3433 \times 299792458$.  
The values of $A$, $wl_0$, and $wl_w$ are the result of a Gaussian fit 
to the solar radii returned by the limb-finder as a function of the wavelength 
position of the images.

\begin{figure}
\centering
\includegraphics[width=\textwidth]{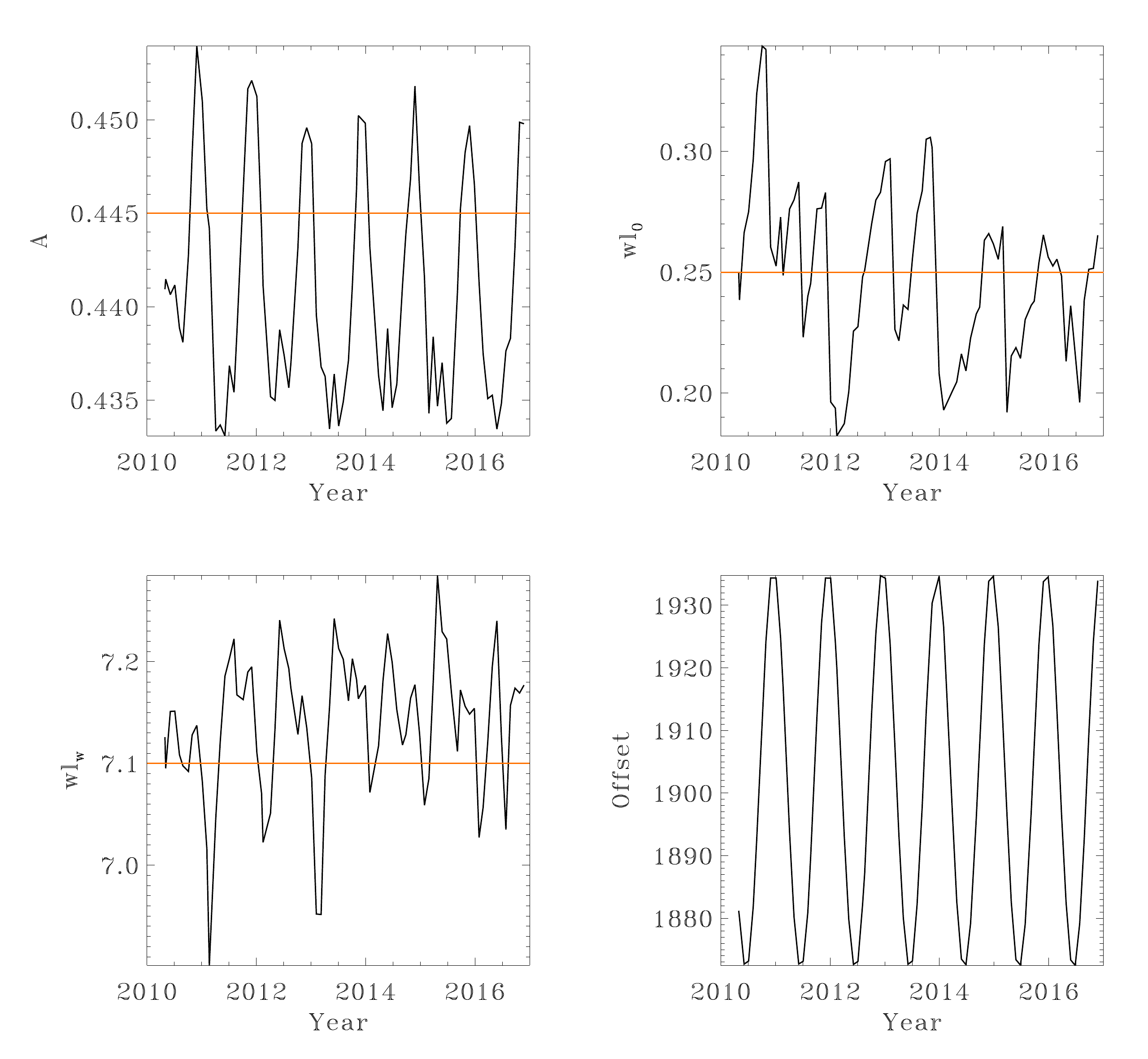}
\caption{Variation with time of the Gaussian-fit parameters that characterize
the height-of-formation correction. The upper-left panel is the scaling
factor [$A$]. The upper-right panel shows $wl_0$; the lower-left is $wl_w$;
and the lower-right is the offset due to distance (not used in the correction).
Eighty one-day fits are shown for months from May 2010 through December 2016. 
The standard values are indicated by the horizontal red lines. See text for
details.} 
\label{fig:Radiusparams} 
\end{figure}

The radius--wavelength relation varies somewhat from day to day depending on 
average velocity and the instrument environment.
Figure \ref{fig:Radiusparams} shows the observed temporal dependence of the 
three fitted parameters as well as the baseline offset due to Sun--spacecraft
distance. The observables pipeline code uses the following 
standard values: $A=0.445$, $wl_0=0.25$, and $wl_w=7.1$.
The standard value of $A$ appears in the plot to be too large by as much as 
0.005 arcseconds (about 35\,km), a significant fraction of the 55\,km error in the reference
solar radius \textsf{RSUN\_REF} discussed in Section \ref{sec:Platescale}.

A single radius and center-position correction is made for each filtergram,
but of course the velocity due to solar rotation also shifts the nominal
line position by a comparable amount. This east--west antisymmetric wavelength 
shift causes an additional position-angle-dependent radius change and image 
offset for which no correction is made.

\subsection{Additional Metadata}
\label{sec:Metadata}

Level-1 filtergrams are associated with a variety of metadata stored as
keywords in the JSOC database.  Information about the status of the instrument
from both spacecraft telemetry and the science data streams is associated with 
the Level-0 filtergrams, and the relevant data are propagated through to Level~1. The
Level-1 processing adds information about the spacecraft state, location,
and pointing, as well as image scale and centering. Information on spacecraft
position and velocity are obtained from spacecraft ephemeris data provided
by the flight operations team. Image coordinate information follows the WCS
standard \citep{Greisen2002} and is computed from a combination of a fit
to the solar limb and the spacecraft ephemeris information. Keywords set
in the Level-1 code are listed in Appendix
\ref{sec:Lev1Keywords} Table \ref{table:level1_wcs_orbit_keywords}.

In addition to these metadata, two keywords are set for the Level-1 filtergrams
that deserve somewhat closer attention: \textsf{QUALITY} and \textsf{CALVER}**.

\subsubsection{Image Quality and the \textsf{QUALITY} Keywords}
\label{sec:Quality}

While nearly all filtergrams taken by HMI over the course of the mission are of
nominal quality and suitable for scientific studies, a few are
taken under non-nominal conditions, are of degraded quality, or are completely
missing. The quality of each filtergram is indicated to the end user by a set
of flags stored bit-wise in a 32-bit integer named \textsf{QUALITY}. At Level
0 a \textsf{QUALITY} bit is set when an error occurs in the data transmission
and capture, or as a result of certain errors from the instrument. 
Appendix \ref{sec:Lev0Qual} Table \ref{table:lev0_quality} 
describes the Level-0 \textsf{QUALITY} bit masks and
meanings. This keyword is propagated to the Level-1 records as \textsf{QUALLEV0}.

At Level 1, a new \textsf{QUALITY} keyword is defined. The bit mask for each
flag and its meaning is shown in Table \ref{table:lev1_quality} in
Appendix \ref{sec:Lev1Qual}. Nominal science-quality filtergrams
have no flags set in the \textsf{QUALITY} keyword, and thus the value will
be zero. The most common reason for a non-zero \textsf{QUALITY} is that the
filtergram was taken as part of a daily or weekly calibration. In fact, many
such filtergrams are no different than those taken in the regular observing
sequence and can be used without concern for computing higher level HMI
observables.

The most common flag indicating a degraded filtergram is the ISS-loop-open flag,
which indicates that HMI's image stabilization system is not correcting for
image jitter. This happens during certain calibration sequences
and updates of the instrument configuration, but is most often due to the
spacecraft not being in its fine-guidance, or ``science'' mode. This is
indicated by the \textsf{ACS\_MODE} flag, and is usually due to spacecraft maneuvers
or lunar or Earth transits. Another \textsf{QUALITY} bit is set to indicate that
the instrument is in thermal recovery after a lunar or Earth transit ---
for a discussion of these intervals see Section \ref{sec:eclipses}.

Bits in the \textsf{QUALITY} keyword can also indicate missing metadata or filtergram
data. These are mostly due to occasional data corruption that occurs in the
instrument electronics; see Section \ref{sec:anomalies}. 

In fact, determining what constitutes a {\em good} measurement depends on
the use to which the observation is put.  The basic quality information for
higher level products, \textit{e.g.} Dopplergrams or magnetograms that are computed
from multiple filtergrams, is also indicated in an observables-level 
\textsf{QUALITY} keyword. These are listed in Tables \ref{table:Obs_quality}, 
\ref{table:Obs_quality2}, and \ref{table:Obs_quality3}
in Appendix \ref{sec:S_ObsQual}.

\subsubsection{Calibration Version and the \textsf{CALVER}** Keywords}

Changes to the instrument observing sequence, processing software, and
calibration constants, which we refer to collectively as the ``calibration
version,'' are rarely made, but each Level-1 filtergram includes a keyword,
\textsf{CALVER32}, that identifies the calibration version used to generate
the data. A longer keyword, \textsf{CALVER64}, is used by higher-level
data products to convey similar information. Unlike the \textsf{QUALITY}
keywords, the \textsf{CALVER}** keywords use nibbles, or 4-bit fields,
to denote various calibration changes. The meaning of each field is
shown in Table \ref{table:calver_keywords}. Currently, seven fields are
defined; more can be employed if and when new changes are introduced
into the processing of HMI data. For Level-1 data, only two of the fields
are used: the height-of-formation-correction version and the 
instrument-rotation-parameter version.  For all currently available Level-1 data,
the height-of-formation correction is Version 2 and the rotation parameter,
which was corrected after the 11 May 2012 Venus transit, is Version 1.

\begin{table}[htb]
\begin{tabular}{crrcl}
Field & Bits & Mask & Name & Note\\
\hline
0 & 0--3   & 0x0F & \textsf{HFCORRVR} & Height-of-formation code version used.\\
1 & 4--7   & 0xF0 & \textsf{CROTA2VR} & Version of \textsf{CROTA2} in Master Pointing Table.\\
2 & 8--11  & 0xF00 & N/A      & If $>$ 0: smooth look-up tables were used.\\
3 & 12--15 & 0xF000 & N/A      & If $>$ 0: a non-linearity correction was applied.\\
4 & 16--19 & 0xF0000 & \textsf{FRAMELST} & {\revtxtb If 0x0: Mod C; if 0x4: Mod L;}\\
 & & & & {\revtxtb  \,\,if 0x2 or 0x3: incorrectly processed Mod L.}\\
5 & 20--23 & 0xF00000 & N/A & If $>$ 0: PSF/scattered light deconvolution applied.\\
6 & 24--27 & 0xF000000 & N/A & If $>$ 0: rotational flat field used.\\
\hline
\end{tabular}
\caption{Key to Values of the CALVER** keyword nibbles.}
\label{table:calver_keywords}
\end{table}
\section{Significant Events and Anomalies}
\label{sec:eventsanomalies}

Through the prime mission, HMI's production of nominal science data was more than 
95\,\% complete. This section discuses that remaining 5\,\%: the events and anomalies that 
take place both routinely and unexpectedly that degrade or interrupt science data from HMI. 
The vast majority of these events are expected and planned for.  The semi-annual series of 
Earth eclipses, as well as occasional lunar transits, obscure HMI's view of the Sun. 
After eclipses, the most common interruptions are caused by planned calibration sequences 
used to ensure that calibration of HMI science data products continues to be as precise 
as possible; these are described in Section \ref{sec:calibobs}. 
Science-quality observations are also interrupted during 
spacecraft maneuvers, which are undertaken for instrument calibrations and for maintaining 
orbit and control.

On rare occasions, data are lost due to unexpected failures in the instrument, spacecraft, 
or ground systems. These anomalies are also discussed in this section. Fortunately, all of 
the anomalies encountered were recovered from 
fully without subsequent adverse effect on instrument health or data quality.

There are four basic ways in which HMI data quality can be affected. First, filtergrams can be taken 
that are not a part of the standard observing sequence; they are generally not used in 
generating the science data products. Second, images may be of degraded quality, due to the 
Sun not being centered, the stabilization system not being on, the instrument being out of 
nominal focus or temperature range, and so on. Third, image data or metadata may be corrupted, 
and finally the data may be missing entirely.

\subsection{Spacecraft Maneuvers}

The SDO spacecraft periodically performs maneuvers that interrupt HMI science-quality 
data. Many of these maneuvers are for instrument calibration: eight yearly off-point 
maneuvers for the EVE instrument, quarterly off-points for AIA and HMI flat fields, quarterly 
rolls for HMI image-quality monitoring, and quarterly maneuvers to calibrate the AIA guide 
telescopes (these are used for SDO fine-guidance). In addition to these regular maneuvers, 
there have been a few special maneuvers: twice to observe the star Regulus for calibration, on 
23 August 2010 and 23 August 2011, and for observations of comets Lovejoy and ISON on 
15 December 2011 and 28 November 2013, respectively. The spacecraft must also periodically 
perform burns of its propulsion system for maintenance of its orbit. These station-keeping 
maneuvers were performed 11 times during the prime mission. Finally, angular momentum must 
periodically be dumped from the reaction wheels by using the reaction control system (RCS) 
thrusters. This was done 21 times during the prime mission. Momentum management maneuvers 
take roughly 14 minutes; station-keeping maneuvers ordinarily take 35 minutes.
When possible, maneuvers are performed together to minimize the number of gaps.
An events table can be found at {\urlurl{aia.lmsal.com/public/sdo\_spacecraft\_events.txt}}.

\subsection{Earth Eclipses}
\label{sec:eclipses}

\begin{figure}
\centering
\includegraphics[width=\textwidth]{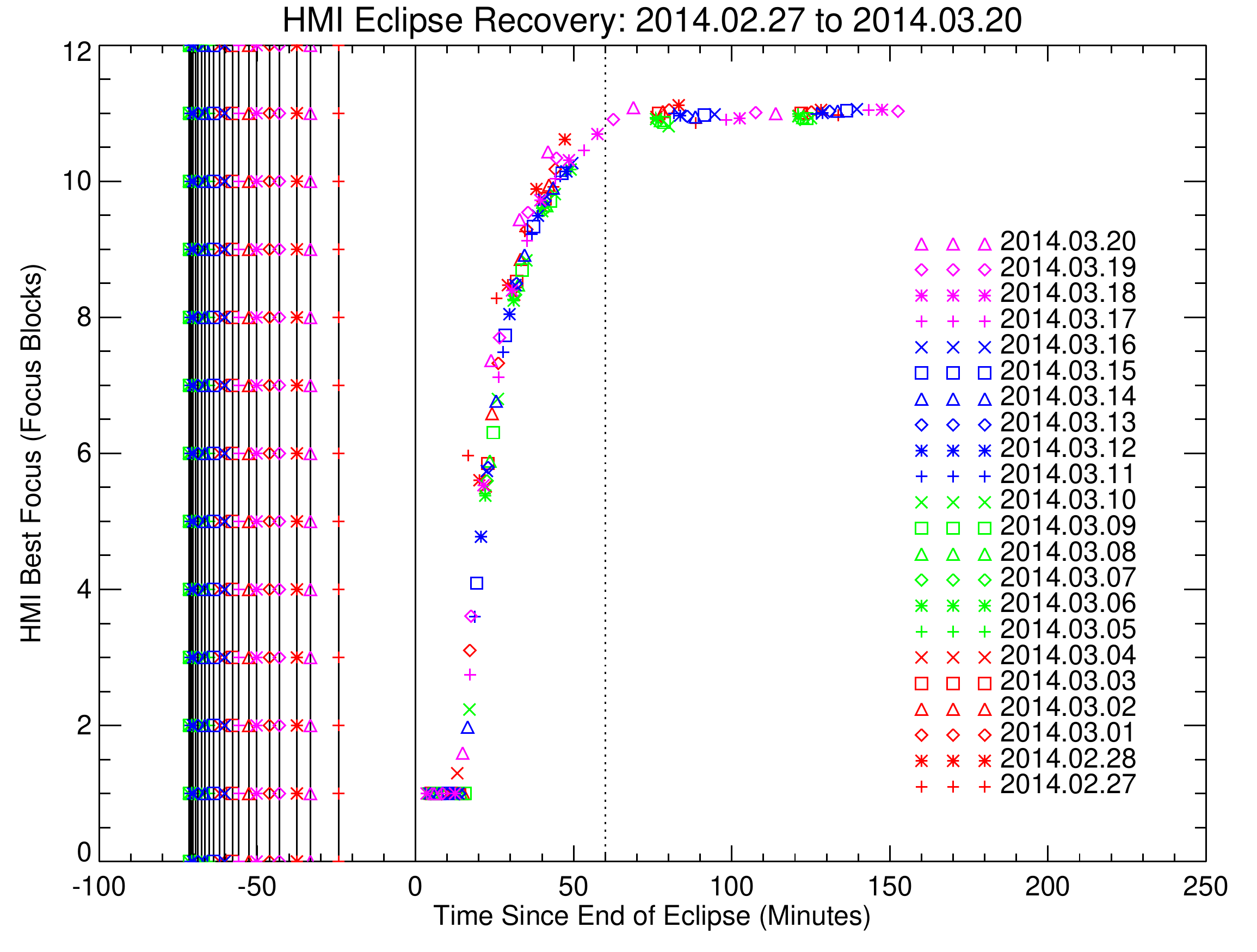}
\caption{HMI post-eclipse focus recovery during the Spring 2014 eclipse season.}
\label{figure:eclipse_recovery}
\end{figure}

Twice yearly, in Spring and Fall, SDO's view of the Sun is obscured by a series of Earth 
eclipses. There are between 22 and 24 such daily eclipses per season, occurring near local Midnight
of the SDO orbit around 06~UT, and they last up to 72 minutes. During the eclipse period 
the front-window temperature drops significantly, causing substantial change in instrument 
focus. After the end of each eclipse there is an extended period while the front-window 
temperature recovers and instrument focus recovers. Throughout the course of the mission 
the team has fine-tuned the use of front-window heaters to minimize this recovery time, which 
currently takes approximately one hour. During this recovery period, periodic focus sweeps 
are taken to monitor the recovery; focus profiles can be seen in Figure \ref{figure:eclipse_recovery}
for the Spring 2014 eclipse season.

\subsection{Lunar and Planetary Transits}

Although they are much less frequent than Earth eclipses, lunar eclipses occur several
times per year and cause interruptions in HMI's science data. Although the
Moon does not fully occult the solar disk, HMI's ISS must be disabled during
these transits, so science-quality data cannot be taken. In addition, the
decrease in solar flux decreases the temperature of the front window, which
causes a change in focus. The duration of these transits are highly variable,
but they typically last between one and three hours.

The planets Mercury and Venus can also pass between the Sun and SDO; this
happened for Mercury in May 2016 and for Venus in June 2012. Transits are
useful for calibrating the instrument roll angle, point spread function, and
distortion correction (Sections \ref{sec:pangle}, \ref{sec:Platescale}, 
and \ref{sec:Distortion}). HMI ran non-standard observing sequences during all of
the transits, which allowed the LoS observables to be produced but not the vector
products.

\subsection{Instrument Anomalies}
\label{sec:anomalies}

Instrument anomalies are caused by occasional and unpredictable problems with 
the operation of the instrument. Most anomalies result in one or two unusable images, 
in certain cases the outages can be hours or days.

\subsubsection{Corrupt Images}

\begin{figure}
\centering
\includegraphics[width=\textwidth]{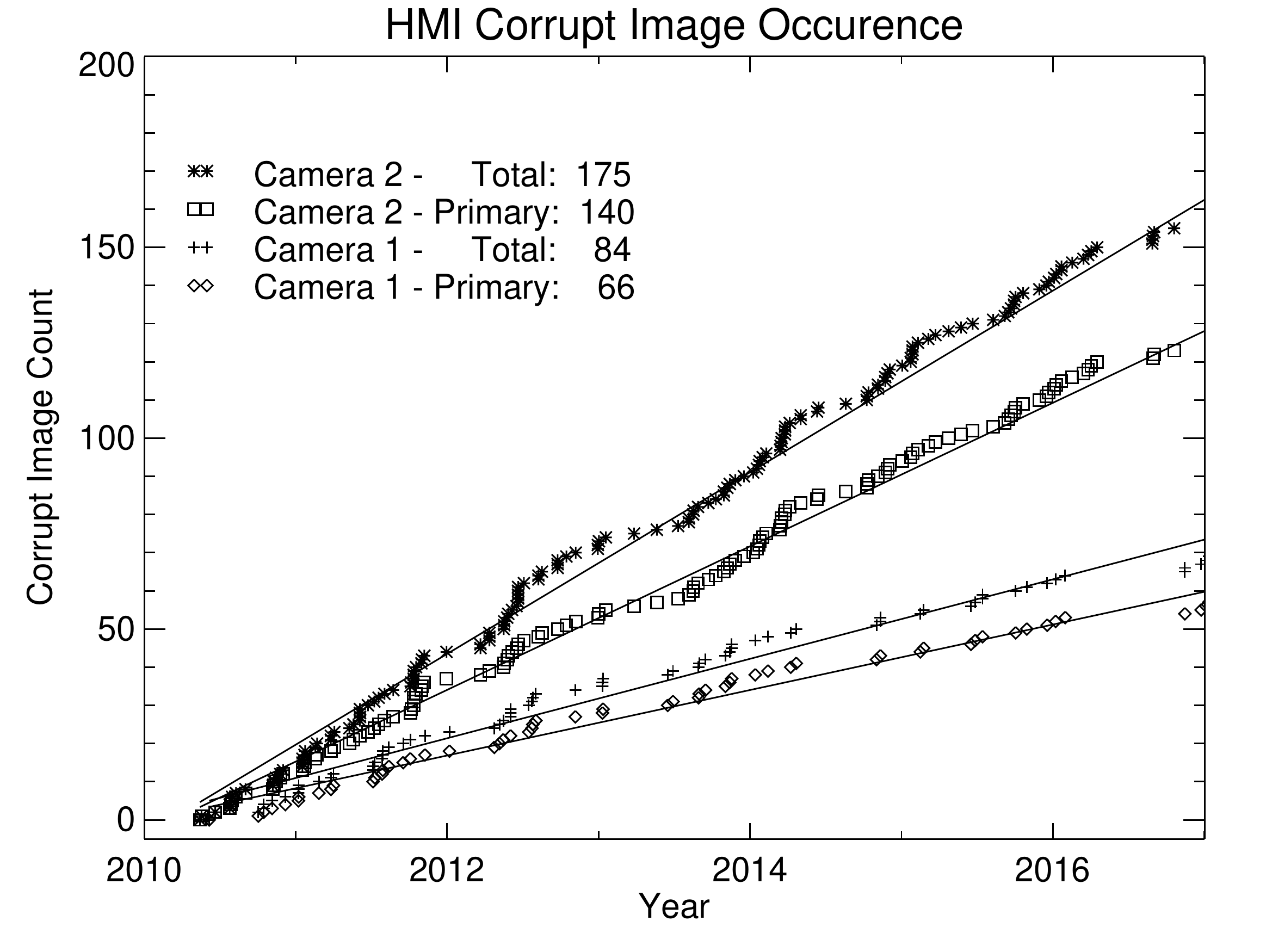}
\caption{Occurrence of corrupt images as a function of time for the two HMI cameras. The larger
total for each camera counts both primary hits and the occasional corruption of the subsequent image.}
\label{figure:corrupt_images}
\end{figure}

On occasion the image file or associated telemetry arrive corrupted at the data-capture system. 
It is believed that most of these occurrences originate in the camera electronics on the spacecraft, 
possibly due to cosmic ray hits. The fraction of images lost this way is roughly one out of every 
million. The front camera suffers from roughly twice as many instances as the side camera. A 
cumulative count of corrupt images for each camera is shown in Figure \ref{figure:corrupt_images}. 
In some instances, corruption of one image affects the data in the following frame, so that the 
total number of corrupted images is somewhat larger than the number of primary hits.

\subsubsection{Camera System Errors}

Persistent data losses can occur due to errors in the HMI electronics Camera Interface (CIF)
cards or in the Data Capture--High Rate Interface (DC-HRI) cards that require
intervention from the ground to clear. Errors on the DC-HRI cards involve
bit-flips to tables loaded into the FPGAs (Field Programmable Gate Arrays) on
the cards that determine how the image data are read out. Every table loaded on
the cards is checked continuously for parity errors, and alerts are generated
when a parity error is detected. The instrument sequencer is then stopped,
and the tables are reloaded to clear the parity error. Two types of
tables have been affected: the crop tables, which define the area of each
image to be stored and downlinked (to save bandwidth, the areas of the image
off the solar limb are not downlinked), and the look-up tables that are used
for data compression. Errors to the crop tables result in garbled images,
while errors to the look-up tables result in one pixel value being changed
to another. Garbled images from crop-table errors can be reconstructed,
although in some cases several rows may have missing values.
Incorrect data due to look-up table errors generally cannot be fixed, but they do
not appreciably affect the quality of the data because only a very few pixels are
affected. CIF card errors result in garbled image-header data. A list of the
camera anomalies experienced by HMI is shown in Table \ref{table:anom}. When
the first anomaly occurred, the error bit set by the parity check was not
being monitored, and the effect on the images was simply one partial row of
bad values that was difficult to see by eye; consequently, the error was not noticed for
almost three weeks. In all subsequent events, the recovery time has been
determined by how quickly HMI and SDO personnel can begin commanding the
instrument. The majority of camera anomalies have been experienced by Camera 2
(the front/Doppler camera) and they have been increasing somewhat in frequency.
Images affected by anomalies are indicated in Level-0 and Level-1 \textsf{QUALITY} 
bits.

\subsubsection{HMI Reboots and Restarts}

The HMI instrument has been rebooted on three separate occasions. The first
occurrence was on 24 April 2013, and it was initiated by an error from the
processor watchdog that halted the processor. Although most instrument
functions were halted, the instrument remained powered on and in the
configuration it was in when the error message was generated. Recovery took
fifteen hours and fifteen minutes. Subsequent analysis of the pre-anomaly
telemetry did not reveal what caused the watchdog error message.  After the
first event, an improperly set sequencer parameter led to errors in the
positions of the tunable elements in the Michelsons and Lyot filter, and
thus incorrect tuning of the instrument. This error was not corrected until
30 April 2013.
A similar event occurred 17 May 2014 with a faster recovery time (eight
hours and twenty minutes).

The third HMI reboot involved a full power-down of the instrument when the
SDO spacecraft entered Sun-acquisition mode on 2 August 2016 and powered down
most of its subsystems including all three instruments. The HMI instrument
was fully powered on and recovered the following day, but science data could
not be taken until all of the observatory's fine-guidance systems were recovered
and calibrated, which did not occur until 4 August 2016.

\begin{table}
\begin{tabular}{rcrcl}
Date & Time & Duration & Camera & Event Type \\
\hline
22 Dec 2011 & 08:41 UT & 20d 12h 34m & 2 & Crop table \\
24 Apr 2013 & 03:47 UT &  6d  9h 46m & - & Processor restart; tuning error \\
22 Jul 2013 & 13:21 UT &      4h 59m & 1 & Look-up table \\
11 Oct 2013 & 04:54 UT &      2h 51m & 2 & Header error \\
30 Mar 2014 & 12:20 UT &      4h 27m & 1 & Look-up table \\
23 Jun 2014 & 09:32 UT &      2h 45m & 1 & Header error \\
23 Mar 2015 & 23:39 UT &      0h 53m & 2 & Header error \\
17 May 2015 & 14:30 UT &      8h 20m & - & Processor restart \\
16 Nov 2015 & 12:02 UT &      2h 39m & 2 & Header error \\
16 Feb 2016 & 19:39 UT &      0h 53m & 2 & Header error \\
06 Apr 2016 & 02:04 UT &      2h 24m & 2 & Look-up table \\
10 Jun 2016 & 08:26 UT &      9h 25m & 2 & Header error \\
13 Jun 2016 & 15:56 UT &      1h 47m & 2 & Look-up table \\
02 Aug 2016 & 11:31 UT &  1d 13h 23m & - & SDO load shed -- intermittent issues until 10 Aug \\
16 Aug 2016 & 03:02 UT &     13h 20m & 2 & Header error \\
16 Nov 2016 & 20:17 UT &      1h 07m & 2 & Header error \\
31 Dec 2016 & 08:49 UT &     10h 49m & 2 & Crop table \\
11 May 2017 & 19:45 UT &      1h 09m & 2 & Header error \\
12 Jun 2017 & 01:10 UT &     14h 41m & 2 & Look-up table \\
15 Dec 2017 & 17:01 UT &      1h 27m & 2 & Header error \\
\hline
\end{tabular}
\caption{Camera system and other anomalies experienced by HMI through December 2017.}
\label{table:anom}
\end{table}
\section{Conclusions}
\label{sec:conclusion}

The HMI instrument has performed nearly flawlessly since the start of
regular SDO operations on 1 May 2010. Nearly 120 million filtergrams have been
collected and more than 98\,\% of all possible 45-second Dopplergrams have
been recovered. The HMI instrument and SDO spacecraft have experienced only 
a very few anomalies, none of which caused extensive data loss.

The HMI team has monitored the instrument continuously to maintain and perfect
the calibration of the instrument. Such activities include long-term trending
of environmental, optical, spectral, and camera characteristics and analysis 
of daily, weekly, and quarterly calibration measurements to verify performance.

Trends in slowly varying parameters, such as the instrument focus, filter
tuning, and exposure time, are regularly evaluated, and in some cases
operation of the instrument is adjusted to maintain uniform data quality. 

For other quantities, such as distortion, wavelength-dependent formation
height, and alignment, values are refined and corrections are made to 
observable quantities as better data become available.

Most of the periodic variations are responses to changes in the thermal environment,
largely due to predictable eclipse seasons, planned events, or daily
and annual orbital variations.  The thermal-control scheme of the instrument
was improved to reduce daily and annual variations inside the instrument.

The goal of all this effort is to provide complete and uniform-quality record
of conditions at the Sun over the solar cycle. The observable quantities
-- Doppler velocity, intensities, and magnetic field -- and downstream
higher-level products -- convection-zone flow maps, internal rotation,
synoptic maps of the photosphere and corona, comprehensive characteristics of
active-region evolution -- all depend on having a well-calibrated instrument
with sufficient information available to eliminate or at least understand the 
sources of uncertainty in the measurements.

\begin{acks}
We thank all of the many team members who have contributed to the success of the
SDO mission and particularly to the HMI instrument. In particular we acknowledge
the contributions of HMI Calibration Team members R.S. Bogart, S. Couvidat, Y. Liu, 
and A.A. Norton. We also thank W. Liu and J. Sommers for assistance with preparation 
of the data and the manuscript.
This work was supported by NASA Contract NAS5-02139 (HMI) to Stanford University.
The German Data Center for SDO is supported by the German Aerospace Center (DLR) 
and the State of Niedersachsen. 
\end{acks}

\section*{Disclosure of Potential Conflicts of Interest}
The authors declare that they have no conflicts of interest.

\newcommand{\adv}{    {\it Adv. Space Res.}}
\newcommand{\annG}{   {\it Ann. Geophys.}}
\newcommand{\aap}{    {\it Astron. Astrophys.}}
\newcommand{\aaps}{   {\it Astron. Astrophys. Suppl.}}
\newcommand{\aapr}{   {\it Astron. Astrophys. Rev.}}
\newcommand{\ag}{     {\it Ann. Geophys.}}
\newcommand{\aj}{     {\it Astron. J.}}
\newcommand{\apj}{    {\it Astrophys. J.}}
\newcommand{\apjl}{   {\it Astrophys. J. Lett.}}
\newcommand{\apss}{   {\it Astrophys. Space Sci.}}
\newcommand{\cjaa}{   {\it Chin. J. Astron. Astrophys.}}
\newcommand{\gafd}{   {\it Geophys. Astrophys. Fluid Dyn.}}
\newcommand{\grl}{    {\it Geophys. Res. Lett.}}
\newcommand{\ijga}{   {\it Int. J. Geomagn. Aeron.}}
\newcommand{\jastp}{  {\it J. Atmos. Solar-Terr. Phys.}}
\newcommand{\jgr}{    {\it J. Geophys. Res.}}
\newcommand{\mnras}{  {\it Mon. Not. Roy. Astron. Soc.}}
\newcommand{\nat}{    {\it Nature}}
\newcommand{\pasp}{   {\it Pub. Astron. Soc. Pac.}}
\newcommand{\pasj}{   {\it Pub. Astron. Soc. Japan}}
\newcommand{\pre}{    {\it Phys. Rev. E}}
\newcommand{\solphys}{{\it Solar Phys.}}
\newcommand{\sovast}{ {\it Soviet  Astron.}}
\newcommand{\ssr}{    {\it Space Sci. Rev.}}

\bibliographystyle{spr-mp-sola}

\bibliography{Performance}


\clearpage 
\appendix

\section{Informational Tables Characterizing HMI Performance}
\label{sec:Appendix1}

The following tables provide more detail than is presented in the main text 
about Dopplergram Recovery Rates and common Frame Lists. 

\subsection{HMI Dopplergram Recovery--72-Day Intervals}
\label{sec:72daytables}

Helioseismology requires long uninterrupted time series to determine precise
oscillation frequencies. A useful way to characterize the instrument
performance is to determine for each 72-day time interval the fraction
of good-quality observations recovered by the instrument. The HMI Level-1
mission success requirement for Data Completeness was to collect at least
95\,\% of all observations during 22 72-day continguous intervals. 

Table \ref{table:72day_recovery} shows the Dopplergram recovery rate
for the first 37 72-day intervals of the HMI mission, from 30 April 2010
to 14 August 2017. Nominal HMI operations began 1 May 2010, but velocity
data were collected early, starting 30 April 2010. For reference 30 April 2010 is
MDI Day 6328.  {\em Perfect} Dopplergrams are those with no \textsf{QUALITY}
bits set, {\em imperfect} Dopplergrams are those with any bit set -- most of
which are usable for helioseismology. The percentage in the table is the 
fraction of all possible time slots for which a Dopplergram was recovered. 
There are 138,240 45-second time slots in each 72-day interval.


\begin{table}[htb]
\begin{tabular}{rrrrrrrr}
72-day & GONG    & Start & End  & \# Perfect   & \# Imperfect & Number  & Percent  \\
Count  & Months  & Date  & Date & Dopplergrams & Dopplergrams & Missing & Recovery \\
\hline
 1 & 153/154 & 30 Apr 2010 & 10 Jul 2010 & 136,034 &  1736 &   470 & 99.66\,\% \\
 2 & 155/156 & 11 Jul 2010 & 20 Sep 2010 & 133,524 &  2404 & 2312 & 98.33\,\% \\
 3 & 157/158 & 21 Sep 2010 & 01 Dec 2010 & 129,886 &  4315 & 4039 & 97.08\,\% \\
 4 & 159/160 & 02 Dec 2010 & 11 Feb 2011 & 135,088 &  1712 & 1440 & 98.96\,\% \\
 5 & 161/162 & 12 Feb 2011 & 24 Apr 2011 & 130,546 &  3176 & 4518 & 96.73\,\% \\
 6 & 163/164 & 25 Apr 2011 & 05 Jul 2011 & 136,141 &  1747 &   352 & 99.75\,\% \\
 7 & 165/166 & 06 Jul 2011 & 15 Sep 2011 & 134,567 &  2054 & 1619 & 98.83\,\% \\
 8 & 167/168 & 16 Sep 2011 & 26 Nov 2011 & 130,699 &  3251 & 4290 & 96.90\,\% \\
 9 & 169/170 & 27 Nov 2011 & 06 Feb 2012 & 134,547 &  2408 & 1285 & 99.07\,\% \\
10 & 171/172 & 07 Feb 2012 & 18 Apr 2012 & 130,278 &  4007 & 3955 & 97.14\,\% \\
11 & 173/174 & 19 Apr 2012 & 29 Jun 2012 & 135,801 &  2244 &   195 & 99.86\,\% \\
12 & 175/176 & 30 Jun 2012 & 09 Sep 2012 & 135,210 &  1754 & 1276 & 99.08\,\% \\
13 & 177/178 & 10 Sep 2012 & 20 Nov 2012 & 131,729 &  3065 & 3446 & 97.51\,\% \\
14 & 179/180 & 21 Nov 2012 & 31 Jan 2013 & 135,983 &  1306 &   951 & 99.31\,\% \\
15 & 181/182 & 01 Feb 2013 & 13 Apr 2013 & 131,594 &  3308 & 3338 & 97.59\,\% \\
16 & 183/184 & 14 Apr 2013 & 24 Jun 2013 & 123,818 & 12,581 & 1841 & 98.67\,\% \\
17 & 185/186 & 25 Jun 2013 & 04 Sep 2013 & 135,469 &  1473 & 1298 & 99.06\,\% \\
18 & 187/188 & 05 Sep 2013 & 15 Nov 2013 & 131,065 &  3331 & 3844 & 97.22\,\% \\
19 & 189/190 & 16 Nov 2013 & 26 Jan 2014 & 136,018 &  1308 &   914 & 99.34\,\% \\
20 & 191/192 & 27 Jan 2014 & 08 Apr 2014 & 131,377 &  3079 & 3784 & 97.26\,\% \\
21 & 193/194 & 09 Apr 2014 & 19 Jun 2014 & 135,026 &  1769 & 1445 & 98.95\,\% \\
22 & 195/196 & 20 Jun 2014 & 30 Aug 2014 & 135,639 &  1438 & 1163 & 99.16\,\% \\
23 & 197/198 & 31 Aug 2014 & 10 Nov 2014 & 131,737 &  3306 & 3197 & 97.69\,\% \\
24 & 199/200 & 11 Nov 2014 & 21 Jan 2015 & 135,665 &  1459 & 1116 & 99.19\,\% \\
25 & 201/202 & 22 Jan 2015 & 03 Apr 2015 & 130,286 &  3182 & 4772 & 96.55\,\% \\
26 & 203/204 & 04 Apr 2015 & 14 Jun 2015 & 135,458 &  1285 & 1497 & 98.92\,\% \\
27 & 205/206 & 15 Jun 2015 & 25 Aug 2015 & 135,361 &  1873 & 1006 & 99.27\,\% \\
28 & 207/208 & 26 Aug 2015 & 05 Nov 2015 & 131,652 &  3053 & 3535 & 97.44\,\% \\
29 & 209/210 & 06 Nov 2015 & 16 Jan 2016 & 133,338 &  2587 & 2315 & 98.33\,\% \\
30 & 211/212 & 17 Jan 2016 & 28 Mar 2016 & 131,605 &  3306 & 3329 & 97.59\,\% \\
31 & 213/214 & 29 Mar 2016 & 08 Jun 2016 & 134,934 &  2221 & 1085 & 99.22\,\% \\
32 & 215/216 & 09 Jun 2016 & 19 Aug 2016 & 128,650 &  4683 & 4907 & 96.45\,\% \\
33 & 217/218 & 20 Aug 2016 & 30 Oct 2016 & 130,822 &  3479 & 3939 & 97.15\,\% \\
34 & 219/220 & 31 Oct 2016 & 10 Jan 2017 & 136,213 &  1842 &   185 & 99.87\,\% \\
35 & 221/222 & 11 Jan 2017 & 23 Mar 2017 & 130,960 &  3820 & 3460 & 97.50\,\% \\
36 & 223/224 & 24 Mar 2017 & 03 Jun 2017 & 135,716 &  1425 & 1099 & 99.21\,\% \\
37 & 225/226 & 04 Jun 2017 & 14 Aug 2017 & 135,014 &  1944 & 1282 & 99.07\,\% \\
\hline
   & Total   & 30 Apr 2010 & 14 Aug 2017 & 4,927,450 & 102,931 & 84,499 & 98.35\,\% \\
\hline
\end{tabular}
\caption{72-Day HMI Dopplergram Recovery: 30 April 2010 -- 18 August 2017}
\label{table:72day_recovery}
\end{table}
\subsection{Primary HMI Observing and Calibration Frame Lists}
\label{sec:MainSequences}

The HMI Framelist Timeline Specification (FTS) specifies the sequence, timing, 
and instrument configuration for exposures in an observation. Each is given a
unique identification number, the FTS~ID.
Table \ref{table:prime_framelists} 
gives information about the primary observing and 
calibration frame lists used during the mission. 

HMI has used two basic observing sequences since commencing regular operations.
The Mod-C sequence (FTS~IDs 1001 and 1021) was used throughout the prime
mission to collect the standard data. On 13 April 2016, after the prime
mission ended, HMI switched to a faster sequence, FTS~ID 1022, also known as
Mod~L. The Mod-L sequence requires that images from both cameras be combined
to determine the vector-field observables. The part of the frame list for
the line-of-sight observables using Camera 2 did not change.

Various calibration sequences are taken on a regular basis to monitor the
evolution of HMI performance.
Note that while the JSOC does not keep most older HMI Level-1 data on-line,
all of the calibration-related Level-1 data are copied into a data series 
{\tt hmi.lev1\_cal}, which is permanently on-line.

A more complete list of HMI framelists appears in 
Tables \ref{table:fts_summary1} and \ref{table:fts_summary2}.
of Appendix \ref{sec:FTSTables}

\begin{table}[htb]
\begin{tabular}{crrp{37mm}p{42mm}}
\multicolumn{5}{l}{Standard Observables Framelists} \\[2pt]
FTS ID & Framelist & Duration & Description & When used \\
\hline
1001 & \textsf{obs\_6Cv01} & 135 & Mod C -- Standard sequence & 1 May -- 13 December 2010 \\
1020 & \textsf{obs\_6Av02} & 90  & Mod A -- Standard sequence & Tested before 1 May 2010 \\
1021 & \textsf{obs\_6Cv02} & 135 & Mod C -- Standard sequence & 13 Dec 2010 -- 13 Apr 2016 \\
1022 & \textsf{obs\_6Lv02} & 90  & Mod L -- Standard sequence & Since 13 April 2016 \\
1026 & \textsf{obs\_10v02} & 150 & Mod A -- Ten wavelengths & Tested 24 Oct 2014 \\[4pt]
\multicolumn{5}{l}{Daily, Weekly or Bi-weekly Calibration Framelists}\\
\hline
2001 & \textsf{cal\_6Cv01} & 135 & Mod-C Darks, continuum, Calmode frames & Daily at 06:00UT and 18:00UT, 1 May -- 13 Dec 2010 \\
2021 & \textsf{cal\_6Cv02}       & 135 & Mod-C Darks, continuum, Calmode frames & Daily at 06 and 18\,UT, 13 Dec 2010 -- 13 Apr 2016 \\
2042 & \textsf{cal\_6Lv02}       &  90 & Mod-L Darks, continuum, Calmode frames & Daily at 06 and 18\,UT, Since 13 April 2016 \\[3pt]
3020 & \textsf{focr\_6Cv02}      & 135 & Mod-C Reduced focus sweep & Run three times every four weeks, Until 13 April 2016 \\
3021 & \textsf{pzt\_def\_6Cv02}  & 135 & Mod-C Obsmode PZT flat & Run twice per week, Until 13 April 2016 \\
3022 & \textsf{pzt\_cal\_6Cv02}  & 135 & Mod-C Calmode PZT flat & Run once per week, Until 13 April 2016 \\
3023 & \textsf{focus\_6Cv02}     & 135 & Mod-C Full focus sweep & Run once every four weeks, Until 13 April 2016 \\
3027 & \textsf{det\_cal\_6Cv02}  & 135 & Calmode detune sequence & Run once every two weeks (too long for a 90-second framelist) \\
3040 & \textsf{focr\_6Lv02}      &  90 & Mod-L Reduced focus sweep & Run 3 times every four weeks, Since 13 April 2016 \\
3041 & \textsf{pzt\_def\_6Lv02}  &  90 & Mod-L Obsmode PZT flat & Run twice per week, Since 13 April 2016 \\
3042 & \textsf{pzt\_cal\_6Lv02}  &  90 & Mod-L Calmode PZT flat & Run once per week, Since 13 April 2016 \\
3043 & \textsf{focus\_6Lv02}     &  90 & Mod-L Full focus sweep & Run once every four weeks, Since 13 April 2016 \\[4pt]
\multicolumn{5}{l}{Eclipse and Calibration-Maneuver Framelists}\\
\hline
3003 & \textsf{focus\_6Cv01}     & 135 & Full focus sweep & After Earth eclipses in 2010 \\
3008 & \textsf{focus\_6Cv01}     & 135 & Full focus sweep, repeating every nine minutes & After Earth eclipses in 2010 \\
3012 & \textsf{focr\_6Cv01}      & 135 & Reduced focus sweep, repeating every 45 minutes & After Earth eclipses in 2010 \\
3028 & \textsf{focus\_6Cv02}     & 135 & Full focus sweep, repeating every nine minutes & After Earth eclipses, 2011 -- March 2016 \\
3031 & \textsf{focr\_6Cv02}      & 135 & Reduced focus sweep, repeating every 33 min 45 sec & After Earth eclipses, 2011 -- March 2016 \\
3128 & \textsf{focus\_6Lv02}     &  90 & Full focus sweep, repeating every nine minutes & After Earth eclipses, since August 2016 \\
3132 & \textsf{focr\_6Lv02}      &  90 & Reduced focus sweep, repeating every 45 minutes & After Earth eclipses, since August 2016 \\[3pt]
4031 & \textsf{focus\_off\_v02}  &  45 & Reduced focus sweep for offpoint maneuvers & During HMI/AIA flat field and EVE FOV maneuvers \\
4033 & \textsf{rolldopic\_v02}   &  45 & Set of continuum filtergrams on side camera & During HMI roll maneuvers \\
\end{tabular}
\caption{HMI Primary Observing and Calibration Framelists}
\label{table:prime_framelists}
\end{table}

\section{Exposure Time and Filter-Wheel Delays}
\label{sec:FigureSupplement}

Figures presented in this section 
provide more detail about the operation of some components.

\subsection{Exposure Time}
\begin{figure}[htb]
\centering
\includegraphics[width=\textwidth]{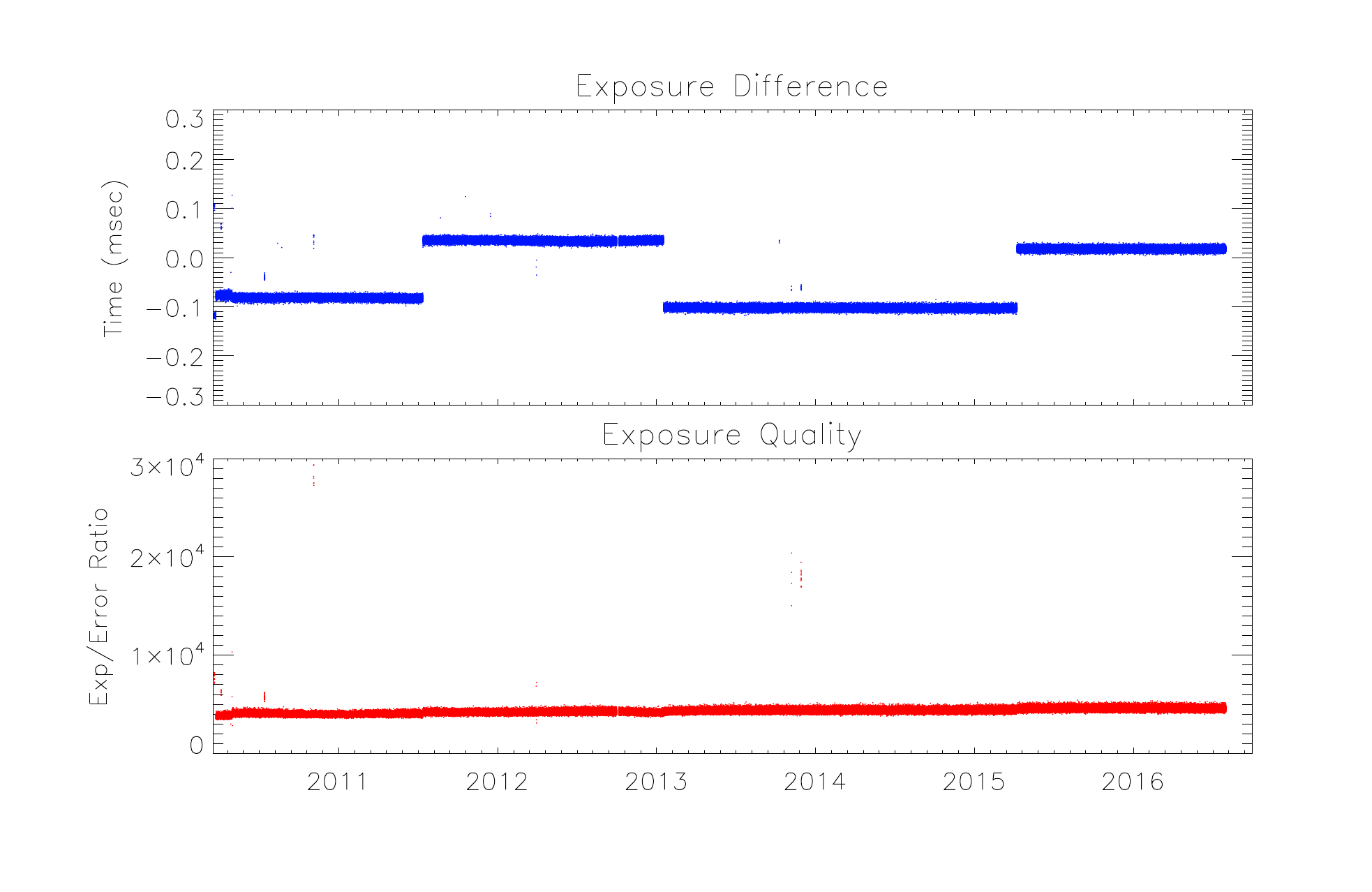}
\caption{Camera~2 exposure error and quality from 1 May 2010 -- 31 July 2016.
The top panel shows the difference between the commanded and measured
exposure times for Camera~2. The bottom panel shows the ratio between the
exposure time and the standard deviation in the measured exposure times.
In each case the values shown are 45-second averages of 12 consecutive
exposures that for display purposes are sampled every 12 minutes.
}
\label{figure:camera2_quality}
\end{figure}

As described in Section \ref{sec:throughput}, the HMI exposure time is controlled
by a mechanical shutter motor that rotates the cut-out sector of an otherwise opaque
disk into place with a pause in the {\it open} position for a commanded time.
The difference between the commanded and actual exposure time is monitored at
three places in the image plane and the average over 12 exposures is displayed
in the upper panel of Figure \ref{figure:camera2_quality} for Camera 2. Camera
1 is similar. Typical exposures are 115 -- 140 milleseconds. The lower panel shows a
measure of the quality of the exposure, which is expressed as the exposure
time divided by the standard deviation of the measured exposure times for
12 consecutive exposures. 
The exposure time can be specified to about 0.12 milleseconds with an observed
rms scatter less than 25 microseconds over 45 seconds and a standard deviation
of 13.2 microseconds, about a part in 10,000 of the typical exposure. The
exposure-time noise is a few times less than the per-pixel photon noise.
Individual exposure times are measured with a precision of 1 microseconds and
an accuracy better than 4 microseconds; the actual exposure time is reported in
the keyword \textsf{EXPTIME} and used in the data analysis pipeline.
Shutter noise contributes directly to uncertainty in the observables, because
the intensities are used to derive them.  The HMI shutter is remarkably
uniform and performs much better than the MDI shutter.  The few outlier
points in the figure occur during non-standard observing sequences.

\begin{figure}[htb]
\centering
\includegraphics[width=\textwidth]{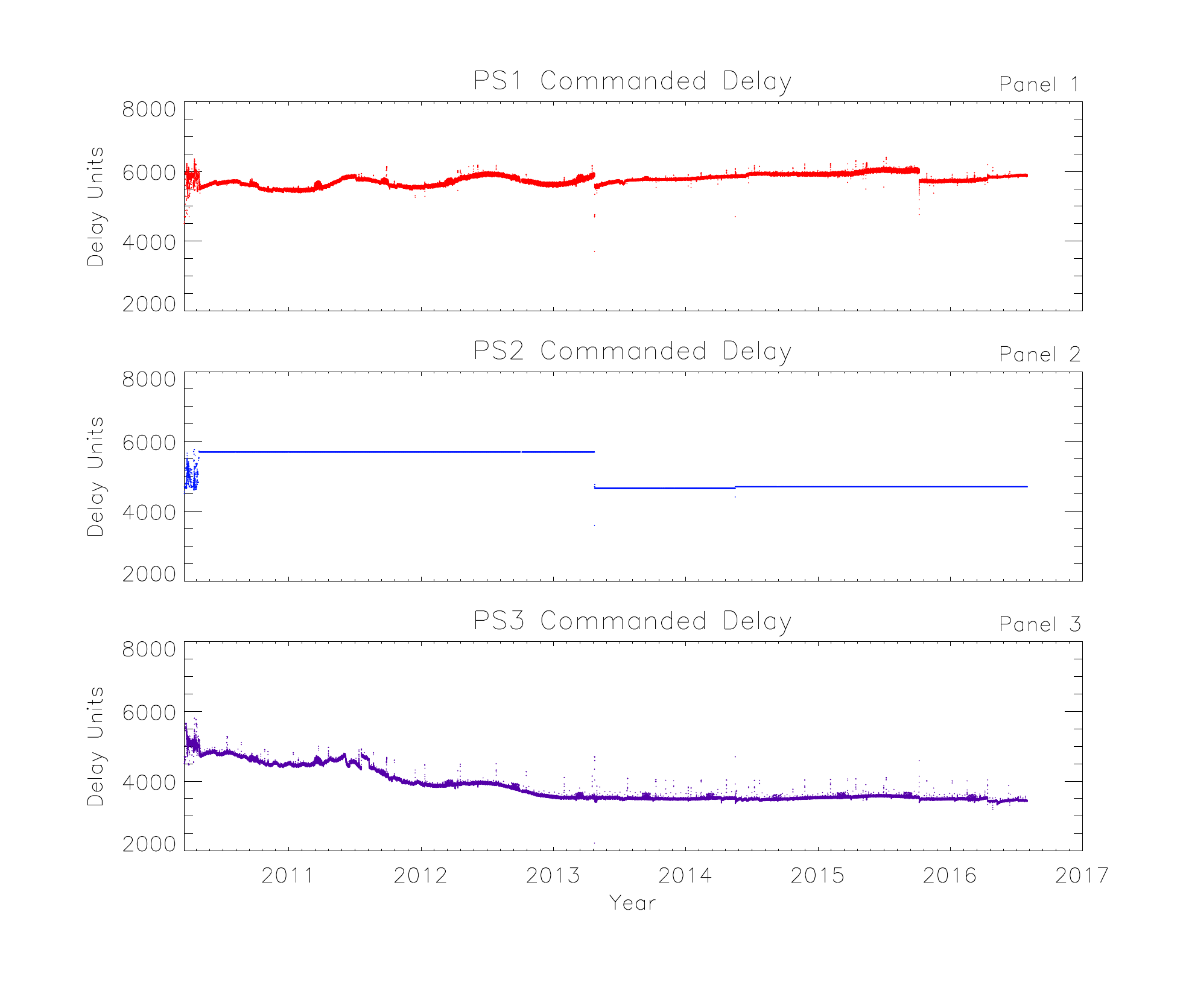}
\caption{Commanded polarization-selector delay times in engineering {\it delay units}.
Delay units are approximately microseconds. 
Values are averaged for 30 minutes. }
\label{figure:ps_delays}
\end{figure}

Figure \ref{figure:ps_delays} shows the commanded delay for each of the three polarization
selector (PS) wheels. The commanded delays indicate how long it takes for each
polarization wheel to move from one selected filter-wheel position to another. Changes
on longer time scales indicate changes in resistance or other mechanical issues. Sudden
changes are indicative of changes in tuning of the instrument.
PS 2 together with wavelength tuning (WT) selector 3 are redundant in case one of the 
wavelength tuners fails. Both are used infrequently.

\begin{figure}[htb]
\centering
\includegraphics[width=\textwidth]{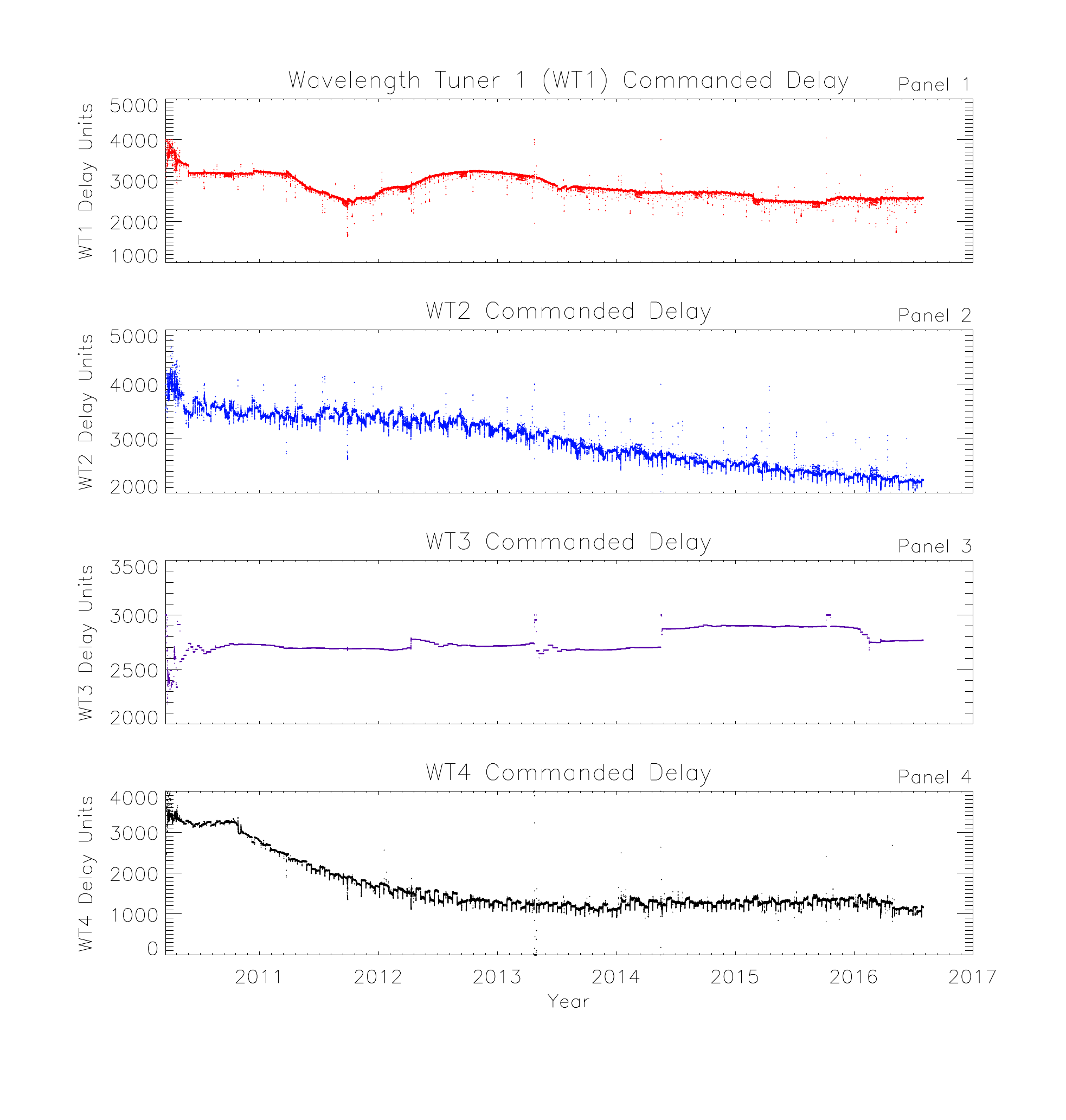}
\caption{Wavelength tuner delays in engineering {\it delay units}.}
\label{figure:wt_delays}
\end{figure}

Figure \ref{figure:wt_delays} shows the commanded delay for the four wavelength tuning
filter wheels. 
WT 1 tunes the Narrow Band Michelson; 
WT 2 tunes the Wide-Band Michelson; 
WT 3 is between the two Michelsons and provides redundancy -- it is seldom used; 
WT 4 tunes the final element of the Lyot Filter.

\section{Framelist and Filtergram IDs and Descriptions}
\label{sec:FTSTables}

Tables \ref{table:fts_summary1} and \ref{table:fts_summary2}
give information about all of the Framelist Timeline Specifications (FTSs) 
used by the HMI instrument. Frame lists define a sequence of filtergrams 
taken for a particular purpose. The duration is indicated in the table.
The standard-observables frame list (\textit{e.g.} 1021 or 1022), as well as most others 
that run only when specifically commanded, repeat continuously unless interrupted. 
Some calibration sequences (\textit{e.g.} 2021 and 2042) operate on timers that interrupt 
the regular sequences and execute once at the scheduled repeat interval indicated 
in the table.

Table \ref{table:fid_summary} identifies the Filtergram ID numbers (FIDs)
for the various standard instrument tunings. Filtergrams from the standard
observing sequence have five-digit FIDs in the range 10000\,--\,10199. The first
(rightmost) digit indicates the polarization setting; the second and third
digits give the wavelength tuning. Standard FIDs can be computed as:
\begin{equation}
FID = 10000 + 10 \times \rm{WL} + \rm{PL}
\end{equation}
where WL runs from 0 (the continuum) to 19 (the continuum on the other side
of the central wavelength)
and PL runs from 0 to 9 as described in the table. Filtergrams used for
calibrations usually have four-digit FIDs, which are listed in the table.

\begin{table}[htb]
\begin{tabular}{clrcl}
FTS ID & FrameList & Duration & Repeat & Description \\
 & & [seconds] & & \\
\hline
1000 & \textsf{obs\_6Av01}      &  90 & cont & Obs framelist, Mod A \\
1001 & \textsf{obs\_6Cv01}      & 135 & cont & Standard Mod-C framelist until 13 Dec 2010 \\
1002 & \textsf{obs\_6Lv01}      &  90 & cont & Obs framelist, Mod L \\
1003 & \textsf{obs\_6Mv01}      &  45 & cont & Obs framelist, Mod M \\
1004 & \textsf{obs\_6Xv01}      &  45 & cont & Obs framelist, Mod X \\
1020 & \textsf{obs\_6Av02}      &  90 & cont & Obs framelist, Mod A \\
1021 & \textsf{obs\_6Cv02}      & 135 & cont & Std. Mod-C framelist 13 Dec 2010 -- 13 Apr 2016 \\
1022 & \textsf{obs\_6Lv02}      &  90 & cont & Std. Mod-L framelist after 13 Apr 2016 \\
1023 & \textsf{obs\_6Mv02}      &  45 & cont & Obs framelist, Mod M \\
1024 & \textsf{obs\_6Xv02}      &  45 & cont & Obs framelist, Mod X \\
1025 & \textsf{obs\_8Av02}      & 120 & cont & Obs framelist, Mod A, 8 wavelengths \\
1026 & \textsf{obs\_10Av02}     & 150 & cont & Obs framelist, Mod A, 10 wavelengths \\[4pt]
2000 & \textsf{cal\_6Cv01}      & 135 & 24\,hr & Daily calibration sequence \\
2001 & \textsf{cal\_6Cv01}      & 135 & 12\,hr & Daily calibration sequence, 1 May - 13 Dec 2010 \\
2002 & \textsf{focr\_6Cv01}     & 135 & 12\,hr & Reduced focus sequence \\
2003 & \textsf{focr\_6Cv01}     & 135 & 24\,hr & Reduced focus sequence \\
2004 & \textsf{focr\_6Cv01}     & 135 & 2.4\,hr & Reduced focus sequence \\
2005 & \textsf{focr\_6Cv01}     & 135 & 1.5\,hr & Reduced focus sequence \\
2020 & \textsf{cal\_6Cv02}      & 135 & 24\,hr & Cal sequence (darks, cont. tuned, Calmode frames) \\
2021 & \textsf{cal\_6Cv02}      & 135 & 12\,hr & Daily cal seq. 13 Dec 2010 - 13 Apr 2016 \\
2022 & \textsf{focr\_6Cv02}     & 135 & 12\,hr & Reduced focus sequence \\
2023 & \textsf{focr\_6Cv02}     & 135 & 24\,hr & Reduced focus sequence \\
2024 & \textsf{focr\_6Cv02}     & 135 & 2.4\,hr & Reduced focus sequence \\
2025 & \textsf{focr\_6Cv02}     & 135 & 1.5\,hr & Reduced focus sequence \\
2042 & \textsf{cal\_6Lv02}      &  90 & 12\,hr & Daily cal sequence after 13 Apr 2016 \\
2043 & \textsf{cal\_6Lv02}      &  90 & 1\,hr & Hourly cal sequence, Mod L \\[4pt]
3000 & \textsf{focr\_6Cv01}     & 135 & cont & Reduced focus sequence \\
3001 & \textsf{pzt\_def\_6Cv01} & 135 & cont & PZT flat-field sequence, Obsmode \\
3002 & \textsf{pzt\_cal\_6Cv01} & 135 & cont & PZT flat-field sequence, Calmode \\
3003 & \textsf{focus\_6Cv01}    & 135 & cont & Full focus sequence \\
3004 & \textsf{lin\_def\_6Cv01} & 135 & cont & Linearity test sequence, Obsmode \\
3005 & \textsf{lin\_cal\_6Cv01} & 135 & cont & Linearity test sequence, Calmode \\
3006 & \textsf{det\_def\_6Cv01} & 135 & cont & Detune sequence, Obsmode \\
3007 & \textsf{det\_cal\_6Cv01} & 135 & cont & Detune sequence, Calmode \\
3020 & \textsf{focr\_6Cv02}     & 135 & cont & Reduced focus sequence, mod C\\
3021 & \textsf{pzt\_def\_6Cv02} & 135 & cont & PZT flat-field sequence, Obsmode \\
3022 & \textsf{pzt\_cal\_6Cv02} & 135 & cont & PZT flat-field sequence, Calmode \\
3023 & \textsf{focus\_6Cv02}    & 135 & cont & Focus sequence \\
3024 & \textsf{lin\_def\_6Cv02} & 135 & cont & Linearity test sequence, Obsmode \\
3025 & \textsf{lin\_cal\_6Cv02} & 135 & cont & Linearity test sequence, Calmode \\
3026 & \textsf{det\_def\_6Cv02} & 135 & cont & Detune sequence, Obsmode \\
3027 & \textsf{det\_cal\_6Cv02} & 135 & cont & Detune sequence, Calmode \\
3040 & \textsf{focr\_6Lv02}     &  90 & cont & Reduced focus sequence, Mod L \\
3041 & \textsf{pzt\_def\_6Lv02} &  90 & cont & PZT flat-field sequence, Obsmode, Mod L\\
3042 & \textsf{pzt\_cal\_6Lv02} &  90 & cont & PZT flat-field sequence, Calmode, Mod L\\
3043 & \textsf{focus\_6Lv02}    &  90 & cont & Full focus sequence, Mod L\\
3048 & \textsf{cal\_6Lv02}      &  90 & cont & Mod L calibration sequence, continuous \\ 
\hline
\end{tabular}
\caption{HMI Framelist Timeline Specification (FTS) Summary -- Part 1}
\label{table:fts_summary1}
\end{table}

\begin{table}[htb]
\begin{tabular}{clrrl}
FTS ID & Frame List & Duration & Repeat & Description\\
 & & [seconds] & \\
\hline
4000 & \textsf{pzt\_def}        &  45 & cont & 45s cadence PZT flat-field sequence, Obsmode \\
4001 & \textsf{pzt\_cal}        &  45 & cont & 45s cadence PZT flat-field sequence, Calmode \\
4002 & \textsf{dop\_ic\_v01}    &  45 & cont & Regular obs seq. on Cam 2; LCP/RCP continuum Cam 1 \\
4003 & \textsf{focus\_off\_v01} &  45 & cont & Focus sequence for offpoint maneuvers \\
4004 & \textsf{pl\_wob\_v01}    & 135 & cont & Polarization wobble sequence \\
4005 & \textsf{wl\_wob\_v01}    & 180 & cont & Wavelength wobble sequence \\
4006 & \textsf{loop\_45}        &  45 & cont & Continuous sequence of default filtergrams \\
4007 & \textsf{loop\_90}        &  90 & cont & Continuous sequence of default filtergrams \\
4008 & \textsf{loop\_135}       & 135 & cont & Continuous sequence of default filtergrams \\
4009 & \textsf{focus\_8\_14d}   &  45 & cont & Reduced focus sequence \\
4010 & \textsf{loop\_led\_45}   &  45 & cont & Pre-launch calibration sequence \\
4011 & \textsf{focus\_off\_v02} &  45 & cont & Reduced focus sequence for offpoint maneuvers \\
4012 & \textsf{regulus\_135}    & 135 & cont & Regulus observing seq.; mix of std. and 3.2s exp. \\
4013 & \textsf{roll\_dop\_ic2}  &  45 & cont & Roll maneuver seq., continuum filtergrams on Cam 1 \\
4014 & \textsf{cruc\_offp\_lin} & 135 & cont & Linearity test sequence for EVE cruciform maneuver \\
4015 & \textsf{regulus\_long}   & 135 & cont & Regulus observing seq. offpoint with 3.2s exp. \\
4020 & \textsf{pzt\_def}        &  45 & cont & 45s cadence PZT flat-field sequence, Obsmode \\
4021 & \textsf{pzt\_cal}        &  45 & cont & 45s cadence PZT flat-field sequence, Calmode \\
4023 & \textsf{focus\_off\_v01} &  45 & cont & Reduced focus sequence for offpoint maneuvers \\
4024 & \textsf{pl\_wob\_v02}    & 135 & cont & Polarization wobble sequence \\
4025 & \textsf{wl\_wob\_v02}    & 180 & cont & Wavelength wobble sequence \\
4026 & \textsf{loop\_45}        &  45 & cont & Continuous sequence of default filtergrams \\
4027 & \textsf{loop\_90}        &  90 & cont & Continuous sequence of default filtergrams \\
4028 & \textsf{loop\_135}       & 135 & cont & Continuous sequence of default filtergrams \\
4029 & \textsf{focus\_8\_14d}   &  45 & cont & Reduced focus sequence \\
4030 & \textsf{loop\_led\_45}   &  45 & cont & Pre-launch calibration sequence \\
4031 & \textsf{focus\_off\_v02} &  45 & cont & Focus sequence for offpoint maneuvers \\
4033 & \textsf{rolldopic\_v02}  &  45 & cont & Roll man. seq., continuum filtergrams on Cam. 1 \\
4034 & \textsf{venus\_2pl\_v01} & 135 & cont & Venus transit -- std. seq. Cam 2, 2 lin. pol. Cam 1 \\
4035 & \textsf{venus\_4pl\_v01} & 135 & cont & Venus transit -- std. seq. Cam 2, 4 lin. pol. Cam 1 \\
4036 & \textsf{comet\_ison}     & 135 & cont & Comet ISON seq., 600\,ms exposures, default tuning \\
4037 & \textsf{rolldopic\_v03}  &  45 & cont & Roll maneuver seq., std. obs Cam 2, continuum Cam 1 \\
4038 & \textsf{rollicscn\_v01}  &  45 & cont & Roll seq., std. Cam 2, scan cont. to line-core Cam 1 \\
4039 & \textsf{merc\_1pl\_v01}  &  45 & cont & Mercury transit, std. obs Cam 2, 1 lin. pol. Cam 1 \\
4040 & \textsf{cont6Lv01}       &  90 & 30 min. & Std. Mod-L seq. with a set of continuum exp. \\ 
\hline
\end{tabular}
\caption{HMI FTS Summary -- Part 2}
\label{table:fts_summary2}
\end{table}

\begin{table}[htb]
\begin{tabular}{l c l}
\multicolumn{3}{l}{{\bf Standard Observing-Program FIDs}} \\[2pt]
FID & PL Index & Polarization State \\
\hline
10**0 & 410 & Mod A pol 1 \\
10**1 & 411 & Mod A pol 2 \\
10**2 & 412 & Mod A pol 3 \\
10**3 & 413 & Mod A pol 4 \\
10**4 & 414 & \textit{I+Q}; linear polarization, 0 deg \\
10**5 & 415 & \textit{I-Q}; linear polarization, 90 deg \\
10**6 & 416 & \textit{I+U}; linear polarization, 45 deg \\
10**7 & 417 & \textit{I-U}; linear polarization, 135 deg \\
10**8 & 418 & \textit{I+V}; left circular polarization \\
10**9 & 419 & \textit{I-V}; right circular polarization \\[4pt]
\end{tabular}

\begin{tabular}{l c r}
FID & WL Index & Wavelength \\
\hline
1000* & \textsf{0xff00} & -344.0\,m\AA \\
1001* & \textsf{0xff01} & -309.6\,m\AA \\
1002* & \textsf{0xff02} & -275.2\,m\AA \\
1003* & \textsf{0xff03} & -240.8\,m\AA \\
1004* & \textsf{0xff04} & -206.4\,m\AA \\
1005* & \textsf{0xff05} & -172.0\,m\AA \\
1006* & \textsf{0xff06} & -137.6\,m\AA \\
1007* & \textsf{0xff07} & -103.2\,m\AA \\
1008* & \textsf{0xff08} &  -68.9\,m\AA \\
1009* & \textsf{0xff09} &  -34.4\,m\AA \\
1010* & \textsf{0xff0a} &    0.0\,m\AA \\
1011* & \textsf{0xff0b} &   34.4\,m\AA \\
1012* & \textsf{0xff0c} &   68.8\,m\AA \\
1013* & \textsf{0xff0d} &  103.2\,m\AA \\
1014* & \textsf{0xff0e} &  137.6\,m\AA \\
1015* & \textsf{0xff0f} &  172.0\,m\AA \\
1016* & \textsf{0xff10} &  206.4\,m\AA \\
1017* & \textsf{0xff11} &  240.8\,m\AA \\
1018* & \textsf{0xff12} &  275.2\,m\AA \\
1019* & \textsf{0xff13} &  309.6\,m\AA \\[4pt]
\end{tabular}

\begin{tabular}{l l}
\multicolumn{2}{l}{{\bf Calibration and Special-Observation FIDs}} \\[2pt]
FID & Description \\
\hline
5000 & Regular dark frame \\
5001 & Throwaway dark frame \\
5002 & General image \\
5003 & Linearity test \\
5004 & Linearity test with darks \\
5101\,--\,5116 & Focus-sequence filtergrams \\
5117 & Calmode filtergram \\[4pt]
\multicolumn{2}{l}{{\bf Wavelength-Calibration FIDs}} \\
\hline
6000-6026 & Standard detune (100\,--\,126) \\
6027-6030 & Extra positions for 31-frame detune \\
6101-6196 & Wobble sequence (\textsf{WLID}=1\,--\,96) \\
6201-6296 & Wobble\,+\,20 sequence \\[4pt]
\multicolumn{2}{l}{{\bf Polarization-Calibration FIDs}} \\
\hline
7101-7172 & Wobble sequence (\textsf{WLID}=1\,--\,72) \\
7201-7272 & Wobble sequence (\textsf{WLID}=175\,--\,246) \\
\end{tabular}
\caption{HMI Filtergram ID (FID) Summary Table}
\label{table:fid_summary}
\end{table} 
\section{Level-0 FITS Keywords}
\label{sec:Lev0Keywords}

Table \ref{table:level0_image_keywords} lists the Level-0 keywords associated
with details of the image.

Table \ref{table:level0_seq_keywords} describes the Level-0 keywords
associated with the status of the onboard Image Status Packet (ISP) sequencer.

Table \ref{table:level0_mech_keywords} describes the Level-0 keywords
associated with the status of the onboard mechanisms.

\begin{table}[hb]
\begin{tabular}{lll}
Keyword & Type & Description\\
\hline
\textsf{BLD\_VERS} & string  & Build Version: from jsoc\_version.h\\
\textsf{ORIGIN}    & string  & Constant: Location where file made - SDO/JSOC-SDP\\
\textsf{DATE}      & time    & Date and time of processing; ISO 8601\\
\textsf{TELESCOP}  & string  & Constant: for HMI: SDO/HMI\\
\textsf{INSTRUME}  & string  & HMI light path: HMI\_SIDE1 or HMI\_FRONT2\\
\textsf{DATE-OBS}  & time    & Date when observation started; ISO 8601\\
\textsf{T\_OBS}    & time    & Observation time\\
\textsf{CAMERA}    & integer & HMI camera numeric identifier: 1 or 2\\
\textsf{IMG\_TYPE} & string  & Image type: LIGHT or DARK \\
\textsf{EXPTIME}   & double  & Exposure duration: shutter open time in seconds\\
\textsf{EXPSDEV}   & float   & Exposure standard deviation in seconds\\
\textsf{WAVELNTH}  & integer & Constant: for HMI = 6173.3 angstrom\\
\textsf{WAVEUNIT}  & string  & Constant: Wavelength unit = angstrom\\
\textsf{FSN}       & integer & FSN - Filtergram Sequence Number\\
\textsf{FID}       & integer & FID - Filtergram ID\\
\textsf{TLMDSNAM}  & string  & Telemety data series based on data packet time\\
\textsf{IMGFPT}    & time    & Time stamp of the first image data packet\\
\textsf{IMGAPID}   & integer & Application ID of the science data packets\\
\textsf{TAPCODE}   & integer & Take A Picture code for the camera readout\\
\textsf{BITSELID}  & integer & Bit select ID; r value for the data compression\\
\textsf{COMPID}    & integer & Compression ID; data compression n and k values\\
\textsf{CROPID}    & integer & Crop table ID used in data downlink\\
\textsf{LUTID}     & integer & Look-up table ID used in data downlink\\
\textsf{NPACKETS}  & integer & Number of packets in image\\
\textsf{NERRORS}   & integer & Number of decompression errors in image\\
\textsf{EOIERROR}  & short   & End Of Image error; Last pixel error occurred in image\\
\textsf{HEADRERR}  & short   & Header error occurred in image\\
\textsf{OVERFLOW}  & short   & Data overflow error occurred in image\\
\textsf{QUALITY}   & integer & Quality keyword\\
\textsf{TOTVALS}   & integer & Expected number of data values (pixels) in image\\
\textsf{DATAVALS}  & integer & Actual number of data values in image\\
\textsf{MISSVALS}  & integer & Missing values: \textsf{TOTVALS\,-\,DATAVALS}\\
\textsf{PERCENTD}  & float   & Percent data; $100\times$\textsf{DATAVALS\,/\,TOTVALS}\\
\textsf{DATAMIN}   & short   & Minimum value of all pixels\\
\textsf{DATAMAX}   & short   & Maximum value of all pixels\\
\textsf{DATAMEDN}  & short   & Median value of all pixels\\
\textsf{DATAMEAN}  & float   & Mean value of all pixels\\
\textsf{DATARMS}   & float   & Rms deviation from the mean value of all pixels\\
\textsf{DATASKEW}  & float   & Skewness from the mean value of all pixels\\
\textsf{DATAKURT}  & float   & Kurtosis of all pixels\\
\hline
\end{tabular}
\caption{HMI Level-0 Keywords -- Image Details}
\label{table:level0_image_keywords}
\end{table}

\begin{table}[ht]
\begin{tabular}{lll}
Keyword & Type & Description\\
\hline
\textsf{ISPSNAME}  & string  & Image Status Packet (ISP) series name\\
\textsf{ISPPKTIM}  & time    & Prime key value for the ISP record\\
\textsf{ISPPKTVN}  & string  & ISP packet version\\
\textsf{HSQFGSN}   & integer & Unique serial number for each image (filtergram) taken\\
\textsf{HSQFGID}   & integer & Filtergram identifier parameters\\
\textsf{HCAMID}    & integer & Current light-path identifier\\
\textsf{HSHIEXP}   & integer & Current shutter-exposure value in milliseconds\\
\textsf{HOBITSEC}  & integer & TAI seconds of the shutter-move start time\\
\textsf{HOBITSS}   & integer & Subseconds field of the shutter-move start time\\
\textsf{HWLTNSET}  & string  & Image Stabilization System (ISS) loop status\\
\textsf{HSQSTATE}  & string  & Sequencer state: IDLE, SELECTING, or PROCESSING\\
\textsf{HSEQERR}   & string  & Sequence error message of the last sequencer error\\
\textsf{HFLREFTM}  & integer & Current framelist reference start time\\
\textsf{HFLRELTM}  & integer & Current frame time: milliseconds from reference time\\
\textsf{HFLID}     & integer & Framelist identification number\\
\textsf{HOBLSTID}  & integer & Observation list identification number\\
\textsf{HFLPSITN}  & integer & Position number of the current frame in framelist\\
\textsf{HSQFCNT}   & integer & Number of frames taken after restarting the sequence\\
\textsf{HFLLNGTH}  & short   & Total number of frames in the current framelist (FTS)\\
\textsf{HFLRPTCT}  & integer & Cadence periods to repeat for the current FTS\\
\textsf{HFLRPTNM}  & integer & Repeat number of the active framelist\\
\textsf{HFLSKPCT}  & integer & Cadence periods to skip for the current FTS\\
\textsf{HFTSACID}  & integer & Identification number of the current active FTS\\
\textsf{HFTSCDMK}  & integer & Number of cadence periods after restarting the sequence\\
\textsf{HFTSINFO}  & integer & FTS status information\\
\textsf{HSQEIDX}   & integer & Current exposure index number\\
\textsf{HIMGCFID}  & integer & Current image configuration identification number\\
\textsf{HCFTID}    & integer & Current focus position identification number\\
\textsf{HPLTID}    & integer & Current polarization selector identification number\\
\textsf{HWLTID}    & integer & Current the wavelength tuning identification number\\
\textsf{HWLSTIDX}  & integer & Current wavelength set index number\\
\textsf{HGP1RGST}  & integer & General purpose register 1 (set by command)\\
\textsf{HGP2RGST}  & integer & General purpose register 2 (set by command)\\
\hline
\end{tabular}
\caption{HMI Level-0 Keywords -- Image Status Packet (ISP) Sequencer Status}
\label{table:level0_seq_keywords}
\end{table}

\begin{table}[ht]
\begin{tabular}{lll}
Keyword & Type & Description\\
\hline
\textsf{HSHMIOPB}  & float   & Shutter timer open value for bottom position\\
\textsf{HSHMIOPM}  & float   & Shutter timer open value for middle position\\
\textsf{HSHMIOPT}  & float   & Shutter timer open value for top position\\
\textsf{HSHMICLB}  & float   & Shutter close timer value for bottom position\\
\textsf{HSHMICLM}  & float   & Shutter close timer value for middle position\\
\textsf{HSHMICLT}  & float   & Shutter close timer value for top position\\
\textsf{HCF1ENCD}  & integer & Encoder value returned from CF1 mechanism\\
\textsf{HCF2ENCD}  & integer & Encoder value returned from CF2 mechanism\\
\textsf{HPS1ENCD}  & integer & Encoder value returned from PS1 mechanism\\
\textsf{HPS2ENCD}  & integer & Encoder value returned from PS2 mechanism\\
\textsf{HPS3ENCD}  & integer & Encoder value returned from PS3 mechanism\\
\textsf{HWT1ENCD}  & integer & Encoder value returned from WT1 mechanism\\
\textsf{HWT2ENCD}  & integer & Encoder value returned from WT2 mechanism\\
\textsf{HWT3ENCD}  & integer & Encoder value returned from WT3 mechanism\\
\textsf{HWT4ENCD}  & integer & Encoder value returned from WT4 mechanism\\
\textsf{HCF1POS}   & integer & Commanded target position for CF1 mechanism\\
\textsf{HCF2POS}   & integer & Commanded target position for CF2 mechanism\\
\textsf{HPL1POS}   & integer & Commanded target position for PS1 mechanism\\
\textsf{HPL2POS}   & integer & Commanded target position for PS2 mechanism\\
\textsf{HPL3POS}   & integer & Commanded target position for PS3 mechanism\\
\textsf{HWL1POS}   & integer & Commanded target position for WT1 mechanism\\
\textsf{HWL2POS}   & integer & Commanded target position for WT2 mechanism\\
\textsf{HWL3POS}   & integer & Commanded target position for WT3 mechanism\\
\textsf{HWL4POS}   & integer & Commanded target position for WT4 mechanism\\
\hline
\end{tabular}
\caption{HMI Level-0 Keywords -- ISP Mechanism Parameters}
\label{table:level0_mech_keywords}
\end{table}

\section{Level-0 \textsf{QUALITY}-Keyword Summary}
\label{sec:Lev0Qual}

HMI uses a \textsf{QUALITY} keyword to describe properties of the data and data processing at
each level of reduction. Generally a bit in the keyword is set only when there is a problem.

Table \ref{table:lev0_quality} gives the meaning of the bits put into the
\textsf{QUALLEV0} keyword for each filtergram record in the Level-0 data series
determined by the \textsf{ingest\_lev0} processing module. Bit~0 is the low bit (0x01)

\begin{table}[hb]
\begin{tabular}{lll}
Quality       & Bit Mask    & Description\\
\hline
\textsf{Q\_OVFL}       & 0x00000001  & Overflow flag set\\
\textsf{Q\_HDRERR}     & 0x00000002  & Header error flag set\\
\textsf{Q\_CMPERR}     & 0x00000004  & Compression error in image\\
\textsf{Q\_LPXERR}     & 0x00000008  & Last pixel error\\
\textsf{Q\_NOISP}      & 0x00000010  & No ISP; FSN $\neq$ HSQFGSN\\
\textsf{Q\_MISSI}      & 0x00000020  & Missing image\\
\textsf{Q\_CORRUPT}    & 0x00000040  & Corrupt image; FSN=469769216 or 0x1c001c00\\
\textsf{Q\_INVALTIME}  & 0x00000080  & HOBITSEC = 0; T\_OBS = 1958.01.01\_00:00:00\_UTC\\
\textsf{Q\_MISS0}      & 0x00000100  & MISSVALS $>$ 0\\
\textsf{Q\_MISS1}      & 0x00000200  & MISSVALS $>$ 0.01*TOTVALS\\
\textsf{Q\_MISS2}      & 0x00000400  & MISSVALS $>$ 0.05*TOTVALS\\
\textsf{Q\_MISS3}      & 0x00000800  & MISSVALS $>$ 0.25*TOTVALS\\
              & 0x00001000  & Unused\\
              & 0x00002000  & Unused\\
              & 0x00004000  & Unused\\
\textsf{Q\_CAM\_ANOM}  & 0x00008000  & Camera anomaly; entered manually\\
\textsf{Q\_DARK}       & 0x00010000  & Dark image\\
\textsf{Q\_ISSOPEN}    & 0x00020000  & ISS loop open; HWLTNSET = 'OPEN'\\
\textsf{Q\_HCF1ENCD}   & 0x00040000  & Focus/Cal Motor 1 Error\\
              &             & HCF1ENCD $\neq$ HCF1POS $\pm$ 1\\
\textsf{Q\_HCF2ENCD}   & 0x00080000  & Focus/Cal Motor 2 Error\\
              &             & HCF2ENCD $\neq$ HCF2POS $\pm$ 1\\
\textsf{Q\_HPS1ENCD}   & 0x00100000  & Polarization Motor 1 Error\\
              &             & HPS1ENCD $\neq$ HPL1POS $\pm$ 1\\
\textsf{Q\_HPS2ENCD}   & 0x00200000  & Polarization Motor 2 Error\\
              &             & HPS2ENCD $\neq$ HPL2POS $\pm$ 1\\
\textsf{Q\_HPS3ENCD}   & 0x00400000  & Polarization Motor 3 Error\\
              &             & HPS3ENCD $\neq$ HPL3POS $\pm$ 1\\
\textsf{Q\_HWT1ENCD}   & 0x00800000  & Wavelength Motor 1 Error\\
              &             & HWT1ENCD $\neq$ HWL1POS $\pm$ 1\\
\textsf{Q\_HWT2ENCD}   & 0x01000000  & Wavelength Motor 2 Error\\
              &             & HWT2ENCD $\neq$ HWL2POS $\pm$ 1\\
\textsf{Q\_HWT3ENCD}   & 0x02000000  & Wavelength Motor 3 Error\\
              &             & HWT3ENCD $\neq$ HWL3POS $\pm$ 1\\
\textsf{Q\_HWT4ENCD}   & 0x04000000  & Wavelength Motor 4 Error\\
              &             & HWT4ENCD $\neq$ HWL4POS $\pm$ 1\\
              & 0x08000000  & Unused\\
\textsf{Q\_GPREGBIT0}  & 0x10000000  & HGP1RGST bit 0 set\\
\textsf{Q\_GPREGBIT1}  & 0x20000000  & HGP1RGST bit 1 set\\
\textsf{Q\_REOPENED}   & 0x40000000  & Image reopened during reconstruction\\
              &             & NPACKETS value may be incorrect\\
\textsf{Q\_MISSALL}    & 0x80000000  & Data is completely missing\\
              &             & High bit\\
\hline
\end{tabular}
\caption{HMI Level-0 Quality Summary}
\label{table:lev0_quality}
\end{table}

\section{Level-1 FITS Keywords}
\label{sec:Lev1Keywords}

Table \ref{table:level1_wcs_orbit_keywords} describes Level-1 keywords associated with WCS
coordinates and SDO orbit parameters.


\begin{table}[ht]
\scalebox{0.87}{
\begin{tabular}{lll}
Keyword & Type & Description\\
\hline
\textsf{T\_OBS\_step}  & double & \textsf{T\_OBS} step (constant); 1.000000 second\\
\textsf{T\_OBS\_epoch} & time & \textsf{T\_OBS} epoch (constant); 1977.01.01\_00:00:00\_TAI\\
\textsf{OSCNMEAN}  & float   & Mean value of removed overscan rows\\
\textsf{OSCNRMS}   & float   & Rms deviation from the mean value of overscan rows\\
\textsf{FLAT\_REC} & string  & Flat field series record pointer\\
\textsf{NBADPERM}  & integer & Count of permanent bad pixels\\
\textsf{NBADTOT}   & integer & Count of total bad pixels\\
\textsf{CTYPE1}    & string  & Typically HPLN-TAN (SOLARX)\\
\textsf{CUNIT1}    & string  & Typically arcseconds\\
\textsf{CRVAL1}    & float   & Image scale in the x direction; arcseconds/pixel\\
\textsf{CDELT1}    & float   & Image scale in the x direction; arcseconds/pixel\\
\textsf{CRPIX1}    & float   & Location of sun center in CCD x direction; pixel\\
\textsf{CTYPE2}    & string  & Typically HPLT-TAN (SOLARY)\\
\textsf{CUNIT2}    & string  & Typically arcseconds\\
\textsf{CRVAL2}    & float   & Image scale in the x direction; arcseconds/pixel\\
\textsf{CDELT2}    & float   & Image scale in the y direction; arcseconds/pixel\\
\textsf{CRPIX2}    & float   & Location of sun center in CCD y direction; pixel\\
\textsf{CROTA2}    & float   & \textsf{INST\_ROT + SAT\_ROT}; degrees\\
\textsf{R\_SUN}    & float   & Radius of the Sun on the CCD detector; pixels\\
\textsf{MPO\_REC}  & string  & Master Pointing series record pointer\\
\textsf{INST\_ROT} & float   & Master pointing CCD rotation wrt SDO Z axis; degrees\\
\textsf{IMSCL\_MP} & float   & Master pointing image scale; arcseconds/pixels\\
\textsf{X0\_MP}    & float   & Master pointing \textsf{X0} sun center in CCD frame; pixels\\
\textsf{Y0\_MP}    & float   & Master pointing \textsf{Y0} sun center in CCD frame; pixels\\
\textsf{RSUN\_LF}  & float   & Limb fit Solar radius; pixels\\
\textsf{X0\_LF}    & float   & Limb fit \textsf{X0} sun center in CCD frame; pixels\\
\textsf{Y0\_LF}    & float   & Limb fit \textsf{Y0} sun center in CCD frame; pixels\\
\textsf{CALVER32}  & integer & Height of formation correction version\\
\textsf{ASD\_REC}  & string  & Ancillary Science Data series record pointer\\
\textsf{SAT\_Y0}   & float   & Position of solar center wrt the SDO -Y axis; arcseconds\\
\textsf{SAT\_Z0}   & float   & Position of solar center wrt the SDO +Z axis; arcseconds\\
\textsf{SAT\_ROT}  & float   & Angle of solar pole wrt the SDO +X axis; degrees\\
\textsf{ACS\_MODE} & string  & ACS pointing mode; \textsf{ACS\_AN\_ACS\_MODE} \\
\textsf{ACS\_ECLP} & string  & ACS eclipse flag; \textsf{ACS\_AN\_FLAG\_CSS\_ECLIPSE} \\
\textsf{ACS\_SUNP} & string  & ACS sun presense flag; \textsf{ACS\_AN\_FLAG\_DSS\_SUNPRES} \\
\textsf{ACS\_SAFE} & string  & ACS safe hold flag; \textsf{ACS\_AN\_FLAG\_ACE\_INSAFEHOLD} \\
\textsf{ACS\_CGT}  & string  & ACS Controlling Guide Telescope ID; \textsf{ACS\_AN\_NUM\_CGT} \\
\textsf{ORB\_REC}  & string  & Orbit vector series record pointer\\
\textsf{DSUN\_REF} & double  & Reference distance to Sun (constant): 149,597,870,691.0 m\\
\textsf{DSUN\_OBS} & double  & Distance from SDO to Sun center; m\\
\textsf{RSUN\_REF} & double  & Reference radius of the Sun (constant): 696,000,000.0 m\\
\textsf{RSUN\_OBS} & double  & Apparent radius of the Sun seen by SDO; arcseconds\\
\textsf{GAEX\_OBS} & double  & Geocentric Inertial X position; m\\
\textsf{GAEY\_OBS} & double  & Geocentric Inertial Y position; m\\
\textsf{GAEZ\_OBS} & double  & Geocentric Inertial Z position; m\\
\textsf{HAEX\_OBS} & double  & Heliocentric Inertial X position; m\\
\textsf{HAEY\_OBS} & double  & Heliocentric Inertial Y position; m\\
\textsf{HAEZ\_OBS} & double  & Heliocentric Inertial Z position; m\\
\textsf{OBS\_VR}   & double  & Speed of observer in radial direction; m/s\\
\textsf{OBS\_VW}   & double  & Speed of observer in solar-west direction; m/s\\
\textsf{OBS\_VN}   & double  & Speed of observer in solar-north direction; m/s\\
\textsf{CRLN\_OBS} & float   & Carrington longitude of the observer; degrees\\
\textsf{CRLT\_OBS} & float   & Carrington latitude of the observer; degrees\\
\textsf{CAR\_ROT}  & integer & Carrington rotation number of \textsf{CRLN\_OBS} \\
\textsf{HGLN\_OBS} & float   & Stonyhurst heliographic longitude of the observer; degrees\\
\textsf{HGLT\_OBS} & float   & Stonyhurst heliographic latitude of the observer; degrees\\
\hline
\end{tabular}}
\caption{HMI Level 1 Keywords - WCS and Orbit Parameters}
\label{table:level1_wcs_orbit_keywords}
\end{table}
\section{Level-1 \textsf{QUALITY}-Keyword Summary}
\label{sec:Lev1Qual}

Table \ref{table:lev1_quality} provides the meaning of the bits put 
into the Level-1 \textsf{QUALITY} keyword for each filtergram in the 
Level-1 data series by the \textsf{build\_lev1} processing module. 
Bit~0 is the low bit (0x01)\

\begin{table}[htb]
\begin{tabular}{lll}
Quality         & Bit Mask    & Description\\
\hline
\textsf{Q\_NOFLAT}       & 0x00000001  & Flat field not available or error\\
\textsf{Q\_NOORB}       & 0x00000002  & Orbit data not available or error\\
\textsf{Q\_NOASD}       & 0x00000004  & Ancillary science data not available or error\\
\textsf{Q\_NOMPD}       & 0x00000008  & Master pointing data not available or error\\
\textsf{Q\_NOLIMB}      & 0x00000010  & Limb fit error\\
                & 0x00000020  & Unused\\
                & 0x00000040  & Unused\\
\textsf{Q\_CAM\_ANOM1}  & 0x00000080  & Camera anomaly\\
\textsf{Q\_1\_MISS0}    & 0x00000100  & MISSVALS $>$ 0\\
\textsf{Q\_1\_MISS1}    & 0x00000200  & MISSVALS $>$ $0.01\times$\textsf{TOTVALS}\\
\textsf{Q\_1\_MISS2}    & 0x00000400  & MISSVALS $>$ $0.05\times$\textsf{TOTVALS}\\
\textsf{Q\_1\_MISS3}    & 0x00000800  & MISSVALS $>$ $0.25\times$\textsf{TOTVALS}\\
\textsf{Q\_NOACS\_SCI}  & 0x00001000  & \textsf{ACS\_MODE} $\neq$ 'SCIENCE'\\
\textsf{Q\_ACS\_ECLP}   & 0x00002000  & \textsf{ACS\_ECLP} = 'YES'; spacecraft eclipse flag\\
\textsf{Q\_ACS\_SUNP}   & 0x00004000  & \textsf{ACS\_SUNP} = 'NO'; no sun presence\\
\textsf{Q\_ACS\_SAFE}   & 0x00008000  & \textsf{ACS\_SAFE} = 'YES'; safemode flag set\\
\textsf{Q\_IMG\_TYPE}   & 0x00010000  & Dark image\\
\textsf{Q\_LOOP\_OPEN}  & 0x00020000  & ISS Loop Open\\
\textsf{Q\_CAL\_IMG}    & 0x00040000  & Calibration image\\
\textsf{Q\_CALM\_IMG}   & 0x00080000  & HMI calibration mode image\\
\textsf{Q\_AIA\_FOOR}   & 0x00100000  & Not used for HMI\\
\textsf{Q\_AIA\_REGF}   & 0x00200000  & Not used for HMI\\
\textsf{Q\_THERM\_RECOV}& 0x00400000  & HMI thermal recovery\\
\textsf{Q\_LUNAR\_TRAN} & 0x00800000  & HMI lunar transit\\
                & 0x01000000  & Unused\\
                & 0x02000000  & Unused\\
                & 0x04000000  & Unused\\
                & 0x08000000  & Unused\\
                & 0x10000000  & Unused\\
                & 0x20000000  & Unused\\
\textsf{Q\_NRT}         & 0x40000000  & Near Real Time mode\\
\textsf{Q\_MISSALL}     & 0x80000000  & Image not available; high bit\\
\hline
\end{tabular}
\caption{HMI Level-1 Quality Summary}
\label{table:lev1_quality}
\end{table}
\section{HMI Observables \textsf{QUALITY}-Keyword Summary}
\label{sec:S_ObsQual}

The keyword \textsf{QUALITY} is provided for each HMI data product at each processing level. 
Observables are constructed from many filtergrams, so \textsf{QUALITY} may depend on information 
from any of the contributing images.

Table \ref{table:Obs_quality} describes \textsf{QUALITY}-bits for the 45-second
Line-of-Sight (LoS) observables that indicate why a data record is missing. The
value of \textsf{QUALITY} will be negative (top bit set) if data are missing
for one of the reasons specified in certain other bits. LoS observables include
45-second Dopplergrams, line-of-sight magnetic field, line depth, and continuum
intensity. Filtergrams at multiple wavelengths and polarization states from
three consecutive 45-second intervals contribute to each data record. 
The 45-second observables depend on filtergrams only from Camera 2.

Data records with no known issues have \textsf{QUALITY}\,=\,0.  Bit\,0 is the
low bit (0x01).

Table \ref{table:Obs_quality2} provides similar information for the bits in
the vector-observable \textsf{QUALITY} keyword. The primary observable is the
Stokes data series \textsf{hmi.S\_720s}.  The temporal interpolation is much longer than
for the 45-second data products, including data collected over a nearly 20-minute
time span from one or both HMI cameras.  The other HMI 720-second observables are
computed from the Stokes observable, so share the same keyword.

Table \ref{table:Obs_quality3} provides information for the additional bits in
the vector-observable \textsf{QUALITY} keyword for data that may be acceptable
in certain circumstances, but may be of lesser quality due to inclusion of
fewer filtergrams or noisier data.

\begin{table}[htb]
\scalebox{0.95}{
\begin{tabular}{llp{65 mm}}
\textsf{QUALITY}-bit Name         & Bit Mask    & Description\\ 
\hline
\textsf{QUAL\_NODATA} & 0x80000000 & 
  No l.o.s. observables image was produced (empty record created, 
  with NO DATA SEGMENT. Most keywords have default value) \\
\textsf{QUAL\_TARGETFILTERGRAMMISSING} & 0x40000000 & 
  No filtergram found near target time \\
\textsf{QUAL\_NOINTERPOLATEDKEYWORDS} & 0x20000000 & 
  Could not interpolate required keywords at target time \\
\textsf{QUAL\_NOFRAMELISTINFO} & 0x10000000 & 
  Could not identify observables framelist used \\
\textsf{QUAL\_WRONGCADENCE} & 0x08000000 & 
  Framelist cadence required time does not match the expected value \\
\textsf{QUAL\_WRONGTARGET} & 0x04000000 & 
  Target filtergram does not belong to the current framelist \\
\textsf{QUAL\_MISSINGLEV1D} & 0x02000000 & 
  Not enough lev1d filtergrams to produce observable \\
\textsf{QUAL\_MISSINGKEYWORDLEV1D} & 0x01000000 & 
  Could not read some required keywords in lev1d data \\
\textsf{QUAL\_WRONGWAVELENGTHNUM} & 0x00800000 & 
  Number of wavelengths in the lev1d records is incorrect \\
\textsf{QUAL\_MISSINGKEYWORDLEV1P} & 0x00400000 & 
  Could not read some required keywords in the lev1p data \\
\textsf{QUAL\_NOLOOKUPRECORD} & 0x00200000 & 
  Could not find a record for look-up tables for the MDI-like algorithm \\
\textsf{QUAL\_NOLOOKUPKEYWORD} & 0x00100000 & 
  Could not read keywords of the look-up tables for the MDI-like algorithm \\
\textsf{QUAL\_NOTENOUGHINTERPOLANTS} & 0x00080000 & 
  Not enough interpolation points for the temporal interpolation at a given wavelength and polarization \\
\textsf{QUAL\_INTERPOLATIONFAILED} & 0x00040000 & 
  Temporal interpolation failed (no lev1d record was produced) \\
\textsf{QUAL\_MISSINGLEV1P} & 0x00020000 & 
  Not enough lev1p records to produce an observable \\
\textsf{QUAL\_NOCOEFFKEYWORD} & 0x00000200 & 
  Could not read keywords of the polynomial coefficient series for the correction of the MDI-like algorithm \\
\textsf{QUAL\_NOCOEFFPRECORD} & 0x00000080 & 
  Could not find a record for the polynomial coefficient
for the correction of the MDI-like algorithm, or could not access the keywords of a specific record \\
\hline
\end{tabular}}
\caption{HMI Line-of-Sight Observable Processing-Failure Quality-Bit Summary}
\label{table:Obs_quality}
\end{table}

\begin{table}[htb]
\begin{tabular}{llp{65 mm}}
\textsf{QUALITY}-bit Name         & Bit Mask    & Description\\ 
\hline
\textsf{QUAL\_NODATA} & 0x80000000 & 
  Not all \textit{I\,Q\,U\,V} filtergrams produced (Some or all data segments missing) \\
\textsf{QUAL\_TARGETFILTERGRAMMISSING} & 0x40000000 & 
 No target filtergram found near target time \\
\textsf{QUAL\_NOINTERPOLATEDKEYWORDS} & 0x20000000 & 
  Could not interpolate some required keywords at target time \\
\textsf{QUAL\_NOFRAMELISTINFO} & 0x10000000 & 
  Could not recognize observables framelist \\
\textsf{QUAL\_WRONGCADENCE} & 0x08000000 & 
 Cadence corresponding to the framelist does not match the expected value provided by user \\
\textsf{QUAL\_WRONGFRAMELISTSIZE} & 0x04000000 & 
  Current framelist size does not match value from the command line \\
\textsf{QUAL\_WRONGNPOL} & 0x02000000 & 
  Current framelist npol does not match value from the command line \\
\textsf{QUAL\_WRONGPOLTYPE} & 0x01000000 & 
  Current framelist does not allow for the production of \textit{I\,Q\,U\,V} \\
\textsf{QUAL\_WRONGTARGET} & 0x00800000 & 
  Target filtergram does not belong to the current framelist \\
\textsf{QUAL\_ERRORFRAMELIST} & 0x00400000 & 
  Filtergrams not where they should be in the framelist \\
\textsf{QUAL\_WRONGWAVELENGTHNUM} & 0x00200000 & 
  Number of wavelengths in the lev1d records is not correct \\
\textsf{QUAL\_NOLOOKUPRECORD} & 0x00100000 & 
  Could not find record for look-up tables for MDI-like algorithm (currently unused) \\
\textsf{QUAL\_NOLOOKUPKEYWORD} & 0x00080000 & 
  Could not read keywords of look-up tables for MDI-like algorithm (currently unused) \\
\textsf{QUAL\_NOTENOUGHINTERPOLANTS} & 0x00040000 & 
  Not enough points for the temporal interpolation at a given wavelength and polarization \\
\textsf{QUAL\_INTERPOLATIONFAILED} & 0x00020000 & temporal interpolation routine failed \\
\hline
\end{tabular}
\caption{HMI Stokes \textit{I\,Q\,U\,V} Observable Processing-Failure \textsf{QUALITY}-bit Summary (any cadence)}
\label{table:Obs_quality2}
\end{table}

\begin{table}[htb]
\begin{tabular}{llp{75 mm}}
\textsf{QUALITY}-bit Name         & Bit Mask    & Description\\ 
\hline
\textsf{QUAL\_LOWINTERPNUM} & 0x00010000 & 
  Too few averaging points or two interpolation points separated by more than the cadence \\
\textsf{QUAL\_LOWKEYWORDNUM} & 0x00008000 &
  Some keywords (\textit{e.g.} \textsf{CROTA2, DSUN\_OBS,} and \textsf{CRLT\_OBS}) could not be interpolated 
  properly, closest-neighbor approximation used \\
\textsf{QUAL\_ISSTARGET} & 0x00004000 & 
  ISS Loop open for one or several filtergrams used to produce observable \\
\textsf{QUAL\_NOTEMP} & 0x00002000 & 
  Cannot read temperatures needed for polarization calibration (default temperature used) \\
\textsf{QUAL\_NOGAPFILL} & 0x00001000 & 
  Code could not properly gap-fill all Lev-1 filtergrams \\
\textsf{QUAL\_LIMBFITISSUE} & 0x00000800 & 
  Some Lev-1 records discarded because \textsf{R\_SUN, CRPIX1}, or \textsf{CRPIX2} were missing or too different from median \\
\textsf{QUAL\_NOCOSMICRAY} & 0x00000400 & 
  Some cosmic-ray hit lists could not be read for level 1 filtergrams \\
\textsf{QUAL\_ECLIPSE} & 0x00000200 & 
  At least one lev1 record taken during an eclipse \\
\textsf{QUAL\_LARGEFTSID} & 0x00000100 & 
  \textsf{HFTSACID} of target filtergram $> 4000$, which adds noise to observable \\
\textsf{QUAL\_TEMPERROR} & 0x00000080 & 
  Code error discovered that will be corrected in later processing versions, 
  see notes at {\urlurl{jsoc2.stanford.edu/doc/data/hmi/Quality\_Bits}} \\
\textsf{QUAL\_POORQUALITY} & 0x00000020 & 
  WARNING poor quality: be careful when using due to eclipse, transit, thermal recovery, open ISS, or other... \\
\end{tabular}
\caption{HMI Stokes \textit{I\,Q\,U\,V}-Observable Poor-\textsf{QUALITY} Bits}
\label{table:Obs_quality3}
\end{table}

\end{article}
\end{document}